\documentclass{article}

\usepackage{PRIMEarxiv}

\usepackage[utf8]{inputenc} 
\usepackage[T1]{fontenc}    
\usepackage{hyperref}       
\usepackage{url}            
\usepackage{booktabs}       
\usepackage{amsfonts}       
\usepackage{nicefrac}       
\usepackage{microtype}      
\usepackage{lipsum}
\usepackage{fancyhdr}       
\usepackage{graphicx}       
\graphicspath{{media/}}     

\usepackage{amsmath}
\usepackage{colortbl}
\usepackage{graphicx}
\usepackage{color}
\usepackage{amsmath,amssymb,amsfonts,amsthm}
\usepackage{graphicx,xcolor}  
\usepackage{algorithm}
\usepackage{algpseudocode}
\usepackage{textcomp}
\usepackage{relsize}
\usepackage{subcaption}
\usepackage{xstring} 
\usepackage{simplekv} 
\usepackage{soul} 
\usepackage{sectsty}
\newtheorem{theorem}{Theorem}[section]
\newtheorem{prop}[theorem]{Proposition} 
\newtheorem{corollary}{Corollary}[theorem]
\newtheorem{lemma}[theorem]{Lemma}
\theoremstyle{definition}

\usepackage{hyperref} 
\hypersetup{
  colorlinks   = true, 
  urlcolor     = blue,
  linkcolor    = red, 
  citecolor   = blue 
}
\usepackage{natbib}
\setcitestyle{authoryear,open={(},close={)}}
\usepackage[english]{babel}
\usepackage[autolanguage]{numprint}
\newcommand{\m}[1]{\boldsymbol{#1}}
\newcommand{\widesim}[2][1.5]{
  \mathrel{\overset{#2}{\scalebox{#1}[1]{$\sim$}}}
}
\usepackage[normalem]{ulem}

\usepackage{cancel}
\usepackage{soul}

\title{Sparse inference in Poisson log-normal model by approximating the $L_0$-norm
}

\author{
  Togo Jean Yves KIOYE,  \\
  Unité Mixte de Recherche sur le Fromage (UMRF) \\
  Université Clermont Auvergne, France  \\
  \texttt{togo\_jean\_yves.kioye@uca.fr} \\
   \And
  Paul-Marie GROLLEMUND \\
  Laboratoire de Mathématiques Blaise Pascal (LMBP)\\
  Unité Mixte de Recherche sur le Fromage (UMRF) \\
  Université Clermont Auvergne, France  \\
  \texttt{paul\_marie.grollemund@uca.fr} \\
   \And
  Jocelyn CHAUVET \\
  Laboratoire Angevin de Recherche en Ingénierie des Systèmes (LARIS) \\
  Centre de recherche de l'ICES, France \\
  \texttt{jchauvet@ices.fr} \\
   \And
  Pierre DRUILHET, Erwan SAINT-LOUBERT-BIE  \\
  Laboratoire de Mathématiques Blaise Pascal (LMBP)\\
  Université Clermont Auvergne, France  \\
  \texttt{\{pierre.druilhet,erwan.saint-loubert-bie\}@uca.fr} \\
   \And
  Christophe CHASSARD \\
  Unité Mixte de Recherche sur le Fromage (UMRF) \\
  INRAE, France  \\
  \texttt{christophe.chassard@inrae.fr} \\
}

\begin{document}
\maketitle

\begin{abstract}
Variable selection methods are required in practical statistical modeling,  to identify and include only the most relevant predictors, and then improving model interpretability.
Such variable selection methods are typically employed in regression models, for instance in this article for the Poisson Log Normal model (PLN, \citealp{chiquet2021poisson}). 
This model aim to explain multivariate count data using dependent variables, and its utility was demonstrating in scientific fields such as ecology and agronomy.
In the case of the PLN model, most recent papers focus on sparse networks inference through combination of the likelihood with a $L_1$-penalty on the precision matrix. 
In this paper, we propose to rely on a recent penalization method (SIC, \citealp{o2023variable}), which consists in smoothly approximating the $L_0$-penalty, and that avoids the calibration of a tuning parameter with a cross-validation procedure.
Moreover, this work focuses on the coefficient matrix of the PLN model and establishes an inference procedure ensuring effective variable selection performance,  so that the resulting fitted model explaining multivariate count data using only relevant explanatory variables.
Our proposal involves implementing a procedure that integrates the SIC penalization algorithm ($\varepsilon$-telescoping) and the PLN model fitting algorithm (a variational EM algorithm). 
To support our proposal, we provide theoretical results and insights about the penalization method, 
and we perform simulation studies to assess the method, which is also applied on real datasets.
\end{abstract}


\section{Introduction} 
\label{sect-introduction}

Multivariate count data analysis is an active component of modern statistical modeling, providing insights across various fields such as among others ecology, microbiology, and accident crash frequency.
Unlike univariate count data analysis, multivariate count data analysis deals with multiple count variables simultaneously, capturing dependent aspects of the studied phenomena. 
Examples include studying the simultaneous abundance of various species in ecological communities, modeling the occurrence of different diseases in a population \citep{zhang2017regression}, or the relative abundances of microorganisms in a specific ecosystem \citep{chiquet2018variational}.
Analyzing multivariate count data presents challenges, including handling dependencies among count variables, addressing over-dispersion, and accommodating complex correlation structures. 
In recent years, advanced statistical methods such as multivariate Poisson regression (PLN, \citealp{chiquet2021poisson}), multivariate negative binomial regression \citep{shi2014multivariate}, and copula-based negative binomial models \citep{nikoloulopoulos2009modeling} have been developed to tackle these challenges.

In this paper, we focus on the PLN model, which offers versatility for considering additional extensions \citep{chiquet2021poisson}, has demonstrated utility across various domains \citep{thierry2023microbial,grollemund2023permanova}, and continues to be studied computationally \citep{stoehr2024composite}.
This model involves a latent layer, complicating inference, particularly computationally. Therefore, its implementation relies on variational approximation and the model is fitted using the Variational Expectation-Maximization algorithm (VEM), according to the \textsf{R} package \texttt{PLNmodels}.
Nevertheless, it is practically feasible to fit this type of model using a Monte Carlo approach \citep{chib2001markov,karlis2005mixed,stoehr2024composite}, or  Laplace's method to approximate the likelihood and to derive penalized maximum likelihood estimates \citep{choi2017poisson}.

Using the PLN model enables the study of network structures through estimated coefficients of the precision matrix in practical contexts. It also allows for investigating the impact of dependent variables on count data by estimating regression coefficients. In such cases, it is natural to design an estimation procedure coupled with a variable selection method to ensure that the fitted model retains only coefficients notably different from 0. 
This facilitates the interpretation of model coefficients, particularly when the primary interest is discerning dependent variables explaining heterogeneity in count data.
For the PLN model, a variable selection procedure can be defined solely on the precision matrix coefficients, focusing on the major connections in a network analysis \citep{choi2017poisson,chiquet2019variational}. 
Alternatively, a variable selection procedure can target only the regression coefficients, particularly when the precision matrix is not the coefficient of interest for addressing the specific issues of the field of application. 
Furthermore, it is also possible to penalize both the precision matrix and the regression coefficients, with the thought that the simplicity of the precision matrix can help achieve good performance in estimating regression coefficients \citep{wu2018sparse,rothman2010sparse,sohn2012joint}.
In this paper, we focus on penalizing regression coefficients to interpret the selected coefficients, while most recent works emphasize sparse estimation of the precision matrix for network analysis.


To perform variable selection, the aim is to balance model fitting and the intensity of the $L_0$-norm of coefficient parameters. Accepting a loss in model fitting can help to minimize the $L_0$-norm, effectively turning off some coefficients. 
Thus, the aim is to adjust the model while minimizing this quantity, yet this leads to a mathematically or computationally complex optimization objective. To address this, a standard approach is to employ $L_1$ penalization (or its variants), which relaxes the optimization problem \citep{tibshirani1996regression}.
Note that these methods typically involve a tuning parameter that drives the trade-off between model fitting and variable selection intensity. To numerically compute this tuning parameter, it is required to rely on an ad-hoc procedure, leading to computational overload that can be significant.

Recently, \cite{o2023variable} introduced a novel penalization approach aiming at approximating the $L_0$-norm to achieve effective variable selection performance without requiring tuning parameter calibration through procedures like cross-validation. This approach is developed within the framework of a distributional regression model, enabling simultaneous assessment of covariate effects on both the mean and variance of the response variable. In this work, we propose to adapt this approach to the context of a PLN model, which is a generalized linear model whose inference is conducted using an EM algorithm and a variational approximation. 
In the following sections, we first recall the PLN model (see Section~\ref{sect-PLN}) and the variable selection procedure (see Section~\ref{sect-SIC}), and we provide additional theoretical insights and interpretation. 
Next, the main methodological contribution is presented in Section~\ref{sect-estimation}, in which we detail how to combine the variable selection algorithm proposed by \cite{o2023variable} ($\varepsilon$-telescoping Algorithm) with the PLN model inference proposed by \cite{chiquet2021poisson} (a VEM Algorithm). 
Then, the developed methodology is evaluated using real and simulated data in Sections~\ref{sect-simulation} and \ref{sect-real_data}.

\section{Multivariate Poisson Log-Normal (PLN) Model}

\label{sect-PLN}


\subsection{Model formulation}
The multivariate PLN model was introduced by \citet{aitchison1989multivariate} with the aim of modelling joint species count data using environmental factors, taking into account the dependency structure between species. 
In contrast to other multivariate Poisson models, the PLN model 
can handle data with a completely general correlation structure as well as overdispersion \citep{park2007multivariate}. 
Each observed count follows a Poisson distribution whose 
log parameter, 
defined on a latent layer, is distributed according to a multivariate Gaussian distribution.
The expectation of this Gaussian distribution takes account of dependent variables, while its variance-covariance matrix characterizes the count data correlation structure. 
Let $n$ be the number of observations (or sites) assumed to be independent, 
and $p$ the number of features (or species) counted for each observation. 
We denote by $\m{Y}\in \mathbb{N}^{n\times p}$ the count matrix 
--- where $Y_{ij}$ is the number of occurrences of the $j^{th}$ feature observed for the $i^{th}$ observation --- 
and by $\m{X} \in \mathbb{R}^{n\times d}$ the matrix of dependent variables. 
The fixed effect $\mu_{ij}$ can be decomposed as
$\mu_{ij} = \m{x}_i^\top \m{b}_j$, where $\m{x}_i \in\mathbb{R}^d$ is the vector of covariates for individual $i$ and $\m{b}_j$ is the vector of regression coefficients associated with feature $j$. 
Finally, $\m{O}\in\mathbb{R}^{n\times p}$ denotes the offset matrix, where the element $o_{ij}$ is the offset of individual $i$ and feature $j$, which represent the sampling effort varying from one individual to another and from one feature to another. The PLN model is then written as follows \citep{chiquet2021poisson}: 
\begin{alignat}{2}
\label{eq:PLN:modele:modele_hierarchique}
Y_{ij}\mid Z_{ij} 
    &\,\widesim{}\, \mathcal{P} \big( \exp(Z_{ij}) \big) 
    && \qquad \text{(observed space)} \\
\m{Z}_{i}         
    &\,\widesim{}\, N_{p} \big( \m{o}_{i}+\m{\mu}_{i}, \m{\Sigma} \big) 
    && \qquad \text{(latent space),} \notag
\end{alignat}
where $\m{o}_{i} = (o_{i1}, \ldots, o_{ip})^\top$ and $\m{\mu}_{i} = (\mu_{i1}, \ldots, \mu_{ip})^\top$.
In the following, $\m{\theta} = (\m{B}, \m{\Sigma})$ denotes the set of model parameters, where $\m{B}$ is the $d \times p$ matrix of regression coefficients.

\subsection{Parameters estimation}

As the PLN model is a latent variable model, computing its log-likelihood is not straightforward and the use of the EM algorithm is required to estimate the parameters. However, to fit a PLN model, the EM algorithm assumes knowledge of the conditional distributions $p_{\m{\theta}}(\m{Z}_i \mid \m{Y}_i)$ for each $i \in \left\lbrace 1, \ldots, n \right\rbrace$, which are intractable. 
Variational approximations have therefore been developed to circumvent this problem \citep{blei2017variational}. 
The combination of variational approaches and the EM algorithm has led to the {variational EM} \cite[VEM,][]{tzikas2008variational}, which maximises a lower bound on the $\log$-likelihood of the PLN model. 
The VEM algorithm can be seen as minimizing the Kullback-Leibler (KL) divergence between 
$p_{\m{\theta}}(\m{Z}_i \mid \m{Y}_i)$ and an approximate distribution $q_i$ chosen from a class of simple distributions.
In most cases, $q_i$ is assumed to be a multivariate Gaussian distribution with expectation $\m{m}_i$ 
and variance-covariance matrix $\m{S}_i = \text{diag}(\m{s}_i^2)$ \citep{hall2011theory}. 
All the variational parameters are gathered in $\m{\psi}=(\m{M},\m{S})$, where $\m{M} = (\m{m}_1^\top, \ldots, \m{m}_n^\top)^\top$ and $\m{S} = ({\m{s}_1^2}^\top, \ldots, {\m{s}_n^2}^\top)^\top$. 
The evidence lower bound (ELBO) of the log-likelihood which is maximized when implementing the VEM algorithm takes the general form \citep{blei2017variational}:
\begin{equation}
    J(\m{Y} ; \m{\theta}, \m{\psi}) 
   =
    \log(p_{\m{\theta}}(\m{Y})) - 
    \text{KL} \Big[ q_{\m{\psi}}(\m{Z}) \;||\; p_{\m{\theta}}(\m{Z} \mid \m{Y}) \Big].
    \label{eq:PLN:estimation:elbo}
\end{equation}
In the case of the PLN model, the ELBO can be expressed as \citep{chiquet2021poisson}: 
\begin{align}
    J(\m{Y} ; \m{\theta}, \m{\psi}) \; 
    =
    \;&\; 
    \m{I}_{n} \Big[ \m{Y} \odot (\m{O} + \m{M})- \m{A} + \frac{1}{2} \log(\m{S}^2) \Big] \m{I}_{p}+
    \frac{n}{2}\log |\m{\Omega}| \notag \\
    \;&\;   
    - \frac{n}{2} 
    \text{trace} \Big( \m{\Omega} 
    \Big[ (\m{M}-\m{XB})^\top(\m{M}-\m{XB})+\text{diag}(\m{I}_{n}^\top \m{S}^2) \Big] \Big)
    + \text{const,} 
    \label{loglikpln}
\end{align}
where $\odot$ is the Hadamard product and $\m{S}^2= \m{S}\odot\m{S}$, 
$\m{\Omega} = \m{\Sigma}^{-1}$ is the precision matrix and $\left\lvert \m{\Omega} \right\rvert$ its determinant, 
and $\m{A}$ is the $n \times p$ matrix of expected count with entries 
$a_{ij} = \exp(o_{ij} + \mu_{ij} + m_{ij} + s_{ij}^2/2)$, 
with $m_{ij}$ the $j^\text{th}$ element of $\m{m}_i$ and $s_{ij}^2$ the $j^\text{th}$ element of $\m{s}_i^2$. 

Typically, all the entries of parameters estimations ${B}_{kj}$ of $\m{B}$ and $\omega_{ij}$ of $\m{\Omega}$ from $\eqref{loglikpln}$ will be non-zero. This will make model interpretation challenging if $p$ and $d$ are large. It is generally assumed that there is a "true model", characterized by a limited number of non-zero parameters ${B}_{kj}$ and $\omega_{ij}$, facilitating interpretation and alleviating overfitting problems. To achieve this, it is necessary to impose constraints or regularization on the estimation process \citep{fan2001variable,rothman2010sparse,hastie2015statistical}.

\subsection{Sparse inference in Poisson log-normal model}
\label{sect-PLN-sparse}

Most studies on variable selection in the context of PLN models have focused on identifying dependency structures in the precision matrix. The main difference between these studies is the method used to compute the likelihood, 
but notice that they involve an $L_1$ penalty as a constraint on the log-likelihood.


\noindent {\bf Sparsity of the precision matrix.}
\citet{choi2017poisson} used the Laplace method to approximate the PLN log-likelihood, and imposed a constraint on the coefficients of precision matrix $\m{\Omega}$. They suggest that the penalized log-likelihood is:
\begin{align*}
    \ell_\text{pen}(\m{\mu}, \m{\Omega})
    = 
    \frac{1}{n} \sum_{i=1}^{n} \ell_i(\m{\mu},\m{\Omega}) 
    - \lambda \sum_{j\ne k}^p p_{jk}|\omega_{jk}|,
\end{align*}
where 
$\ell_i$ is the log-likelihood for observation $i$ 
and $\m{\mu}$ is a vector representing the mean parameter of a multivariate Gaussian distribution. 
The regularization is controlled by parameter $\lambda>0$ and weights $p_{jk}$, the latter allowing differential amounts of penalty on the entries of $\m{\Omega}$ based on prior knowledge or anticipated network properties.
Next, \citet{chiquet2019variational} focused on network inference and applied variational approximation methods to maximize a penalized lower bound of the log-likelihood in order to identify direct interactions among features. 
The corresponding objective function they suggest to maximize is defined as \citep{chiquet2018variational}:
\begin{align*}
    J_{\text{struct}}(\m{Y} ; \m{\theta}, \m{\psi})
    =
    J(\m{Y} ; \m{\theta}, \m{\psi})
    - \lambda||\m{\Omega}||_{1,\text{off}},
\label{loglikplnpenastruct}
\end{align*}
where $J(\m{Y} ; \m{\theta}, \m{\psi})$ is the ELBO \eqref{loglikpln}, 
$||\m{\Omega}||_{1,\text{off}}=\sum_{j \ne j'}|\omega_{jj'}|$ 
is the off-diagonal $L_1$-norm of $\m{\Omega}$
and $\lambda>0$ is a tuning parameter controlling the amount of sparsity.


\noindent {\bf Sparsity of the coefficient matrix and the precision matrix.}
\citet{wu2018sparse} proposed a Monte Carlo EM algorithm for parameter estimation in the PLN model. To induce sparsity in the parameters, they imposed an $L_1$ constraint on both matrixes $\m{B}$ and $\m{\Omega}$.
The penalized log-likelihood they maximize is defined as:
\begin{align*}
    \ell_\text{pen}(\m{B},\m{\Omega})
    =
    \ell(\m{B},\m{\Omega})
    - \lambda_1 ||\m{B}||_1
    - \lambda_2||\m{\Omega}||_1,
\end{align*}
where $\ell(\m{B},\m{\Omega})$ is the log-likelihood of the PLN model, $||\cdot||_1$ denote the $L_1$ matrix norm and $\lambda_1 >0$, $\lambda_2>0$ are two tuning parameters.
Lastly, to our knowledge, there are no existing studies addressing sparse inference solely for the coefficient matrix $\m{B}$ in the case of the PLN model relying on variational approximation.


\noindent {\bf A remark on predicting on a validation dataset.}
In the aforementioned references, the best values of $\lambda$, $\lambda_1$ and $\lambda_2$ are selected by
maximizing a pseudo-information criterion, either the BIC \citep{schwarz1978estimating} 
or its high-dimensional-adapted extension known as EBIC \citep{chen2008extended}, 
in which the true log-likelihood is replaced by an approximation.
In standard practice, sparsity can also be reached thanks to a penalization method, for which it is essential to properly calibrate parameter $\lambda$, as it determines the trade-off between goodness of fit and intensity of penalization. 
Typically, $\lambda$ is calibrated through cross-validation, by using a predefined sequence of values and assessing prediction performance on test data not used in the learning process.
To predict count data values with PLN model, two formulas are available:
\begin{align*}
    \m{\hat{Y}} & = \exp(\m{O} + \m{XB} + \m{M} + \m{S}^2/2) \tag{Variational PLN prediction} \\
    \m{\hat{Y}} & = \exp(\m{O} + \m{XB} + \m{\Sigma}^{\frac{1}{2}}/2) \tag{PLN prediction}    
\end{align*}
The first one involves variational parameters, while the other depends solely on PLN model parameters. 
It should be noted that the variational parameters are individual-specific, and to estimate the variational parameters of the $i$-th individual, count data and covariates concerning the $i$-th individual are required.
Therefore, the formula involving variational parameters cannot be used to predict new data, 
which are not present in the training dataset.
Additionally, an empirical study (see Section~\ref{sect-simulation-comparison_prediction_formulae} and Figure~\ref{fig-simulation-comparison_prediction_formulae}) reveals that prediction performance of the formula without variational parameters is relatively weak compared to the other formula. 
Consequently, robust calibration of tuning parameter $\lambda$ may not be achieved through cross-validation procedure. This limitation is an obstacle to the use of a penalization method requiring calibration of a tuning parameter, such as LASSO estimation.


\section{Variable selection with SIC}
\label{sect-SIC}

Numerous variable selection methods have already been developed; 
we refer the reader to \citet{heinze2018variable} for a recent review of the literature. 
Standard methods consist in regularizing the regression coefficients 
(for example, the LASSO \citep{tibshirani1996regression} and its extensions \citep{zou2006adaptive}), 
or to optimize an information criterion (such as AIC \citep{akaike1974new} or BIC \citep{schwarz1978estimating}).
Unfortunately, in practice, these methods can be computationally intensive: 
LASSO-type methods require the intensity of the regularisation to be calibrated by cross-validation, 
while AIC/BIC-type methods require a very large number of models to be fitted. 
Recently, a penalization method called Smooth Information Criterion 
\cite[SIC,][]{o2023variable} was introduced in order to achieve variable selection using a numerical procedure with reduced computational cost. 
It is based on an approximation of the $L_0$-norm of the regression coefficients, so that 
it offers a relaxed version of BIC.
It is expressed as a penalized $\log$-likelihood $\ell_{\text{SIC}}$ by:
\begin{equation}
    \ell_{\text{SIC}}(\m{\vartheta}) 
    = 
    \ell(\m{\vartheta}) 
    - 
    \frac{\log(n)}{2} 
    \Big[ \lVert \widetilde{\m{\vartheta}} \rVert _{0,\varepsilon} + k \Big],
	 \label{eq:selection_variables:sic_regression_lineaire:SIC}
\end{equation} 
where $\ell(\m{\vartheta})$ denotes a log-likelihood depending on parameter $\m{\vartheta}$, 
$\widetilde{\m{\vartheta}}$ is a sub-vector of $\m{\vartheta}$ corresponding to the parameters to be regularized,
$\lVert \cdot \rVert _{0,\varepsilon}$ is an approximation of the $L_0$-norm (see section \ref{sect-SIC-formulations}) and $k$ is the number of the unpenalized parameters.

This approach simplifies the model selection process by avoiding to estimate the model coefficients for different values of a tuning parameter $\lambda$, or to fit many models for evaluating each of them with regards to an information criterion \citep{o2023variable}. 
In other words, this method enables variables to be selected in a single step, eliminating the need to calibrate a regularization parameter by cross-validation. 
This new variable selection method seems  advantageous in the case of the PLN model, since a cross-validation procedure is challenging to implement in this context.
However, the proposed method do not avoid the use of an iterative algorithm to fit the model, although the procedure is numerically less costly.
The specific iterative algorithm called $\varepsilon$-telescoping  involves varying the form of the penalty, to smoothly approximate the $L_0$-norm in order to ensure robust results 
(see Section \ref{sect-SIC-formulations-optimization} for a few details or \cite{o2023variable} for a complete description).

\subsection{Formulations and interpretations}
\label{sect-SIC-formulations}

\subsubsection{Optimization framework}
\label{sect-SIC-formulations-optimization}

In order to provide valuable insights in Section~\ref{sect-SIC-formulations-geometry} and  Section~\ref{sect-SIC-formulations-bayes}, we firstly present below the SIC approach in the context of a Multiple Linear Regression model:
\begin{equation}
    y_{i} = \beta_0 + \m{x}_{i}^{\top}\m{\beta} + \eta_{i}, 
    \quad 
    i = 1, \ldots ,n,
    \label{eq:selection_variables:sic_regression_lineaire:equation_modele}
\end{equation}
where $y_i$ is the response for individual $i$, 
$\m{x}_i = (1,x_{i1},\ldots,x_{id})^\top$ the vector of $d$ covariates, 
and $\m{\beta} = (\beta_1,\ldots,\beta_d)^\top$ the vector of regression coefficients. 
In the following, we consider that the covariates are centered and reduced and we assume that $\eta_i \sim N(0,\sigma^2)$, 
where the variance $\sigma^2$ is known.
In this context, the SIC procedure targets the solution 
to the optimization problem: 
\begin{equation}
    \underset{\m{\beta}\,\in\,\mathbb{R}^{d}} {\text{min}} 
    \left\{  \sum_{i=1}^{n} 
    \left(y_i-\beta_0-\m{x}_{i}^{\top}\m{\beta}\right)^2 
    \right\}
    \quad
    \text{subject to} \; \lVert \m{\beta} \rVert_{0,\varepsilon} \leqslant t, 
\label{SIC:contraint}
\end{equation}
where $\lVert \m{\beta} \rVert_{0,\varepsilon} = \sum_{j=1}^{d}\phi_{\varepsilon}(\beta_j)$, with
\begin{equation}
\label{eq:Variable_selection_with_SIC:Optimization_framework:sicnorm}
\phi_{\varepsilon}(x)= \frac{x^2}{x^2+\varepsilon^2}\cdot
\end{equation}
Note that $\lVert \cdot \rVert _{0,\varepsilon}$ 
can be seen as a smooth approximation of the $L_0$-norm since 
$\lim_{\varepsilon \to 0} \phi_{\varepsilon}(\m{\beta}) = \lVert \m{\beta} \rVert _{0}$.
Moreover, the constrained minimization problem \eqref{SIC:contraint} 
can be written in the so-called Lagrangian or penalized form: 
\begin{equation}
    \underset{\m{\beta}\,\in\,\mathbb{R}^{d}} {\text{min}} 
    \left\{ 
     \sum_{i=1}^{n} 
    \left( y_i-\beta_0-\m{x}_{i}^{\top}\m{\beta}\right)^2 
    + \lambda \, \lVert \m{\beta} \rVert_{0,\varepsilon}
    \right\}
    \label{SIC:lagrange}
\end{equation}
where the shrinkage parameter $\lambda \geqslant 0$ in (\ref{SIC:lagrange}) 
is in bijection with the radius $t$ in (\ref{SIC:contraint}).

For a Generalized Linear Model, statistical inference relies usually on maximizing the likelihood function. 
To obtain sparse parameters in this context, the penalized maximum likelihood method can be used \citep{fan2001variable,tibshirani1996regression} and we denote 
the $\log$-likelihood of Model \eqref{eq:selection_variables:sic_regression_lineaire:equation_modele} by $\ell(\m{\beta})$. 
Then the penalized log-likelihood involving SIC can be written as: 
\begin{align*}
    \ell_{\text{SIC}}(\m{\beta}) 
    = 
    \ell(\m{\beta}) 
    - \frac{\lambda}{2} \Big[ \lVert \m{\beta} \rVert _{0,\varepsilon} + k \Big] 
    =
    -\frac{n}{2} \log \left( 2 \pi \sigma^2 \right) - 
    \frac{1}{2 \sigma^2} \sum_{i=1}^{n} \left( y_i-\beta_0-\m{x}_{i}^{\top}\m{\beta} \right)^2
    - 
    \frac{\lambda}{2} \Big[ \lVert \m{\beta} \rVert _{0,\varepsilon} + k \Big],
\label{eq:selection_variables:sic_regression_lineaire:SIC:loglik_multi}
\end{align*} 
where $k$ is the number of unpenalized parameters (here, $k=1$) 
and $\lambda$ is fixed at $\lambda=2$ or $\lambda= \log(n)$ for AIC and BIC respectively. 
The parameters maximizing the penalized log-likelihood are obtained using an iterative algorithm 
such as Newton-Raphson or Fisher scoring.

The Smooth Information Criterion depends on parameter $\varepsilon$, 
which influences the way variables are selected. 
Small values of $\varepsilon$ lead to a better approximation of the $L_0$-norm, 
but induce instability in the estimation process. 
Conversely, large values of $\varepsilon$ provide a stable estimation process, 
but do not allow some coefficients to be set to zero because the level of shrinkage is too low. 
\citet{o2023variable} propose an algorithm, called $\varepsilon$-telescoping, 
which stabilises the variable selection by iteratively decreasing the values of $\varepsilon$ in the optimization problem.
As $\varepsilon$ tends towards 0 in this algorithm, and a small $\varepsilon$ implies that the approximation function $\phi_\varepsilon$ is close to the $L_0$-norm, the estimators are encouraged to be sparse at the end of the procedure.

\subsubsection{Geometry of SIC}
\label{sect-SIC-formulations-geometry}

\begin{figure}[h]
    \centering
    \includegraphics[width=0.49\linewidth]{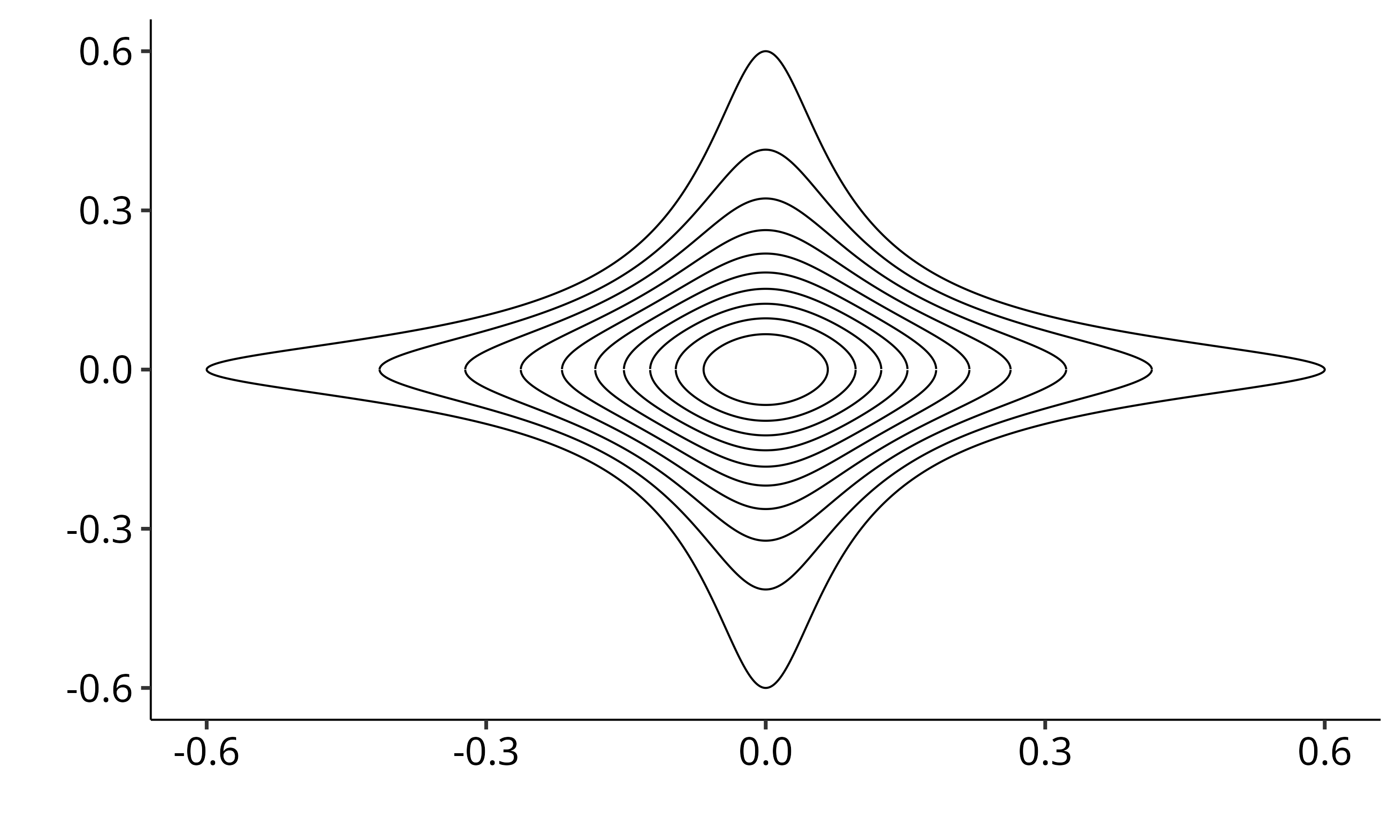}
    \includegraphics[width=0.49\linewidth]{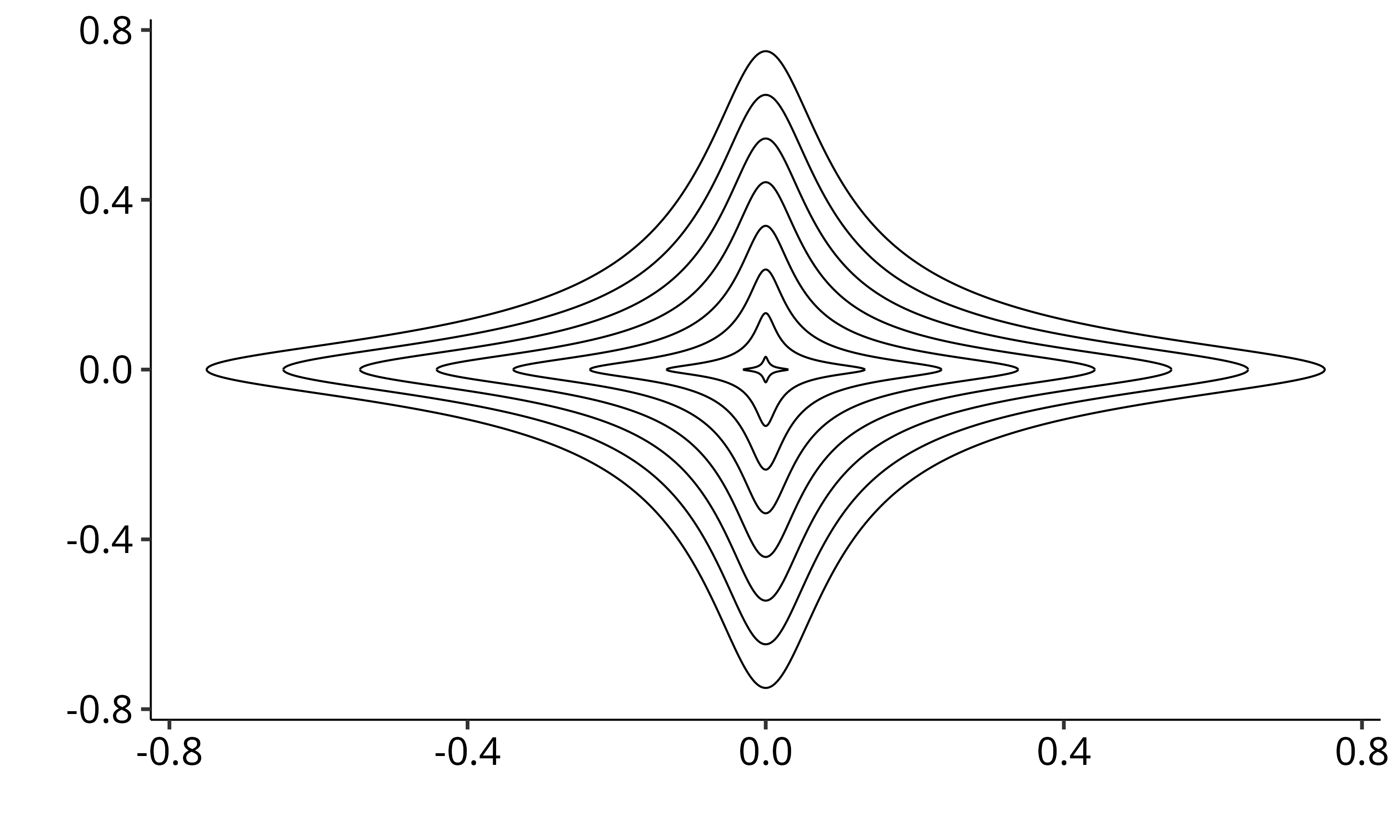}
    \caption{\textsc{Illustration of the ball associated to the $\|.\|_{0,\varepsilon}$-norm.} 
    \textit{Left plot shows balls for different radius varying from 0.9 to 0.1 and right plot considers $\varepsilon$ varying from 0.25 to 0.01.}
    } 
    \label{fig:point_of_view:geometry:balls} 
\end{figure}

In what follows, for simplicity, function $\phi_{\varepsilon}$ is treated as a norm, although it is not the case. 
Indeed, $\phi_{\varepsilon}$ is not a norm since it does not satisfy the property of absolute homogeneity because 
$\phi_{\varepsilon}(sx) \neq |s| \phi_{\varepsilon}(x)$, and in particular : 
$\phi_{\varepsilon}(sx) = \phi_{\varepsilon/s}(x)$.
Notice that "$L_0$-norm" is also a misnomer as it does not respect the property of absolute homogeneity either.
To provide additional insights into the interpretation of SIC, 
we present two plots depicting the balls related this penalty. 
The left plot in Figure~\ref{fig:point_of_view:geometry:balls} shows contour lines of a ball, 
for different radius and for a fixed $\varepsilon$. 
The right plot illustrates contour lines of different balls with same radius as $\varepsilon$ decreases.
Mathematical details are given in Appendix~\ref{sect-appendix-geometry}.

\begin{figure}[h]
    \centering
    \includegraphics[width=0.49\linewidth]{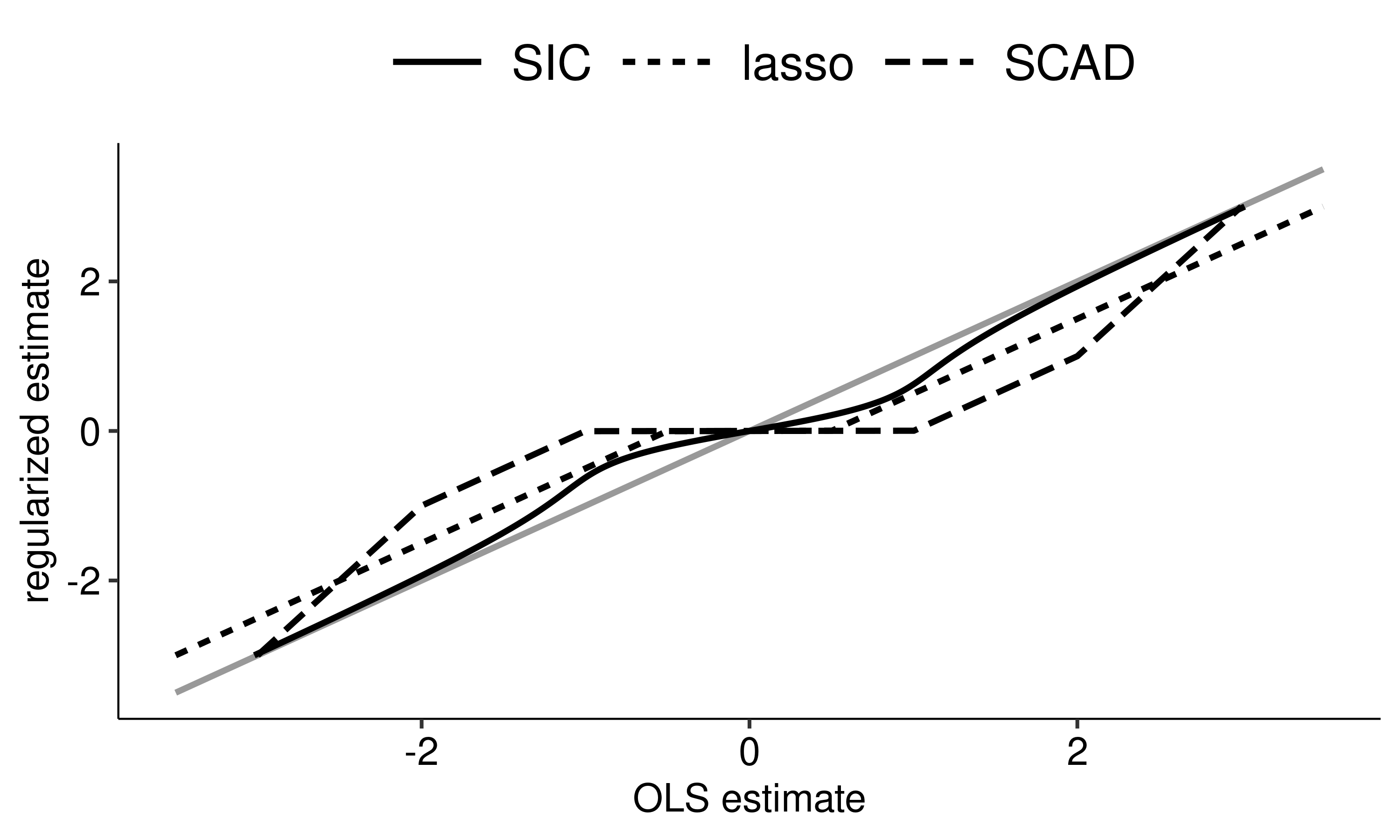}
    \includegraphics[width=0.49\linewidth]{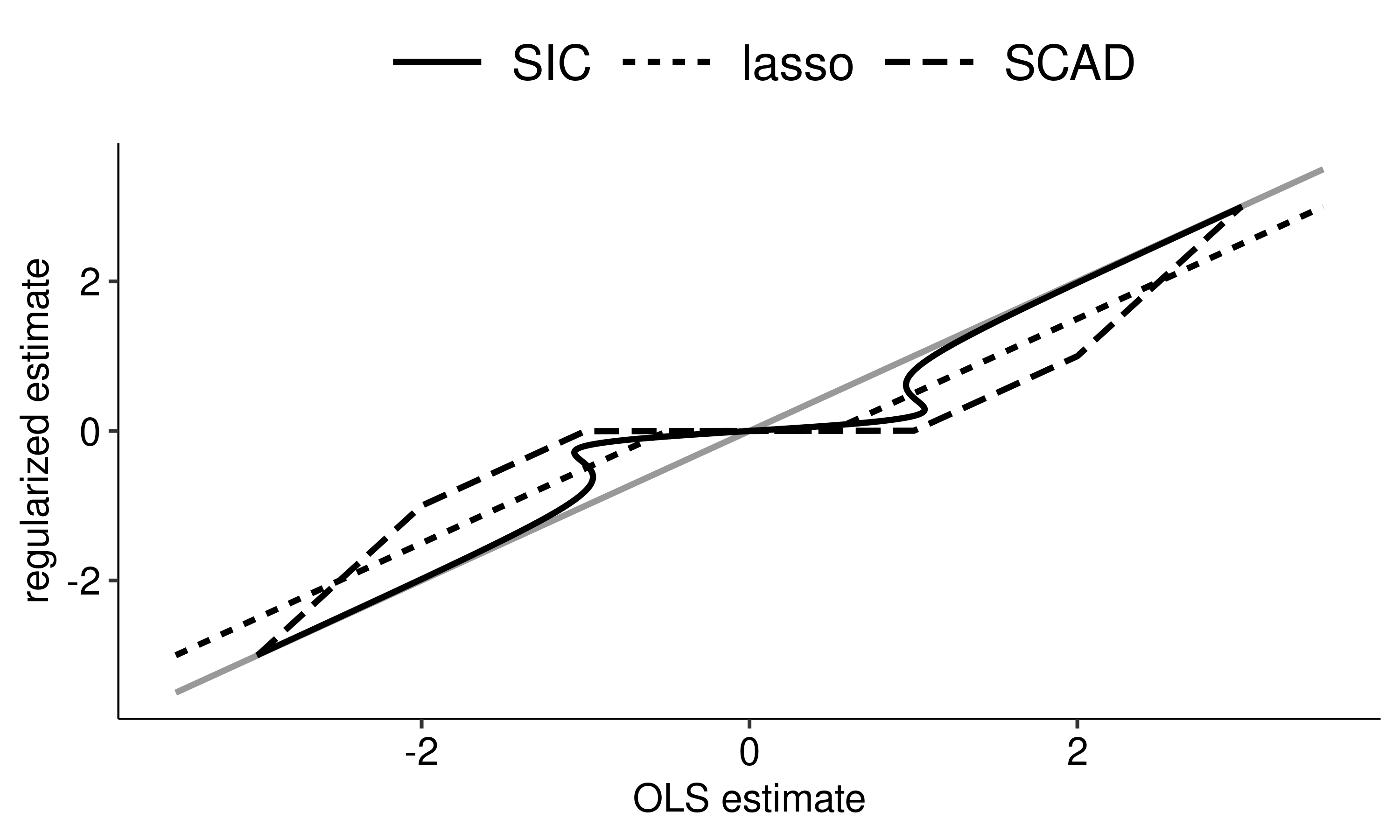}
    \includegraphics[width=0.49\linewidth]{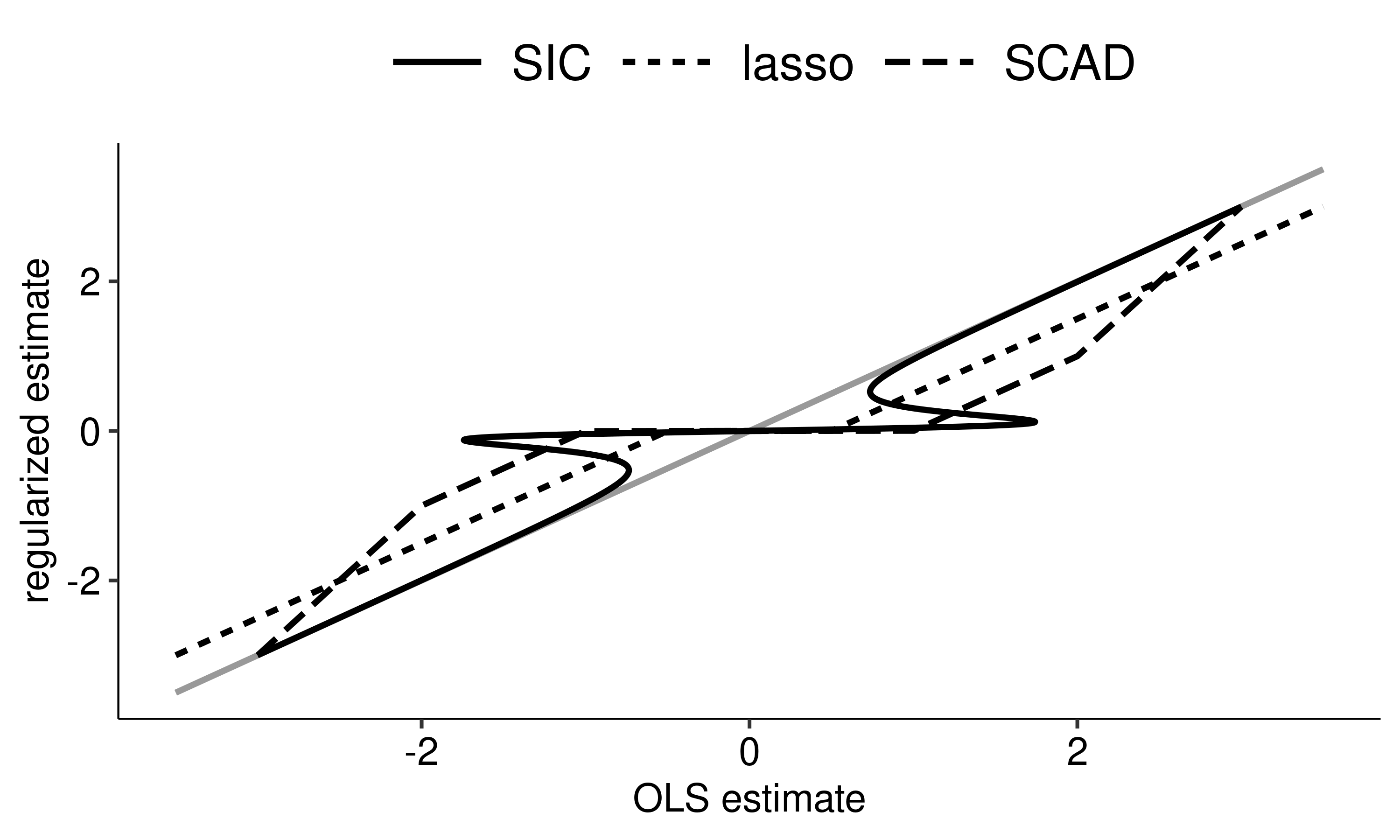}
    \caption{\textsc{Plot of thresholding estimators against the ordinary least-squares estimate.} 
    The solid gray line is $y=x$.
    The tuning parameter $\lambda$ is fixed to 1 for all thresholding estimators. 
    SCAD method involves an additional tuning hyperparameter, which is fixed as $a=3$ in this context.
    The approximation level $\varepsilon$ varies across the three plots:  $0.8$, $0.4$ and $0.2$, from the first plot to the last one.}
    \label{fig:point_of_view:geometry:thresholding}     
\end{figure}

To compare the SIC penalization with other well-known penalization methods 
such as LASSO \citep{tibshirani1996regression} or SCAD \citep{fan2001variable}, 
it can be interpreted as a soft-thresholding rule. 
To illustrate with a simple yet informative plots, 
consider a model on $\m{s} = \m{X}^T \m{y}$,  
where $\m{y} = (y_1, \dots, y_n)$
and
where the design is supposed to be orthonormal, so that $\m{s}$ coincides with the OLS estimate of $\m{\beta}$ for Model \eqref{eq:selection_variables:sic_regression_lineaire:equation_modele}.
For what follows, according to \cite{fan2001variable}, tThe Gaussian white noise model is assumed for $s_i$, so that 
$s_i = \theta_i + \omega_i$, where $\omega_i \overset{\text{iid}}{\sim} \mathcal{N}(0,\sigma^2)$.
Therefore, using the SIC approach to estimate the $\theta_i$'s leads to the minimization of the following criterion:
\begin{equation*}
    \sum_{i=1}^p (s_i - \theta_i)^2 + \lambda \sum_{i=1}^p \phi_{\varepsilon}(\theta_i).
\end{equation*}
Note that minimizing this criterion reduces to minimizing it component by component, 
leading to the equation:
\begin{equation*}
    \frac{\partial}{\partial\theta_i} (s_i - \theta_i)^2 + \lambda  \frac{\partial}{\partial\theta_i}\phi_{\varepsilon}(\theta_i) = 0
    \qquad \text{leading to} \qquad 
    s_i = \theta_i + \lambda \frac{\theta_i \varepsilon^2}{(\theta_i^2 + \varepsilon^2)^2}\cdot
\end{equation*}
Equations related to LASSO and SCAD penalizations can be determined in a similar way, 
and all these conditions are depicted in Figure~\ref{fig:point_of_view:geometry:thresholding}, 
which helps understanding the bias induced in the OLS estimator by adding such penalization terms.
For instance, it is observed that SIC penalization exhibits an interesting property, similar to SCAD penalty: for large OLS estimator values, the penalization does not induce bias on the estimator (unlike LASSO penalty).
Moreover, it can be observed from Figure~\ref{fig:point_of_view:geometry:thresholding} that when $\varepsilon$ is close to 0, the perturbation of the OLS estimator does not have a unique solution. This is related to the instability induced by this penalization, as previously identified by \cite{o2023variable}, which justified the use of the $\varepsilon$-telescoping algorithm to ensure a robust estimation procedure.

\subsubsection{SIC as Bayes estimate}
\label{sect-SIC-formulations-bayes}

\noindent\textbf{The prior distribution related to the SIC penality.}
Penalization methods in Gaussian regression models can be interpreted in a Bayesian context. For example, Ridge penalization corresponds to the MAP estimate associated to a Gaussian prior, while LASSO penalization corresponds to a Laplace prior \citep{park2008bayesian}.  
Similarly, SIC estimators can be obtained as the Bayesian MAP estimate of $\m{\beta}$ for the prior :
\begin{equation}
    \label{SIC:Bayesian1}
   \pi({\m{\beta}}) \propto \prod_{j=1}^{d} \pi_\text{SIC}(\beta_j) 
\end{equation}
where

\begin{equation*}
\label{eq.piSICbetaj}
    \pi_\text{SIC}(\beta_j) =  \exp \left\{ -\frac{\lambda}{2\sigma^2} \, \frac{\beta_j^2}{\beta_j^2 + \varepsilon^2} \right\}.
\end{equation*}
This can be seen by noting that the estimator $\widehat{\m{\beta}}$ obtained by  maximising \eqref{SIC:lagrange}  can be represented as
\begin{equation}
 \widehat{\m{\beta}} \in   \underset{{{\m{\beta} \in \mathbb{R}^{d} }} }{{\text{argmax}}} \; \exp \left \{ \frac{-\left\lVert \m{Y} - \beta_0 \mathbf{1}_n -\m{X \beta}\right\rVert^2-\lambda \lVert {\m{\beta}} \rVert _{0,\varepsilon}}{2\sigma^2} \right\}.
\label{SIC:Bayesian}
\end{equation}
When $\beta_j$ goes to $\pm\infty$, $\pi_\text{SIC}(\beta_j)$ converges to the constant $\exp\{-\lambda/2\sigma^2\}>0$. Therefore $\pi(\m{\beta})$ is an improper distribution whose tail is equivalent to that of a flat prior.
The posterior distribution is
$$
\pi(\m{\beta} \mid \m{Y})\propto \exp \left \{ \frac{-\left\lVert \m{Y} - \beta_0 \mathbf{1}_n-\m{X \beta}\right\rVert^2-\lambda \lVert {\m{\beta}} \rVert _{0,\varepsilon}}{2\sigma^2} \right\} 
$$
which is proper when $X$ is a full-rank matrix.

\begin{figure}[h]
    \centering
    \includegraphics[width=0.8\linewidth]{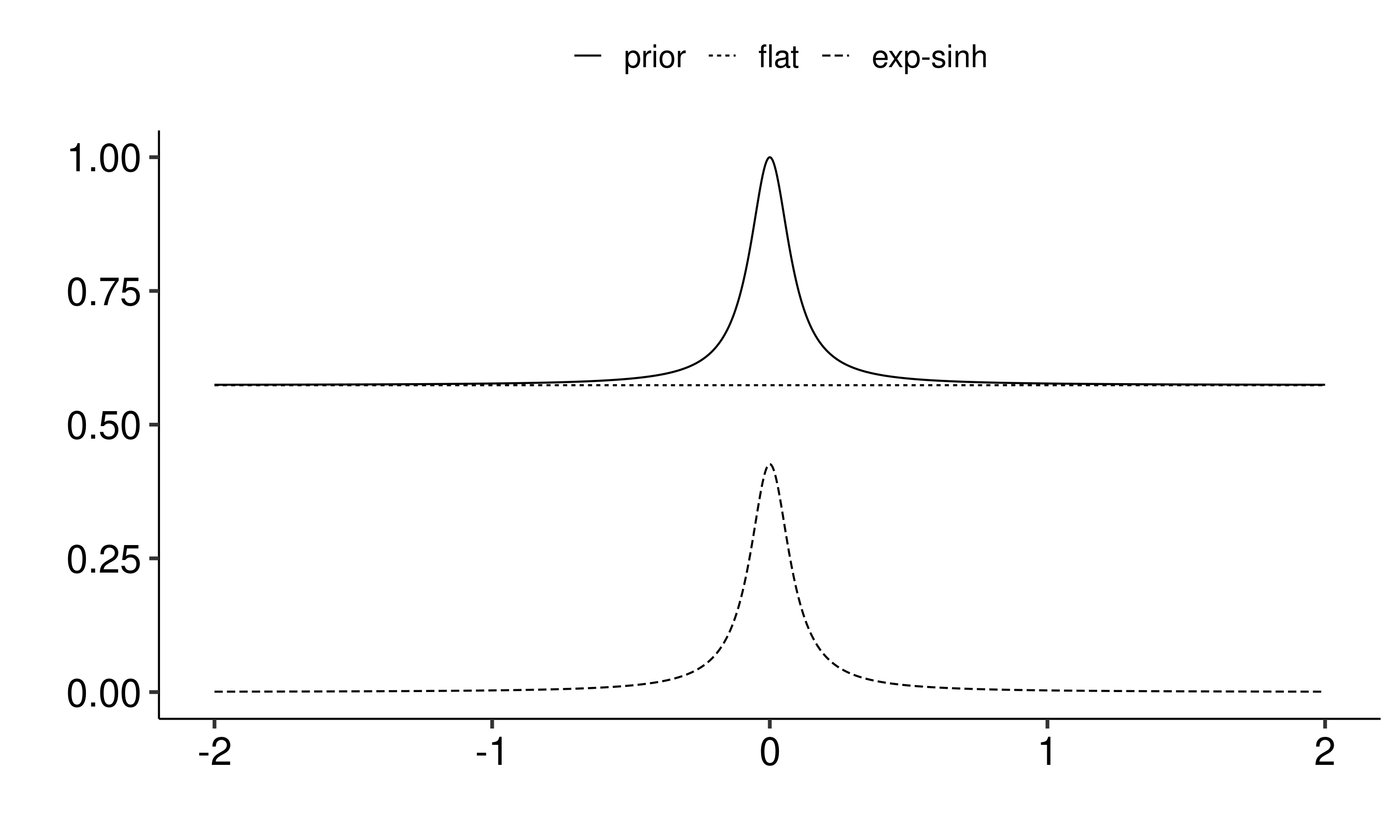}
    \caption{\textsc{Illustration of the prior $\pi_\text{SIC}$ and its two components.}}
    \label{fig:theoritical_result:bayes:mixture_decomposition} 
\end{figure}

\noindent\textbf{Two additive components and interpretations.}
To get a better understanding of  the prior $\pi(\m{\beta})$,  let decompose  $\pi_\text{SIC}(\beta_j) $   into two additive parts: a flat prior and a proper prior denoted   $\tilde \pi(\beta_j)$ and defined   by
\begin{align}
\label{eq:theoritical_result:bayes:mixture_decomposition_intermediaire1}
 \tilde \pi(\beta_j)&= c(\lambda,\varepsilon)\left(\exp\left\{\frac{ \lambda }{2\sigma^2} \times\frac{\varepsilon^2}{ \beta_j^2 + \varepsilon^2} \right\}-1 \right) \\
\label{eq:theoritical_result:bayes:mixture_decomposition_intermediaire2}
& =\;2 c(\lambda,\varepsilon)
 \exp\left\{ \frac{\lambda \varepsilon^2}{4\sigma^2(\beta_j^2 + \varepsilon^2)} \right\}  \sinh \left(\frac{\lambda \varepsilon^2}{4\sigma^2(\beta_j^2 + \varepsilon^2)} \right)   
\end{align}
where Equation~\eqref{eq:theoritical_result:bayes:mixture_decomposition_intermediaire2} is obtained since $e^x - 1 = 2 e^{-x/2} \sinh (x/2)$ and
 where $c(\lambda,\varepsilon)$  is  the normalizing constant:
  $$
 c(\lambda, \varepsilon)=  \left\{\int \left(  \exp\left\{\frac{ \lambda }{2\sigma^2} \times\frac{\varepsilon^2}{ \beta_j^2 + \varepsilon^2} \right\}-1 \right)d\beta_j\right\}^{-1}.
 $$
 Up to the scalar factor  $c(\lambda,\varepsilon) \exp\{-\lambda/2\sigma^2\}$, the prior $\pi_\text{SIC}(\beta_j)$ can be reformulated as 
\begin{equation}  
\label{eq:theoritical_result:bayes:mixture_decomposition}
  \pi_\text{SIC}(\beta_j)\propto  {c(\lambda,\varepsilon)} +  \tilde \pi(\beta_j)
\end{equation}
See Figure~\ref{fig:theoritical_result:bayes:mixture_decomposition} for a visual representation of the prior distribution $\pi_\text{SIC}$ and its two components.  
The scalar  $c(\lambda, \varepsilon) $ appears as the relative weight between  the flat distribution and  the proper distribution  $ \tilde \pi(\beta_j)$.
To see that $\tilde \pi(\beta_j)$   is proper and therefore that $c(\lambda,\varepsilon)$ is well defined, we note that the tail  behaviour of $\pi_\text{SIC}(\beta_j) -1$   is equivalent to  $\frac{\lambda}{2\sigma^2\beta_j^2}$ which is integrable. 
Note that $ \tilde \pi(\beta_j)$ can be seen as  a smoothed approximation of  $\delta_0$, the Dirac measure in 0,  for large values of $\lambda$. Indeed, it can be shown that  $ \tilde \pi(\beta_j)$ converges narrowly to $\delta_0$ 
as $\lambda $ goes to $+\infty$. So, the prior  $\pi_\text{SIC}(\beta_j)$ has some similarity with  another prior distribution consisting of two additive components:
the prior used in the spike and slab regression \citep{MittchelBeauchamp1988, ishwaran2005spike} where the  heavy tailed distribution is replaced by a flat prior and the Dirac measure is approximate by $\tilde \pi$.

\begin{figure}[h]
    \centering
    \includegraphics[width=0.49\linewidth]{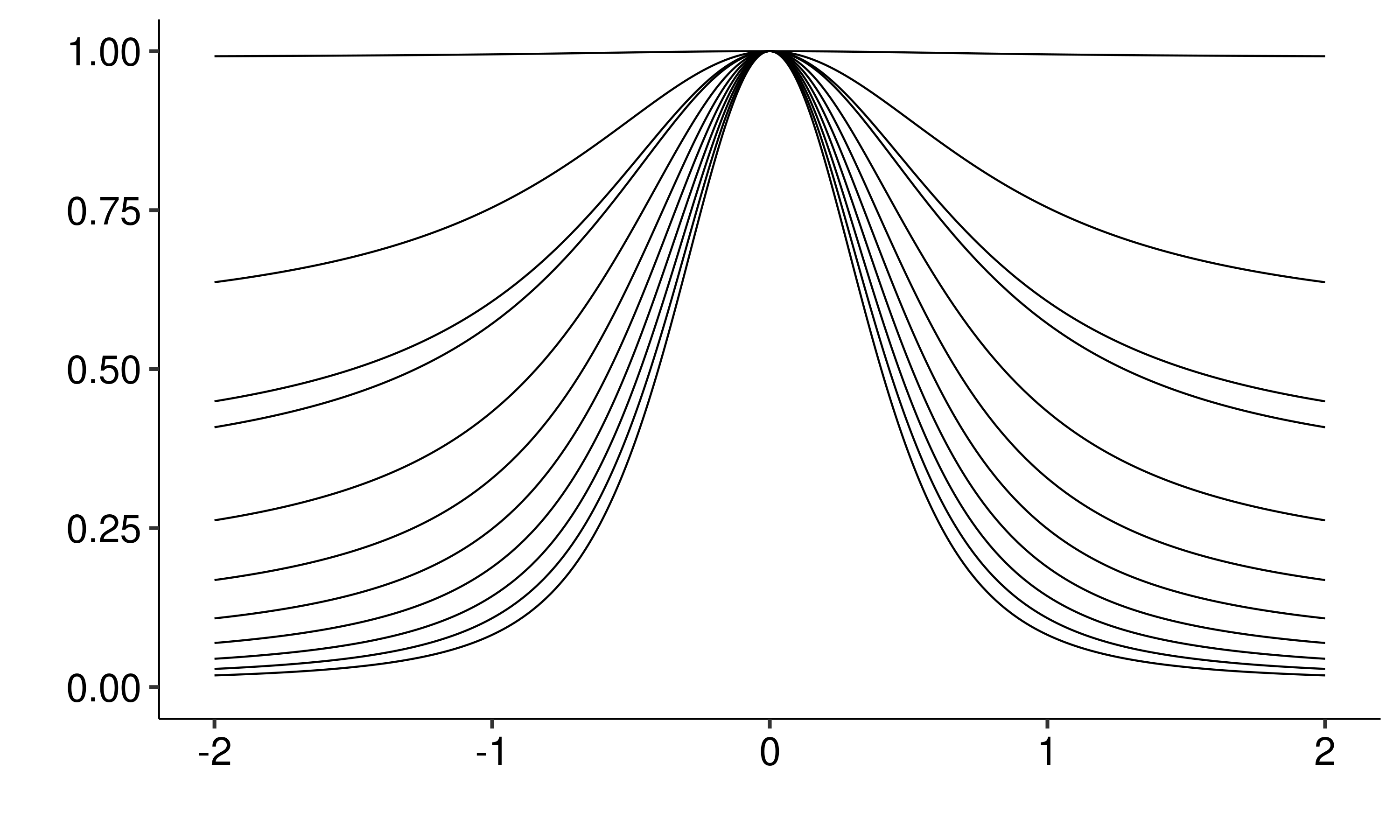}
    \includegraphics[width=0.49\linewidth]{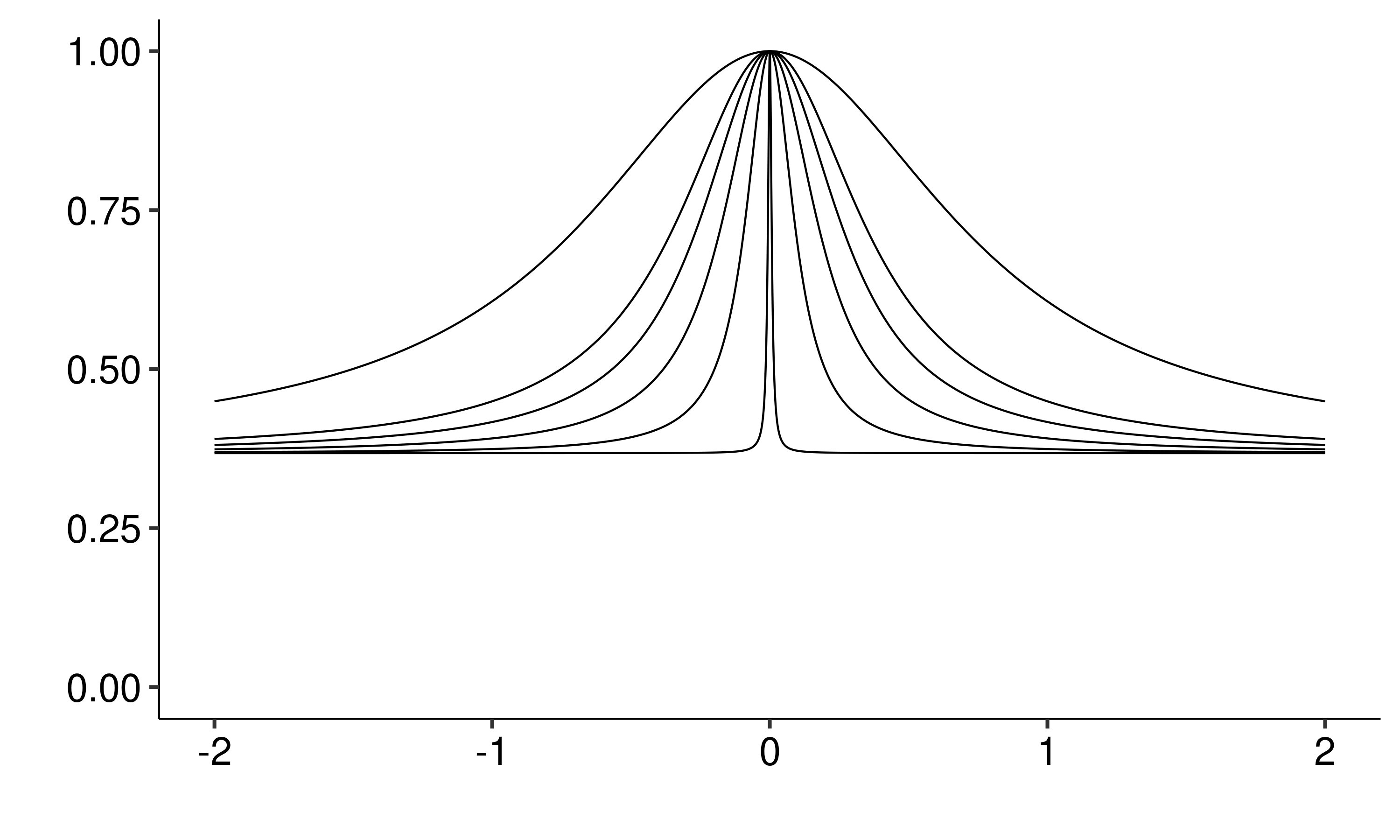}
    \caption{\textsc{Variation of the prior distribution $\pi_\text{SIC}$ according to $\lambda$ and $\varepsilon$.} Left Plot shows to the variation of $\lambda$, from 5 (close to a flat prior) to 0.01 (spiky prior). Right Plot corresponds to the variation of $\varepsilon$, from 0.5 (large scale) to 0.01 (sharp).}
    \label{fig:theoritical_result:bayes:prior_variation}
\end{figure}

Moreover, note that for spike-and-slab regression, it is the flat part that is approximated by a distribution, not the Dirac measure. In other words, both the SIC prior and the spike-and-slab regression prior propose working with a distribution that approximates a mixture of a flat prior and a Dirac measure, but they do not approximate the same term of the mixture. 
In addition, Equation~\eqref{eq:theoritical_result:bayes:mixture_decomposition} reveals that the influence  of the flat prior is determined by $\lambda$.
In addition, the "$\exp-\sinh$" terms is proportional to a distribution that approximates the Dirac measure,  with $\varepsilon$ tuning the bias of the approximation.
Figure~\ref{fig:theoritical_result:bayes:prior_variation} indicates how variations in $\lambda$ or $\varepsilon$  affect  the prior density.
Typically, a flat prior is employed to allocate substantial prior mass to large values, contrary to most prior distributions used for variable selection purposes (Gaussian, Laplace and Horseshoe priors).
Then, the prior distribution \eqref{SIC:Bayesian1} allows to rely on a flat prior (non-informative in some sense) for large values, and also to simultaneously promote values close to 0.

For SIC penalization, the tuning parameter $\lambda$ is suggested to be fixed as $\log(n)$, as proposed by \cite{o2023variable}, suggesting an analogy with $L_0$ regularization and the BIC procedure as a rationale behind the definition of $\lambda$. 
However, according to the interpretation of the prior distribution $\pi_\text{SIC}$, we suggest that $\lambda$ could be tuned based on prior beliefs regarding the expected number of null $\m{\beta}$ coordinates.
Thus, the default choice "$\lambda \propto \log n$" implies that confidence in potentially assigning certain coordinates to 0, is primarily driven by the amount of available data.

\noindent\textbf{Asymptotic behavior.}
To have a better understanding of the prior, we consider   the limit behaviour of $ \pi_\text{SIC}(\beta_j)$ when $\lambda $ is large and/or $\varepsilon$ is small.  Since we deal with   improper distribution,  usual convergence modes cannot be used. That is why \cite{bidr2016} proposed a convergence mode  adapted to the Bayesian framework where prior distributions are defined up to a scalar factor. When the priors are Radon measures, that is distributions  finite on compact sets, they derive a quotient topology   called  q-vague topology.  A sequence $\pi_n$ converge q-vaguely to $\pi$ if there exists  positive scalars  $a_n$   such that $a_n\pi_n$ converges weakly to $\pi$, or equivalently, if for any real-valued continuous function $f$ with compact support, we have  
	$a_n \int  f(\beta)\pi_n(\beta)\;d\beta \xrightarrow[n\rightarrow +\infty]{}\int a_n\pi(\beta)\;d\beta.$
The q-vague limit of $\pi_n$, when it exists, is unique  up to a scalar factor. Moreover,  when the likelihood is continuous, the  posterior distributions $\pi_n(\cdot \mid x)$ also converge q-vaguely to $\pi(\cdot \mid x)$. Proofs are detailed in Appendix~\ref{sect:appendix:proofs}.

\begin{prop}
\label{prop-SIC-formulations-bayes-lambda}
 Let $\lambda_n$ be a sequence that converges   to $+\infty$. Then, for    $\varepsilon$ fixed, the related sequence $\pi_\text{SIC}^{(n)}(\beta_j)$ converges q-vaguely to a Dirac measure in $0$.     
\end{prop}  

\begin{prop}
\label{prop-SIC-formulations-bayes-epsilon}
Let  $\varepsilon_n$ be a sequences that converge  to $0$. Then, for $\lambda$ fixed, the related sequence $\pi_\text{SIC}^{(n)}(\beta_j)$ converges q-vaguely to a flat prior. 
\end{prop}

 We now study the convergence of $\pi_\text{SIC}^{(n)}(\beta_j)$  when $\lambda_n$ goes to $+\infty$ and $\varepsilon_n$ goes to $0$ simultaneously. The limit will depend on the relative rate of convergence of $\lambda_n$ and $\varepsilon_n$. When $\varepsilon_n$ goes to $0$ much faster  than $\lambda_n$ goes to $+\infty$, then  $\pi_\text{SIC}^{(n)}(\beta_j)$ converges to a flat prior. 
 Similarly, when $\varepsilon_n$ goes to $0$ much   much slower  than $\lambda_n$ goes to $+\infty$ then  $\pi_\text{SIC}^{(n)}(\beta_j)$ converges to the Dirac measure   $\delta_0$. To give a precise meaning of "much faster" or much slower", the following proposition give the relative rate of convergence corresponding to the phase transition, where the limit is between a flat distribution and a Dirac measure.

\begin{prop}
\label{prop-SIC-formulations-bayes-lambda_and_epsilon}
  Assume that   $\lambda_n$ goes to $+\infty$ and $\varepsilon_n$  goes to $0$ with $\varepsilon_n =K \sqrt{\lambda_n/2\sigma^2}   \exp\{-\lambda_n/2\sigma^2\}  $ for some constant $K$. Then, the  related sequence $\pi_\text{SIC}^{(n)}(\beta_j)$ converges q-vaguely to the prior $\pi(\beta_j) \propto \frac{1}{\sqrt \pi K}+\delta_0$.
\end{prop}

\subsection{Oracle inequality}
\label{sect-SIC-theoretical_results}

The aim of this section is to demonstrate that the SIC penalty satisfies the required frequentist theoretical properties.
In particular, we target to highlight that prediction error performance is similar to that of other penalization methods.
The theoretical argument we obtain provides reliable guarantees for using a SIC penalization approach, akin to an oracle inequality.
To achieve this, we examine squared error loss under a fixed design.

For the following theoretical results, we carry out a standard  mathematical procedure used to demonstrate the oracle inequality of the LASSO \citep{tibshirani1996regression}. For more details, we refer the reader to \cite{zou2005regularization,buhlmann2011statistics,hastie2015statistical,tibshirani2015sparsity}. 
For what follows, assume the linear model:
\begin{equation*}
    \m{y}=\m{X \beta^{0}+ \omega}, \quad \m{\omega} \sim \mathcal{N}(0,\sigma^2 I ).
\end{equation*}
To ensure that these procedures are valid within the framework of the SIC penalty, we introduce Lemma \ref{lemma:theoritical_result:equivalence_norms}, which shows how to upper bound the SIC penalty with usual norms.
\begin{lemma}[{\bf Links between $\|.\|_{0,\varepsilon}$ and norms}]
    \label{lemma:theoritical_result:equivalence_norms}
    For $\m{x}$ a $p$-dimensional vector, and for $\varepsilon>0$, we have: 
\begin{equation}
    \left\lVert \m{x} \right\rVert_{0,\epsilon} \leq \frac{1}{\varepsilon^2} \left\lVert \m{x} \right\rVert_2^2,  
    \qquad 
    \left\lVert \m{x} \right\rVert_{0,\epsilon} \leq \frac{1}{2\varepsilon} \left\lVert \m{x} \right\rVert_1    
    \quad \text{ and } \quad 
    \left\lVert \m{x} \right\rVert_{0,\epsilon} \leq \frac{\sqrt{p}}{2\varepsilon} \left\lVert \m{x} \right\rVert_2
    \label{majoration:SIC}
\end{equation}
\end{lemma}

\begin{lemma}[{\bf Basic Inequality}]
    \label{theorical:Basic_inequality}
    Let $\widehat{\m{\beta}}$ be a solution of \eqref{SIC:lagrange}. An optimality condition for the SIC  estimator is given by
    \begin{equation*}
        \frac{\lVert \m{X}(\widehat{\m{\beta}}-\m{\beta}^0) \rVert _{2}^{2}}{n}+ \frac{\lambda}{n} \lVert\widehat{\m{\beta}} \rVert _{0,\varepsilon}  \le \frac{2 \m{\omega}^{\top}\m{X}(\widehat{\m{\beta}}-\m{\beta}^{0})}{n}+ \frac{\lambda}{n} \lVert{\m{\beta^{0}}} \rVert _{0,\varepsilon}.
      \end{equation*}
\end{lemma}
The dominant term of the inequality is random, and to ensure a non-random upper bound of the left term, the following relies on a concentration result.
First, using Holder’s inequality, an upper bound on the random term is written as,
\begin{equation*}
    {2 \m{\omega}^{\top}\m{X}(\widehat{\m{\beta}}-\m{\beta}^{0})} \le 2\lVert  \m{\omega}^{\top}\m{X} \rVert_{\infty} \rVert \widehat{\m{\beta}}- \m{\beta}^{0} \rVert_{1}, \quad\text{with} \; \|2\m{\omega}^{\top}\m{X}\|_{\infty} = \max_{j=1,\dots,d} \left\lvert 2\m{\omega}^{\top}\m{X}^{j}\right\rvert. 
\end{equation*}
Following \citet{buhlmann2011statistics} we introduce the set $\zeta = \left\{\lambda_0 \ge \left\lVert 2\m{\omega}^{\top}\m{X}\right\rVert_{\infty} \right\}$ 
and choose $\lambda \ge \lambda_0$. For a suitable value of $\lambda$, the set $\zeta$ has large probability. Note by $\widehat{\sigma}_j^2$ the diagonal elements of the normalized Gram matrix: $\widehat{\sigma}_j^2=\widehat{\m{\Sigma}}_{j,j}$, for $j=1,\dots,d$.
\begin{lemma}
    \label{lemme:theoritical_results:gram_condition}

    Suppose that for $j=1,\dots,d$, $\widehat{\sigma}_j^2=1$. For all $t > 0$, we pose $\lambda_0= 2\sigma\sqrt{\frac{t^2+ 2\log(d)}{n}}$, 

    \begin{equation*}
        \mathbf{P}(\zeta)=\mathbf{P}\left(\left\{\lambda_0 \ge \left\lVert \m{\omega}^{\top}\m{X}\right\rVert_{\infty} \right\}\right)  \ge 1-2 \exp\left({-t^2}/{2}\right)  
    \end{equation*}

\end{lemma}

\begin{corollary}{\bf (Inequality of SIC consistency)}
\label{cor:theoritical_results:consistency}
Assume that for $j=1,\dots,d$ we have $\widehat{\sigma}=1$. For $t>0$, let 
 $\lambda= 4 \widehat{\sigma} \sqrt{\frac{{t^2 + 2 \log (d)}}{n}}$, where $\widehat{\sigma}$ is an estimator of $\sigma$. Then, with a probability of at least $1- 2 \exp\left({-t^2}/{2}\right)$ we have 
 \begin{align*}
    \frac{2\lVert \m{X}(\widehat{\m{\beta}}-\m{\beta}^0) \rVert _{2}^{2}}{n}\le \left(1+\frac{1}{n\varepsilon}\right)\lVert{\m{\beta}^0} \rVert _{1}
  \end{align*}
\end{corollary}

Then,  $\mathcal{S}_0 =  \left\{j \, ; \,  \beta_j^0 \ne 0\right\} $ denotes the support for the unknown parameter vector $\m{\beta}^0$, in other words: the set of active variables. 
For any vector $\m{\beta} \in \mathbb{R}^d$, the following vectors are also defined $\m{\beta}_{\mathcal{S}_0}={\beta}_j \m{I}_{j \in {\mathcal{S}_0}}$ and $\m{\beta}_{\mathcal{S}_0^c}={\beta}_j \m{I}_{j \notin {\mathcal{S}_0}}$,
this means that if $j$ is in ${\mathcal{S}_0}$ then the component $j$ of the vector $\m{\beta}_{\mathcal{S}_0}$ is non-zero.
Then, $\m{\beta}$ can be expressed as a function of $\m{\beta}_{\mathcal{S}_0}$ and $\m{\beta}_{\mathcal{S}_0^c}$:
\begin{equation*}
    \m{\beta}=\m{\beta}_{\mathcal{S}_0}+\m{\beta}_{\mathcal{S}_0^c}
\end{equation*}

\begin{lemma}
    \label{lemma:theoritical_results:sparsity}
    Assume that $ \zeta = \left\{\lambda_0 \ge \left\lVert 2\m{\omega}^{\top}\m{X}\right\rVert_{\infty} \right\} $, and $\lambda \ge \lambda_0$,  then with large probability we have
    \begin{align*}
        \frac{2\lVert \m{X}(\widehat{\m{\beta}}-\m{\beta}^0) \rVert _{2}^{2}}{n} +\lambda \left(\frac{1}{n\varepsilon}-1\right)\lVert\widehat{\m{\beta}}_{\mathcal{S}_0^c}\rVert_{1} & \le \lambda\left(\frac{1}{n\varepsilon}+1\right) \lVert \m{\widehat{\beta}}_{\mathcal{S}_0}-\m{\beta}_{\mathcal{S}_0}^0\rVert_{1}
      \end{align*}
      and 
      \begin{equation*}
        \lVert{\m{{\beta}}^0_{\mathcal{S}_0}-\m{\widehat{\beta}}_{\mathcal{S}_0}} \rVert _{1} \le \sqrt{s_0}{\lVert{\m{{\beta}}^0_{\mathcal{S}_0}} - \m{\widehat{\beta}}_{\mathcal{S}_0} } \rVert _{2}.
      \end{equation*}
\end{lemma}

By exploiting the compatibility condition related to $\mathcal{S}_0$ in \cite{buhlmann2011statistics}, and for some $\eta \ge 0$, for all $\m{\beta}$ such that, $\left(\frac{1}{\varepsilon}-1\right)\lVert{\m{\beta}}_{\mathcal{S}_0^c}\rVert_{1} \le \left(\frac{1}{\varepsilon}+1\right) \lVert \m{\beta}_{\mathcal{S}_0}\rVert_{1}$ we have
\begin{equation*}
    \lVert\widehat{\m{\beta}}_{\mathcal{S}_0^c}\rVert_{1}^2 \le \frac{\left( \m{\beta}^\top\m{\widehat{\Sigma}}\m{\beta}\right)s_0}{\eta^2}
\end{equation*}
which leads to 
\begin{align*}
   &\lVert (\m{\widehat{\beta}}_{\mathcal{S}_0}-\m{\beta}_{\mathcal{S}_0}^0) \rVert _{1}^{2} \le \frac{\left(\m{\widehat{\beta}}_{\mathcal{S}_0}-\m{\beta}_{\mathcal{S}_0}^0\right)^\top\widehat{\Sigma}\left(\m{\widehat{\beta}}_{\mathcal{S}_0}-\m{\beta}_{\mathcal{S}_0}^0\right) s_0}{n \eta^2}\\
   &\lVert (\m{\widehat{\beta}}_{\mathcal{S}_0}-\m{\beta}_{\mathcal{S}_0}^0) \rVert _{1} \le \frac{\lVert \m{X}(\m{\widehat{\beta}}_{\mathcal{S}_0}-\m{\beta}_{\mathcal{S}_0}^0) \rVert _{2} \sqrt{s_0}}{\sqrt{n}\eta}
\end{align*}

\begin{theorem}
    \label{theorem:theoritical_results:sparsity}
    Assuming the compatibility condition on $\mathcal{S}_0$, and for $\lambda > 2\lambda_0$, we can rewrite the lemma basic inequality in the following form:
    \begin{equation*}
        \label{rewrite_basic_inq_compatibilty}
        \frac{2\lVert \m{X}(\widehat{\m{\beta}}-\m{\beta}^0) \rVert _{2}^{2}}{n} + \lambda \left(\frac{1}{n\varepsilon}-1\right)\lVert \m{\widehat{\beta}}-\m{\beta}^0\rVert_{1} 
        \le
         \frac{2\lambda}{n\varepsilon}\frac{\lVert \m{X}(\m{\widehat{\beta}}_{\mathcal{S}_0}-\m{\beta}_{\mathcal{S}_0}^0) \rVert _{2} \sqrt{s_0}}{\sqrt{n}\eta} \le\frac{2\lambda^2{s_0}}{(n\varepsilon \eta)^2}
      \end{equation*}
\end{theorem}

\begin{corollary} {\bf (Oracle Inequality)}
    \label{cor:theortical_result:oracle}
    Assume $\widehat{\sigma}_j = 1$ for $j=1,\dots,p$, and that the compatibility condition is satisfied on the set $\mathcal{S}_0$, then it exists a constant $c>0$ for $\lambda=4\widehat{\sigma}\sqrt{\frac{t^2+2 \log (d)}{n}}$ and $t=\sqrt{2 \log(d)}$ such that:
    \begin{equation*}
        \label{oracle_inequality}
        \frac{\lVert \m{X}(\widehat{\m{\beta}}-\m{\beta}^0) \rVert _{2}^{2}}{n} 
        \le
        c \frac{ \sigma^2 \log(d){s_0}}{ n^3 \varepsilon^2}.
    \end{equation*}
\end{corollary}


\section{Sparse covariate estimation}
\label{sect-estimation}

In the following, we consider the context of the PLN model~\eqref{eq:PLN:modele:modele_hierarchique}, focusing on the parameter of interest, matrix $\m{B}$, for which we aim to achieve sparse estimation.
To this end, we adopt a regularized log likelihood lower bound \eqref{eq:PLN:estimation:elbo} with a SIC penalty as our loss function. This can also be viewed as a PLN model with a SIC constraint on the regression coefficients.  
The ELBO of the log-likelihood is an objective function widely used in the estimation of regression models when the true log-likelihood is intractable \citep{you2014variational,carbonetto2012scalable,zhang2019novel}. 
As detailed in Section~\ref{sect-SIC-theoretical_results}, the SIC penalty is able to constrain and capture sparse patterns in the model with guaranteed convergence to the $L_0$-norm.
The objective function to be optimized is then:
\begin{align*}
 J_{\text{pen}}(\m{Y}; \m{\theta}, \m{\psi}) \; 
    =
    J(\m{Y} ; \m{\theta}, \m{\psi})   - \frac{\lambda}{2}\Big[ \lVert \m{B} \rVert _{0,\varepsilon} + k \Big],
\label{eq-selection_variable-loglikplnpena}
\end{align*}
where  $\lVert \m{B} \rVert _{0,\varepsilon}$ denote the SIC penalty matrix which is derived as $\lVert \m{B} \rVert _{0,\varepsilon}=\sum_{i,j}\phi_{\varepsilon}({B}_{kj})$ and $k$ is the number of unpenalized parameters. Here $\lambda=\log(n)$ is a fixed tuning parameter which maximizes a pseudo-BIC \citep{chen2018use} due to the replacement of the true log-likelihood in the SIC criterion \eqref{eq:selection_variables:sic_regression_lineaire:SIC} by the ELBO \eqref{eq:PLN:estimation:elbo}.

\noindent {\bf Optimization algorithm.}
To sparsly estimate the support of the entries in the regressor matrix, we combine the PLN estimation algorithm proposed by \citet{chiquet2021poisson} and the $\varepsilon$-telescoping algorithm proposed by \citet{o2023variable}. 
The resulting algorithm we propose is described in Algorithm~\ref{algo-estimation-SICPLN}, and first we provide details about steps of the procedure.
\begin{itemize}
    \item[-] The pseudo-Fisher information matrix denoted by $\m{\mathcal{I}}_{\text{pen}}(\m{B})$ is given from the matrix of negative second derivatives of $J_{\text{pen}}(\m{Y}; \m{\theta}, \m{\psi})$ with respect to $\m{B}$ and the pseudo penalized scoring matrix denoted by $\m{\mathcal{S}}_{\text{pen}}$ is given from the gradient of $J_{\text{pen}}(\m{Y}; \m{\theta}, \m{\psi})$ with respect to $\m{B}$. 
    The penalized pseudo-Fisher information matrix  is:
\begin{align*}
\m{\mathcal{I}}_{\text{pen}}(\m{B})&= \m{\mathcal{I}}_{n}(\m{B}) +\frac{\log(n)}{2} \m{\Lambda} = -\mathbb{E}_{\m{B}} \Big[\frac{ \partial^2 J(\m{Y} ; \m{\theta}, \m{\psi}) }{\partial \m{B}^2}\Big] +\frac{\log(n)}{2}\m{\Lambda},\\
    &=\frac{1}{n}\left( (\m{I}_p \otimes  \m{X}^\top )\text{diag}(\operatorname{vec}(\m{A}))( \m{I}_p \otimes\m{X})\right) + \frac{\log(n)}{2}\m{\Lambda},
\end{align*}
where $\m{\mathcal{I}}_{n}(\m{B})$ is the observed information of the unpenalized ELBO;  $\m{A} = \exp(\m{O} +\m{XB} +\m{M} +\m{S}^2/2)$ and $\m{\Lambda}$ is a $p \times d$ diagonal matrix that arises due to penalization, where each entry is given by $\phi''_{\varepsilon}({B}_{kj})$, except for the first $p$ diagonal terms which are zero because the intercepts of the $p$ count columns are not penalized. 
Note that the computation of the inverse of $\m{\mathcal{I}}_{\text{pen}}(\m{B})$ relies on  a QR decomposition \citep{stewart1973introduction} in order to avoid numerical problems. 
    \item[-] The penalized scoring matrix expression is given by: 
\begin{align*}
   \m{\mathcal{S}}_{\text{pen}}(\m{B})&=\frac{ \partial J_{\text{pen}}(\m{Y}; \m{\theta}, \m{\psi}) }{\partial \m{B}}-\frac{\log(n)}{2} \m{\Gamma} = \frac{1}{n}(\m{I}_p \otimes  \m{X}^\top\text{diag}(\operatorname{vec}(\m{Y}-\m{A})))-\frac{\log(n)}{2} \m{\Gamma}
\end{align*}
where $\m{\Gamma}$ is a matrix of row $p\times d$ and column $p$ whose components come from $\phi^{'}_{\varepsilon}({B}_{kj})$.
    \item[-] The update of $\m{B}$ during the optimization procedure is obtained by the $\varepsilon$-telescoping approach on a penalized pseudo-Fisher scoring algorithm.
    In particular at te $t$-th iteration, $\m{B}^{t-1}$ is updated according to: $\m{B}^{t}= \m{B}^{t-1} + \left( \m{\mathcal{I}}_{\text{pen}}(\m{B}^{t-1})\right)^{-1} \left(\m{\mathcal{S}}_{\text{pen}}(\m{B}^{t-1}) \right)$.    
\end{itemize}

\begin{algorithm}[H]
\caption{SICPLN algorithm}
\begin{algorithmic}
\State \textbf{Initialisation:} 
\\ \quad - $\pi^{0}=(\m{B}^{0}, \m{\Sigma}^{0}, \m{M}^{0}, \m{S}^{0})$,  
\\ \quad - $\m{E}=(\varepsilon_1,\cdots,\varepsilon_T)$ a vector of decreasing values, with $\varepsilon_t = \varepsilon_1 r^{t-1}$ where $r \in ]0,1[$.

   \For {$t$  in $1$ to  $T$} \Comment{$\varepsilon$-telescoping} 
  
        \Repeat \Comment{VEM}
        \State \textbf{VE Step:} to update $\m{B}^{t-1}$ and $\m{\Sigma}^{t-1}$
        \Repeat \Comment{Fisher scoring step}
            \begin{align*}
    \m{B}^{t}= \m{B}^{t-1} + \left( \m{\mathcal{I}}_{\text{pen}}(\m{B}^{t-1})\right)^{-1} \left(\m{\mathcal{S}}_{\text{pen}}(\m{B}^{t-1}) \right). 
\end{align*} 
\Until convergence; \Comment{End Fisher scoring step}
\begin{align*}
                \m{\Sigma}^{t} & =  \frac{1}{n}(\m{M}^{t-1}-\m{X} \m{B}^{t})^\top(\m{M}^{t-1}-\m{X} \m{B}^{t})+\frac{1}{n} \text{diag}(\m{I}_n^\top (\m{S}^{t-1})^{2}).
            \end{align*}
            \State \textbf{VM Step:} to update $\m{M}^{t-1}$ and $\m{S}^{t-1}$ according to the maximization of
            \begin{align*}
                \frac{\partial J_{\text{pen}}(\m{Y},\m{\psi},\m{\theta})}{\partial\m{M}} &= (\m{Y}- \m{A} - (\m{M}-\m{X}\m{B}^{t}) (\m{\Sigma}^{t})^{-1}), \\
                \frac{\partial J_{\text{pen}}(\m{Y},\m{\psi},\m{\theta})}{\partial\m{S}} &= \frac{1}{\m{S}}- \m{S} \odot \m{A} -\m{S}D_{((\m{\Sigma}^{t})^{-1})}. 
            \end{align*}
        \Until convergence; \Comment{End VEM}
        \State \textbf{Update} ${\pi}^t= (\m{B}^t, \m{\Sigma}^t, \m{M}^t, \m{S}^t)$.

\EndFor \Comment{ End $\varepsilon$-telescoping} 

\State \textbf{Post-treatment:} Set at $0$ all entries of $\m{B}^{T}$ below $w$ ($w=10^{-5}$), and update $\pi^T$.
\State \textbf{Return} ${\pi}^T$.
\end{algorithmic}
\label{algo-estimation-SICPLN}
\end{algorithm}

\section{Simulation study}
\label{sect-simulation}

\subsection{Setup}
\label{sect-simulation-setup}

To evaluate the proposed approach, we perform a simulation study, on different synthetic databases and with different methods. 

\noindent{\bf Compared methods.} To provide a context for the performance of the proposed method compared to existing methods, we apply the following methods in this simulation study.
\begin{itemize}
    \item[-] GLMNET: a univariate Poisson regression model using Lasso regularization (implemented in the \verb|R| \verb|glmnet| package). As this modeling is univariate, we duplicate it separately for each column of the count matrix $\m{Y}$. While this approach disregards the dependence between columns of $\m{Y}$, at least it  includes a variable selection procedure.
    \item[-] PLN: the standard Poisson Log Normal model implemented in the \verb|R| package \verb|PLNmodels|. Unlike the GLMNET approach, this is a multivariate approach that takes the dependence between columns of Y into account, but it does not perform variable selection.
\end{itemize}
 
\noindent{\bf Simulation protocol.} Below we  describe how synthetic data are simulated 
according to PLN model \eqref{eq:PLN:modele:modele_hierarchique}, to assess the performance of different models in relevant configurations.
\begin{itemize}
    \item[-] \textit{Dimensions.} We consider diverse sample sizes $(n=30,\;50,\; 100,\; 1000)$ and species numbers $(p=10,\; 20,\; 30,\; 40)$. 
    \item[-] \textit{To generate the design matrix $\m{X}$.} The design matrix $\m{X}$ consists of six covariate column vectors $\m{x_1}$, $\m{x_2}$, $\m{x_3}$, $\m{x_4}$, $\m{x_5}$, $\m{x_6}$,  where the entries of each covariate are drawn independently from a uniform distribution $\mathcal{U}_{[0.5,1.5]}$.
    \item[-] \textit{Model parameters.} The coefficient matrix $\m{B}$ is the parameter of interest for the proposed approach, and in particular we choose to make this matrix sparse by setting some coefficients to zero, enabling us to assess how well the proposed method recovers coefficients that are exactly equal to zero. 
    For coefficients different from zero, we choose to set some at 0.5 and others at 1 to assess if this induces a difference in the ability to estimate them.
    Concerning the covariance matrix $\m{\Sigma}$, two scenarios are considered (and a third one is also provided with Supplementary Materials), in order to  evaluate the impact on the ability to accurately estimate coefficients and effectively recover coefficients equal to zero. 
    \begin{enumerate}
    \item Full: indicating that the counts are dependent across species. 
    In this case, we derive the covariance matrix by $\m{\Sigma}=\m{\Psi}^\top \m{\Psi}$ 
    where each entry of $\m{\Psi}$ is drawn independently from a $\mathcal{U}_{[-1.5,1.5]}$. In this case, all entries of $\m{\Sigma}$ are non-negative.
    \item Diagonal: 
    indicating that $\m{\Sigma}$ is a diagonal matrix whose the terms are different and are drawn independently from a $\mathcal{U}_{[0,5]}$.
    In this case, the species are uncorrelated from each other and they have a specific variance coefficient. 
    \end{enumerate} 
    \item[-] \textit{To generate the count data matrix $\m{Y}$.}  Using the design matrix $\m{X}$ and model parameters $\m{B}$ and $\m{\Sigma}$, we derive coefficients $Z_{ij}$ from the latent layer of model~\eqref{eq:PLN:modele:modele_hierarchique}, enabling the simulation of count data matrix $\m{Y}$ according to model~\eqref{eq:PLN:modele:modele_hierarchique} .
\end{itemize}
Databases are generated for all possible combinations of the following elements: sample size $n$, number of species $p$, and dependence structure in the matrix $\m{\Sigma}$.
The process to generate a database is repeated 100 times and the results are averaged over these 100 repetitions, in order to effectively compare the performance of different methods.

\noindent {\bf Performance indicator.}  Given the focus of this article, we propose evaluating the performance of the different methods on databases using three indicators, detailed in the following.
\begin{itemize}
    \item[-] \textit{Estimation error}: the relative error based on the Frobenius norm between the true value of coefficients matrix $\m{B}$ and the estimated coefficient matrix $\widehat{\m{B}}$. In particular, the indicator is the same as in \cite{wu2018sparse}, which is:
\begin{align*}
    \frac{\left\lVert \m{B}- \widehat{\m{B}} \right\rVert_{F}}{\left\lVert \m{B}\right\rVert_{F}}.
\end{align*}
\item[-] \textit{Support error}: the true negative rate (TNR), which measures how coefficients equal to zero (called negative) are accurately estimated as being equal to zero (called true negative).
\begin{align*}
    \text{TNR}(\widehat{\m{B}},\m{B}{)}=\frac{ \text{Card}\left\{(k,j): \hat{{B}}_{kj}= 0 \;\text{and} \; {{B}}_{kj}=0 \right\}}{ \text{Card}\left\{(k,j):{{B}}_{kj}=0\right\}}.
\end{align*}
\item[-] \textit{Prediction error}: the mean square error between the true value of $\m{Y}$ and $\hat{\m{Y}}$.
\end{itemize}

\subsection{Comparison of prediction formulas}
\label{sect-simulation-comparison_prediction_formulae}

\begin{figure}[!h]
\centering
    \includegraphics[width=0.6\textwidth]{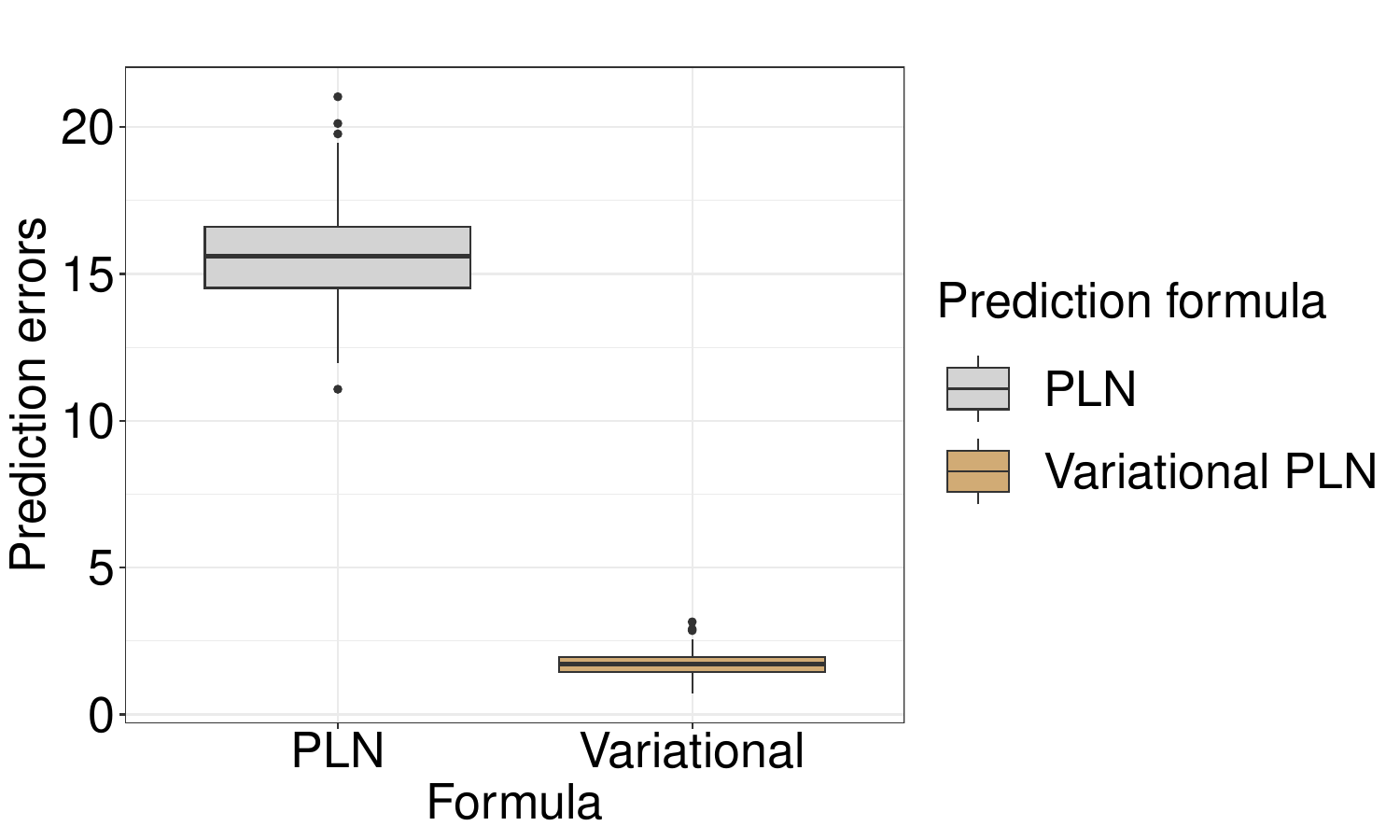}
\caption{\textsc{Comparison of the prediction errors of both prediction formulas described in Section~\ref{sect-PLN-sparse}.}
Both prediction formula performances are evaluated on 100 data sets. 
}
 \label{fig-simulation-comparison_prediction_formulae} 
\end{figure}

To illustrate the performance of the two prediction formulae outlined in Section~\ref{sect-PLN-sparse}, we simulate 100 datasets on which the PLN model is fitted. Specifically, each database is simulated under the scenario of a full covariance structure. The results of this simulation are presented in Figure~\ref{fig-simulation-comparison_prediction_formulae}, showing that the prediction formula not involving variational parameters generally yields weak results. In the following sections on the simulation study of prediction and variable selection performance, PLN performances are evaluated based on predictions made by the formula yielding the best results, which is the formula involving variational parameters.

\subsection{Simulation results}

\noindent{\bf Numerical results for one configuration.} To straightforwardly present numerical results of competing methods, we propose highlighting the estimated coefficients of each method in the configuration where: $n=10000$, $p=4$, and the structure of $\m{\Sigma}$ is the Full case.
Table \ref{tab:Simulation_results:Uncorrelated_variables:coefficients_extimation_fixed_sample} provides the results of coefficients estimation.
We observe that SICPLN outperforms GLMNET for accurate coefficient estimation, 
probably because GLMNET uses a LASSO penalty, which can introduce bias into the estimates.
In addition, unlike GLMNET, 
SICPLN exploits dependency structures between multidimensional responses to improve estimates. 
Finally, with SICPLN, coefficients associated with irrelevant variables 
are estimated as exactly zero, which is not always the case with GLMNET (see species 4).

 \begin{table}[H]
       \centering
        \caption{\textsc{Estimated coefficients with PLN, GLMNET and SICPLN}.}
        \begin{tabular}{cccccccc}
            \hline
             Species & Estimation method  & $\m{x_1}$&  $\m{x_2}$& $\m{x_3}$ & $\m{x_4}$ & $\m{x_5}$ & $\m{x_6}$\\
             \hline
            \rowcolor{gray!20}{Species 1} & True coefficient & 0&1&1&1&1&0 \\
             & PLN   &0.08&1.04&1.09&1.07&1.10&0.09\\
            & GLMNET   &0&0.91&0.99&0.93&0.96&0\\
             & SICPLN  & 0&0.95&1&0.98&0.92&0 \\
            \hline
             \rowcolor{gray!20}Species 2 & True coefficient & 0.5&0&0&1&1&0 \\
             & PLN   &0.55&0.07&0.15&1.04&0.99&0.13\\
            & GLMNET &0.43&0&0&0.98&0.87&0\\
             & SICPLN  & 0.47&0&0&0.98&0.92&0\\
              \hline
           \rowcolor{gray!20} Species 3 & True coefficient & 1&0.5&0.5&1&1&0 \\
             & PLN &1.10&0.58&0.54&1.05&1.06&0.10\\
            & GLMNET  &0.97&0.39&0.43&1.05&0.84&0\\
             & SICPLN  & 1&0.48&0.44&0.96&0.97&0 \\
            \hline
             \rowcolor{gray!20}Species 4 & True coefficient & 1&1&0&0&0.5&0 \\
             & PLN   &0.91&0.95&0.05&0.10&0.52&0.02\\
            & GLMNET   & 0.64&1.14&0.10&-0.14&0.14&-0.21 \\
             & SICPLN  & 0.94&0.98&0&0&0.54&0 \\
            \hline
        \end{tabular}
        \label{tab:Simulation_results:Uncorrelated_variables:coefficients_extimation_fixed_sample}
        \end{table}

\begin{figure}[!h]
\centering
\begin{subfigure}{0.47\textwidth}
   \includegraphics[width=\textwidth]{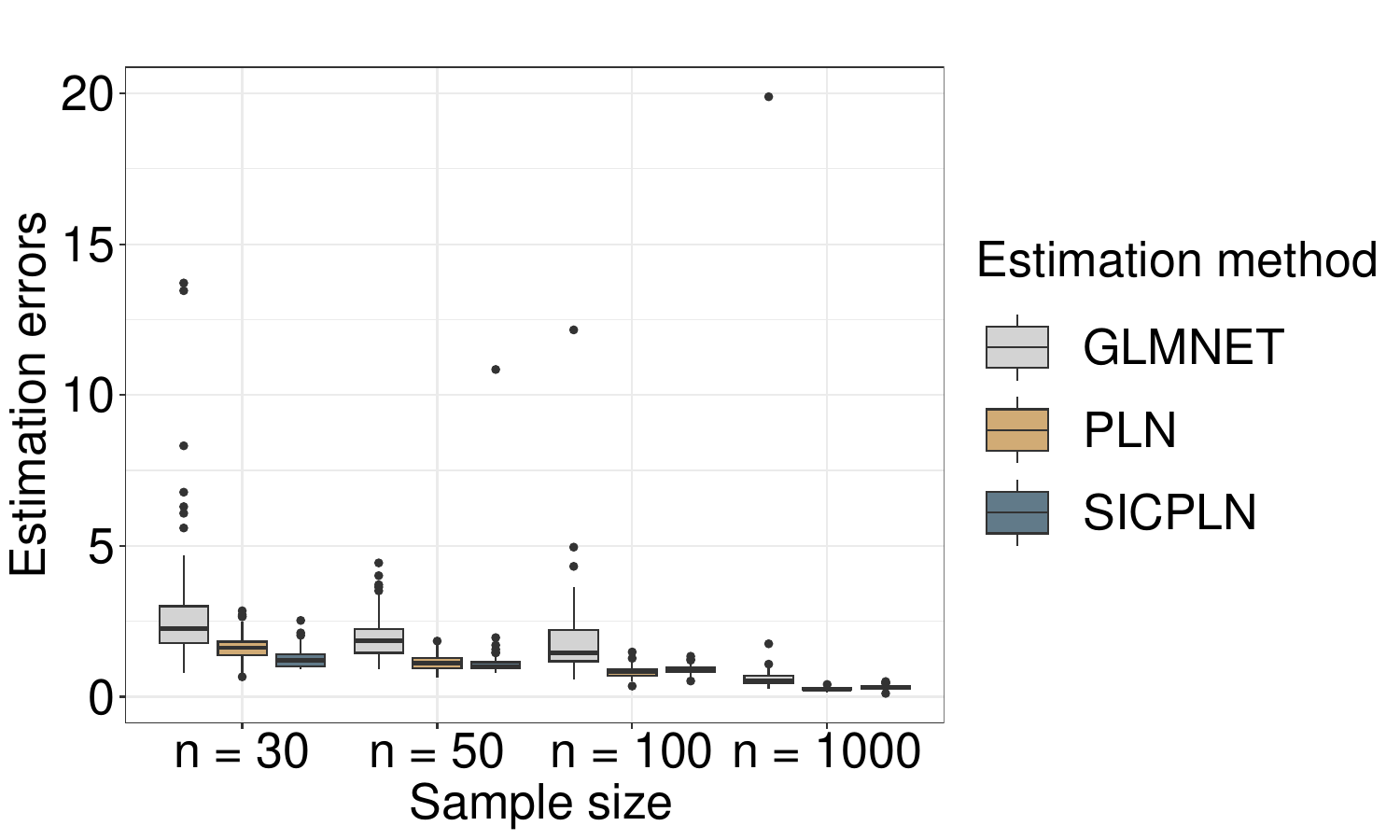}
\caption{}
    \label{fig:Simulation_results:Uncorrelated_variables:full_covariance:coefficients_error:sample}
\end{subfigure}
\hfill
\begin{subfigure}{0.47\textwidth}
    \includegraphics[width=\textwidth]{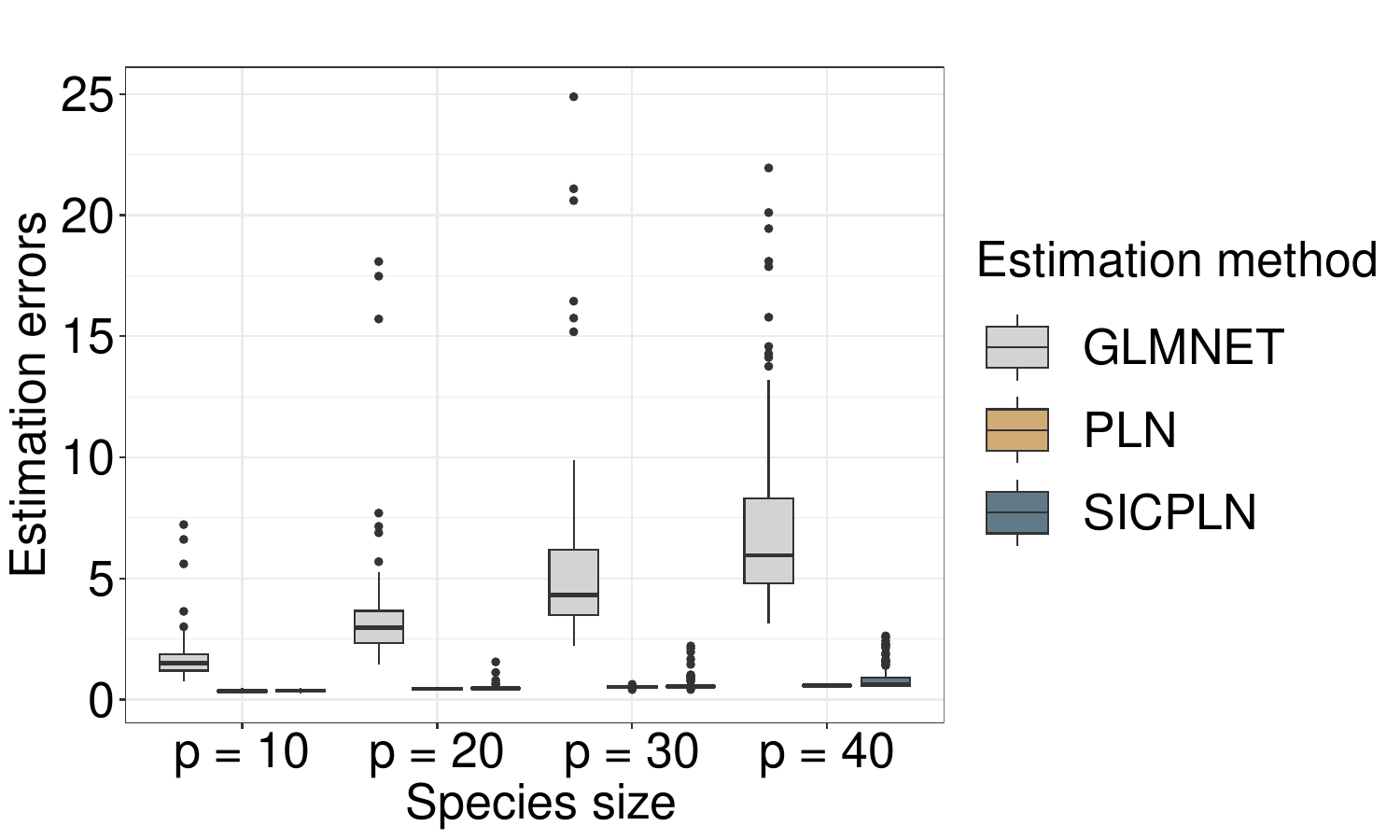}
    \caption{}
    \label{fig:Simulation_results:Uncorrelated_variables:full_covariance:coefficients_error:species}
\end{subfigure}        
\caption{\textsc{Boxplots of 100 replication values depicting the difference between the true coefficient matrix and the estimated coefficients by GLMNET, PLN, and SICPLN with full covariance}. Left plot shows the boxplots with variations in sample size $n$ (30, 50, 100, 1000). Right plot presents the boxplots for various numbers of columns in the counting matrix from 10 to 40.}
\label{fig:Simulation_results:Uncorrelated_variables:full_covariance:coefficients_error}
\end{figure}

\begin{figure}
\centering
\begin{subfigure}{0.47\textwidth}
   \includegraphics[width=\textwidth]{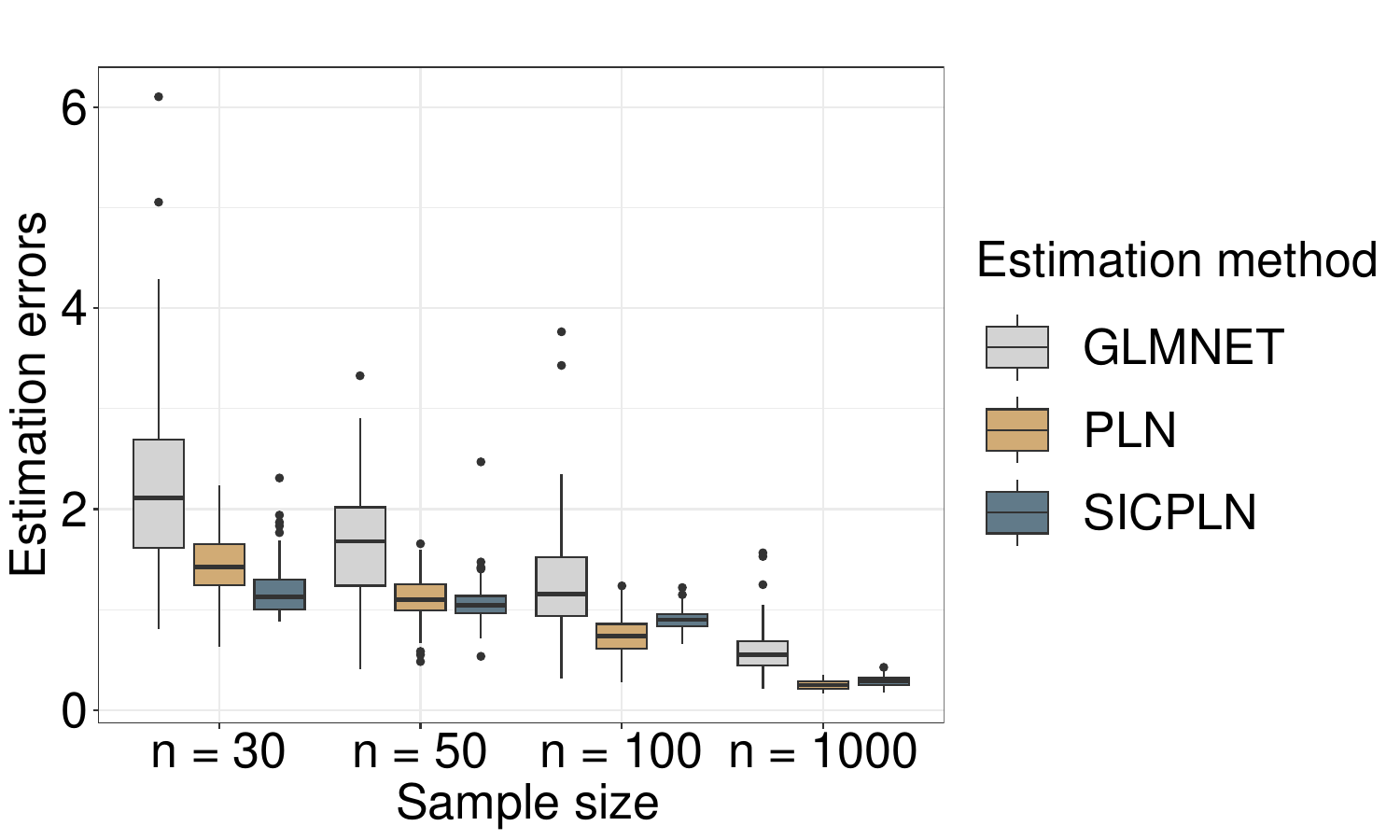}
 \caption{}
    \label{fig:Simulation_results:Uncorrelated_variables:diagonal_covariance:coefficients_error:sample}
\end{subfigure}
\hfill
\begin{subfigure}{0.47\textwidth}
    \includegraphics[width=\textwidth]{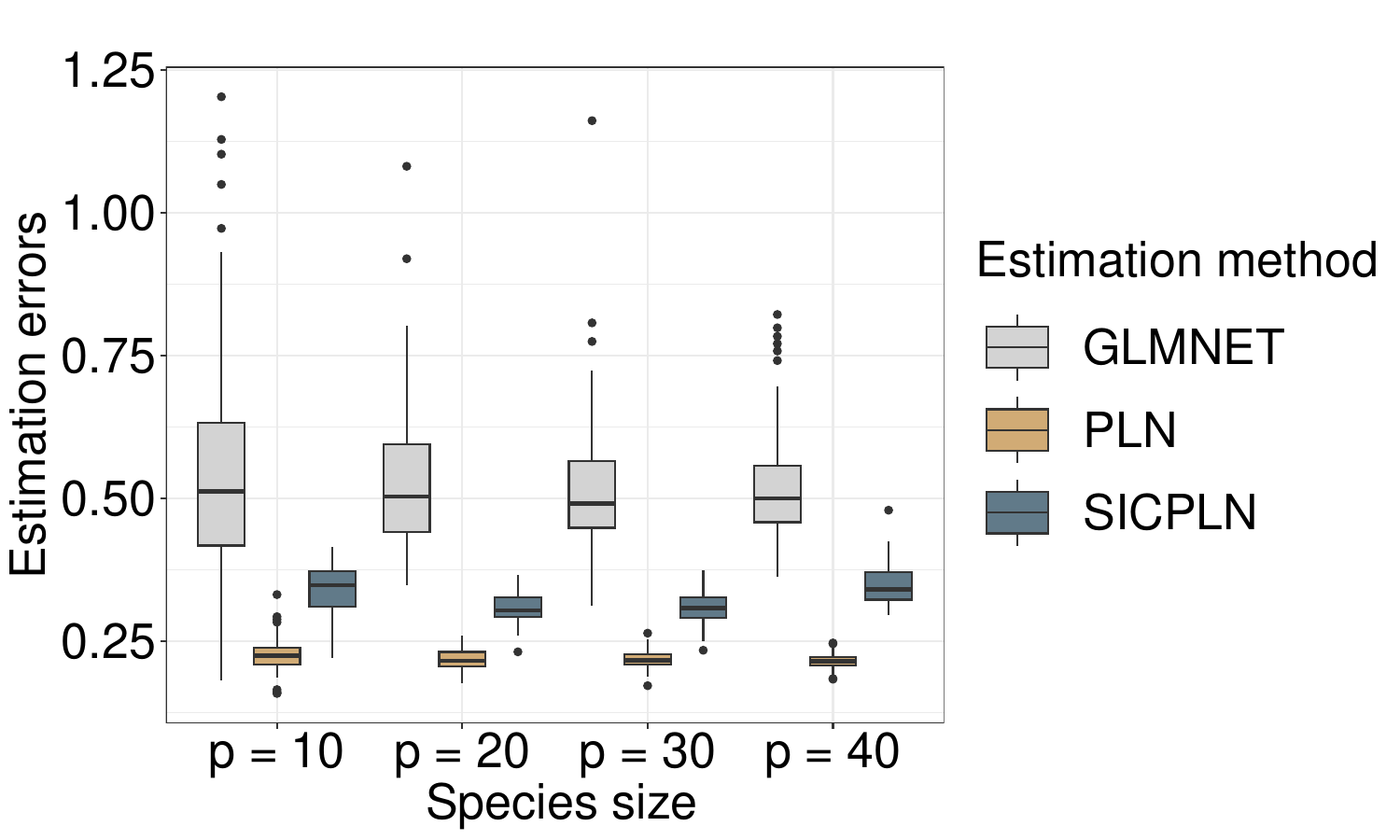}
    \caption{}
    \label{fig:Simulation_results:Uncorrelated_variables:diagonal_covariance:coefficients_error:species}
\end{subfigure}        
\caption{\textsc{Boxplots of 100 replication values depicting the difference between the true coefficient matrix and the estimated coefficients by GLMNET, PLN, and SICPLN with diagonal covariance}. Left plot shows the boxplots with variations in sample size $n$ (30, 50, 100, 1000). Right plot presents the boxplots for various numbers of columns in the counting matrix from 10 to 40.}
\label{fig:Simulation_results:Uncorrelated_variables:diagonal_covariance:coefficients_error}
\end{figure}

\noindent{\bf Results in terms of estimation error.}
Figure \ref{fig:Simulation_results:Uncorrelated_variables:full_covariance:coefficients_error:sample} shows that coefficient estimation accuracy is improved when the dependency structure between counts is taken into account. In addition, as Table \ref{tab:Simulation_results:Uncorrelated_variables:coefficients_extimation_fixed_sample} shows, SICPLN leads to sparser and more accurate estimates. In Figure \ref{fig:Simulation_results:Uncorrelated_variables:full_covariance:coefficients_error:species}, when the number of columns in the count matrix increases, indicating complexity in the dependency structure between counts, the estimation error with GLMNET significantly increases  compared to PLN and SICPLN. The estimation errors of SICPLN also increase as the number of columns in the count matrix increases. 
What we observe here can be explained by two phenomena described in the following.
The first one is related to the tendency of SICPLN to overly shrink or to set exactly to zero all coefficients associated with species with multiple irrelevant variables. The second one could be linked to the non-convergence of the algorithm during the Fisher scoring step. However, in the less complex situations represented in figures \ref{fig:Simulation_results:Uncorrelated_variables:diagonal_covariance:coefficients_error}, i.e. when there is no dependency between the count columns, the estimation errors with GLMNET remain very high compared with SICPLN and PLN when the sample size becomes large and the number of columns in the count matrix increases.

\noindent{\bf Results in terms of true negative rate.}
The averaged true negative rates on 100 replications are presented in Table \ref{tab:Simulation_results:Uncorrelated_variables:performance:sample} for various sample sizes. Similarly, Table \ref{tab:Simulation_results:Uncorrelated_variables:performance:species} displays these rates for different numbers of columns in the count matrix. 
We observe that the true negative rates increases for each method when the sample size increases. SICPLN performs better than GLMNET in terms of sparsity measurement, showing a tendency to accurately estimate exactly zero entries in the regression coefficient matrix. The increase in the number of columns in the count matrix leads to a degradation of GLMNET and SICPLN true negative rates, especially in presence of dependencies between counts. The degradation of the average true negative rate of SICPLN is primarily due to non-convergence, resulting in coefficients that are not exactly set to zero. In the presence of a simpler dependency structure, SICPLN exhibits significantly higher true negative rates compared to GLMNET.

             \begin{table}[H]
       \centering
        \caption{ \textsc{True negative rate (TNR) for various sample sizes in function to covariance structures, averaged over 100 replications}. TNR is always 0 for PLN since the estimation procedure is not designed to impose sparsity.}
        \begin{tabular}{ccccc}
            \hline
              Covariance structure& Sample size   &  GLMNET &  PLN  &  SICPLN  \\
            \hline
            Full  &  $ 30$ &  0.006 & 0  & 0.79\\
            Diagonal  & $ $ &   0.09 & 0  & 0.82  \\
            \hline
            Full  &  $50 $ &  0.007 & 0  & 0.79 \\
             Diagonal  & $  $ &  0.08 & 0  & 0.9 \\
             \hline
             Full  & $ 100$ &   0.07 & 0  & 0.87 \\
             Diagonal  &  $ $ &   0.13 & 0  & 0.90 \\
              \hline
            Full  &  $1000 $ &  0.18 & 0  & 0.98 \\
             Diagonal  & $ $ &   0.22 &0  & 0.95 \\
            \hline
        \end{tabular}
        \label{tab:Simulation_results:Uncorrelated_variables:performance:sample}
        \end{table}

             \begin{table}[H]
       \centering
        \caption{ \textsc{True negative rate (TNR) for various columns of the counting matrix as a function of dependence  structures on 100 replications.} TNR is always 0 for PLN since the estimation procedure is not designed to impose sparsity.}
        \begin{tabular}{ccccc}
            \hline
             Covariance structure  & Species size &   GLMNET  &  PLN &  SICPLN  \\
            \hline
            Full  & $10 $ &   0.01 & 0  & 0.55 \\
           Diagonal  &  $  $ &   0.17 & 0  & 0.91\\
            \hline
            Full  & $ 20$ &   0.007 & 0  & 0.19 \\
           Diagonal  &  $  $ &   0.11 & 0 & 0.92 \\
             \hline
            Full  & $30 $ &   0.006 & 0 &  0.04 \\
            Diagonal  & $ $ &   0.09 & 0  & 0.93 \\
              \hline
           Full  &  $40 $ &   0.005 & 0  & 0.07 \\
            Diagonal  & $ $ &   0.09& 0 & 0.90 \\
             \hline
        \end{tabular}
        \label{tab:Simulation_results:Uncorrelated_variables:performance:species}
        \end{table}

\begin{figure}
\centering
\begin{subfigure}{0.47\textwidth}
   \includegraphics[width=\textwidth]{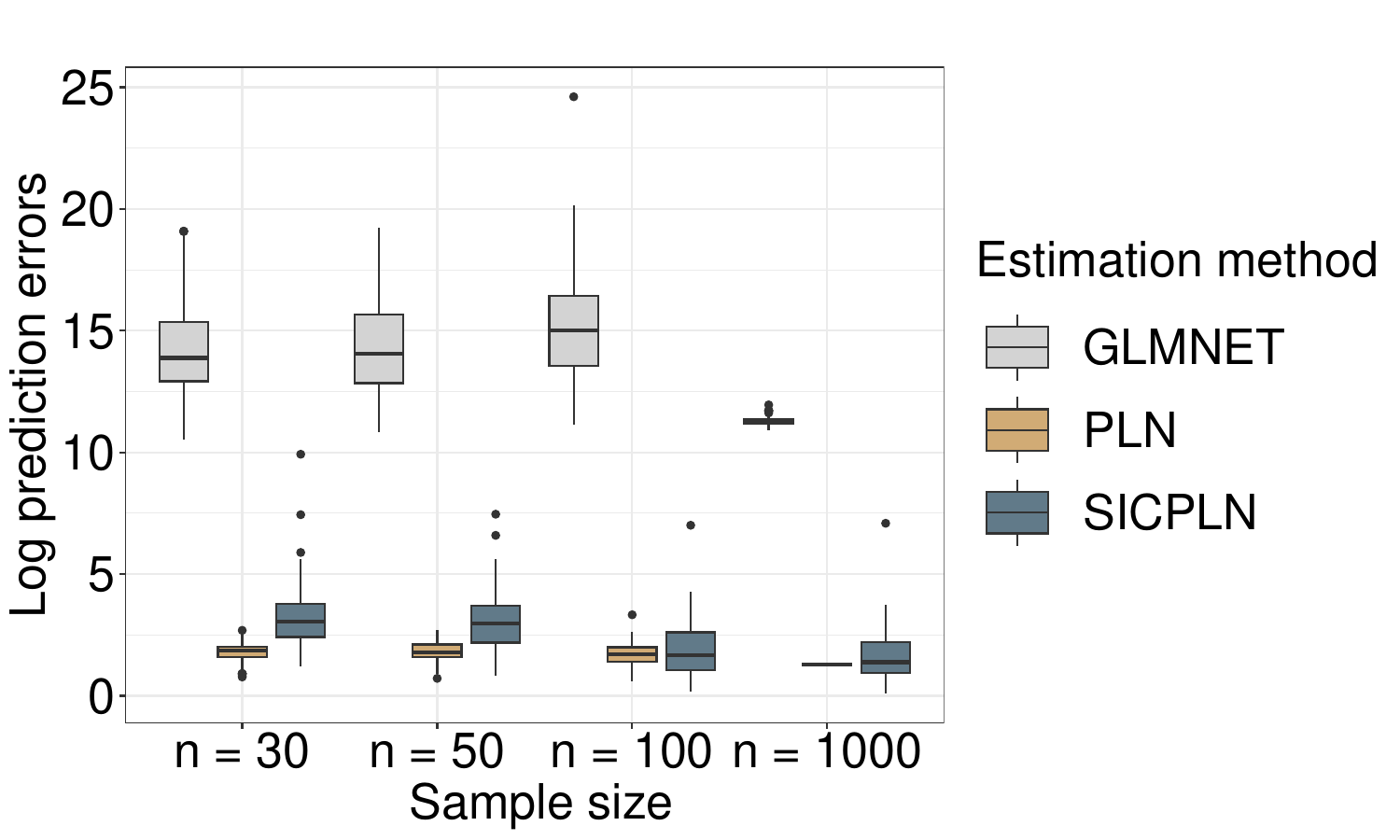}
\caption{}
\label{fig:Simulation_results:Uncorrelated_variables:full_covariance:prediction_error:sample}
\end{subfigure}
\hfill
\begin{subfigure}{0.47\textwidth}
    \includegraphics[width=\textwidth]{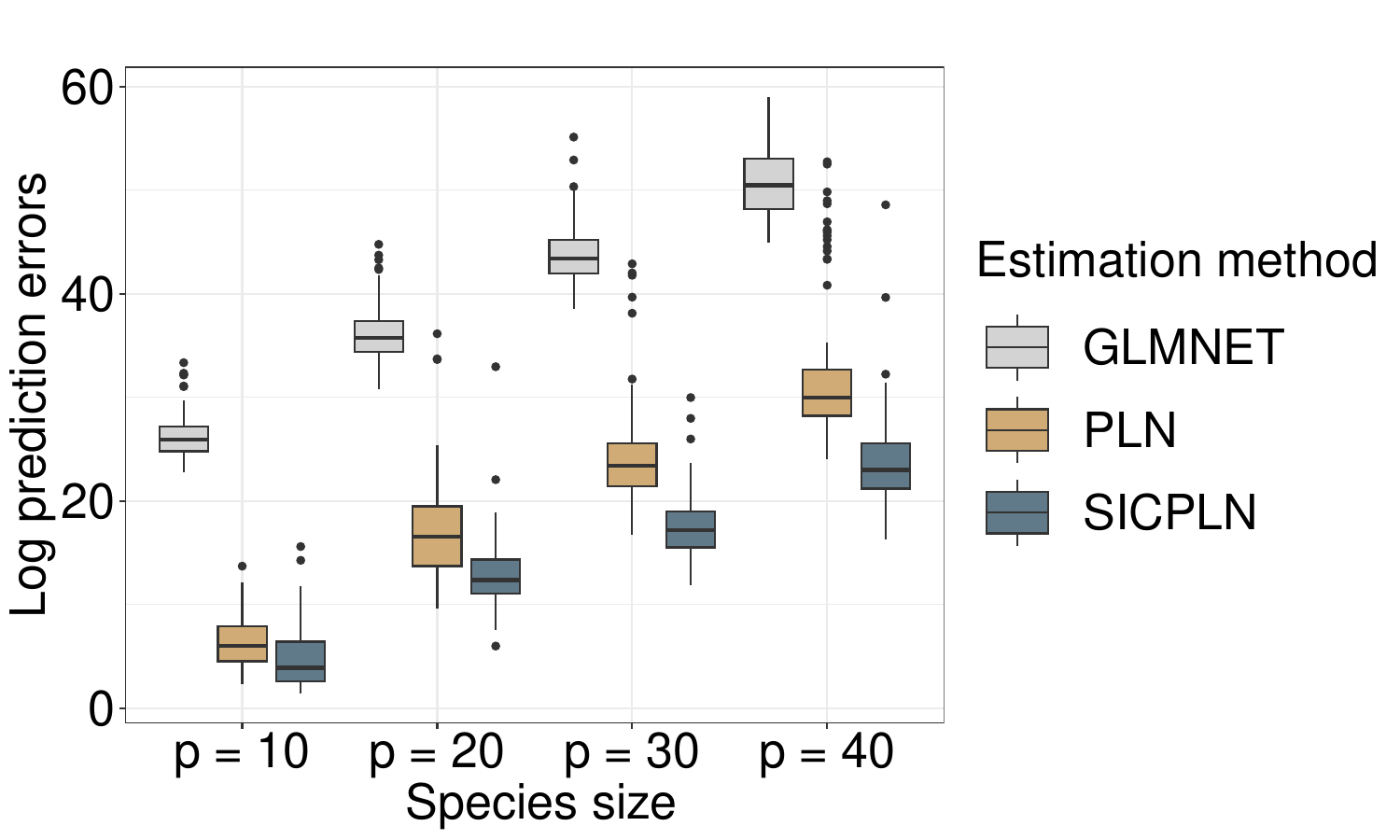}
    \caption{}
\label{fig:Simulation_results:Uncorrelated_variables:full_covariance:prediction_error:species}
\end{subfigure}        
\caption{\textsc{Prediction error when the sample size and the number of columns in the counting matrix varying with full covariance averaged on 100 replication.} Left plot shows the variations by changing the sample size from 30 to 1000. Right plot  shows the variations in the number of columns in the counting matrix from 10 to 40.}
\label{fig:Simulation_results:Uncorrelated_variables:full_covariance:prediction_error}
\end{figure}

\begin{figure}
\centering
\begin{subfigure}{0.47\textwidth}
    \includegraphics[width=\textwidth]{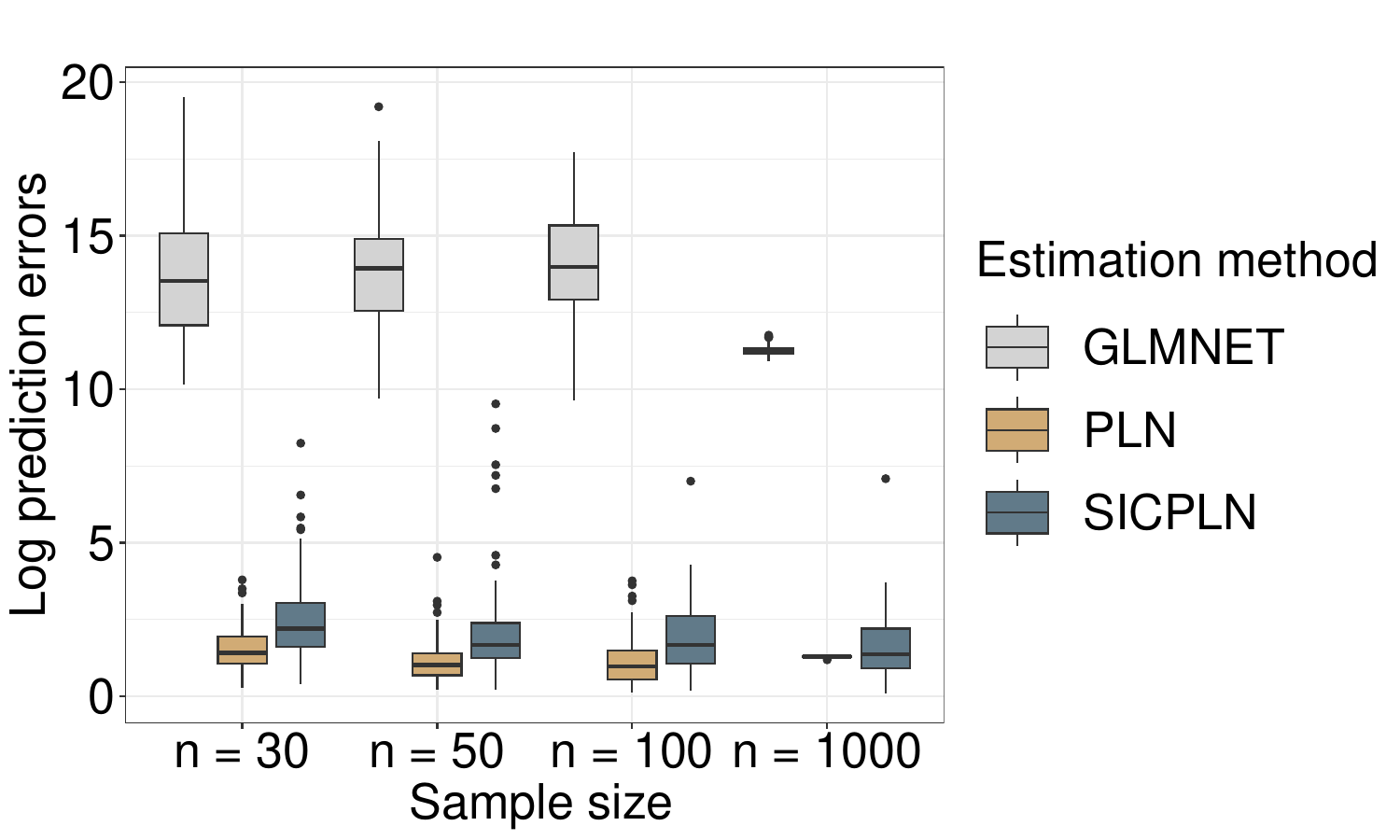}
 \caption{}
\label{fig:Simulation_results:Uncorrelated_variables:diagonal_covariance:prediction_error:sample}
\end{subfigure}
\hfill
\begin{subfigure}{0.47\textwidth}
    \includegraphics[width=\textwidth]{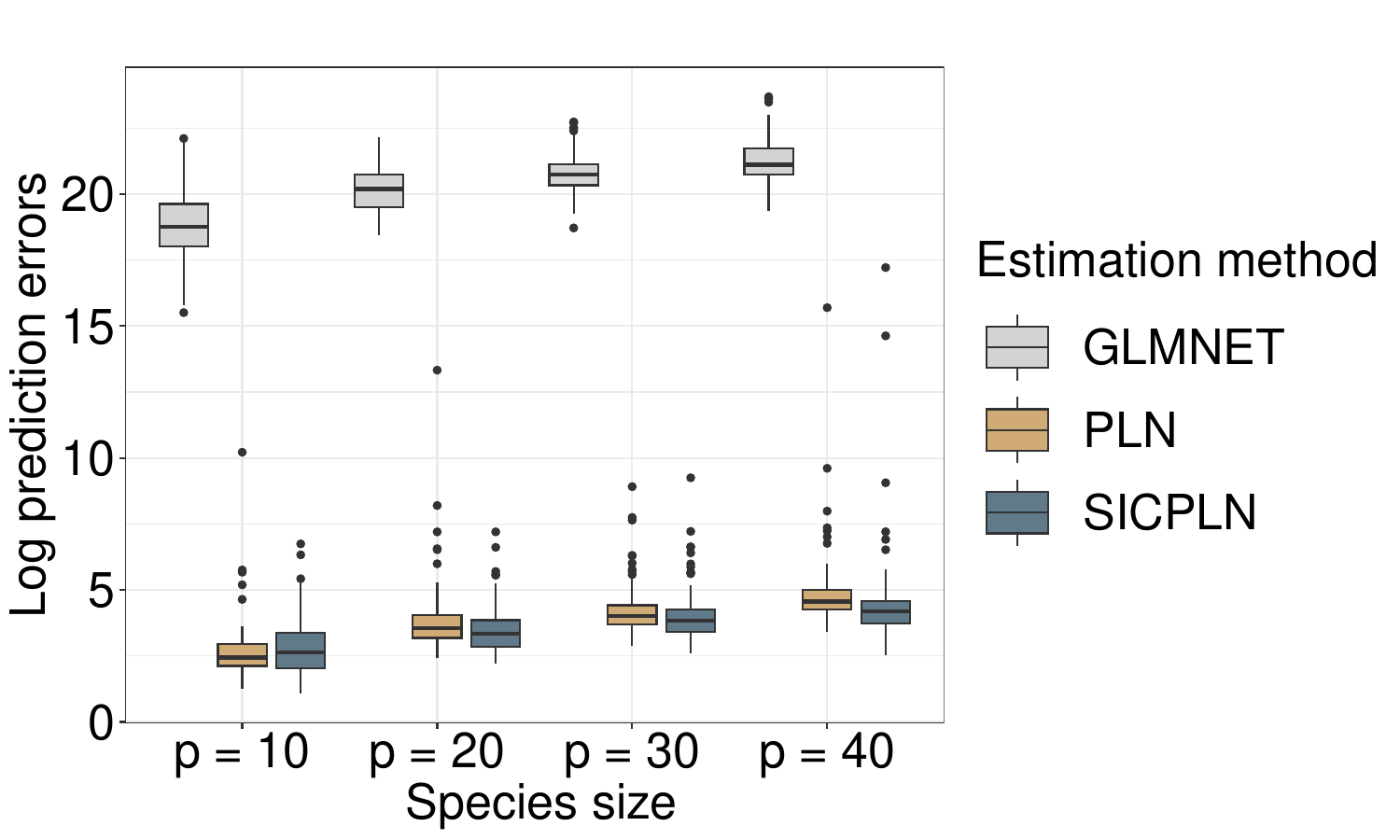}
   \caption{}
\label{fig:Simulation_results:Uncorrelated_variables:diagonal_covariance:prediction_error:species}
\end{subfigure}        
\caption{\textsc{Prediction error when the sample size and the number of columns in the counting matrix varying with diagonal covariance over 100 simulation}. Left plot shows the variations by changing the sample size from 30 to 1000. Right plot  shows the variations in the number of columns in the counting matrix from 10 to 40.}
\label{fig:Simulation_results:Uncorrelated_variables:diagonal_covariance:prediction_error}
\end{figure}

\noindent{\bf Results in terms of prediction error.}
Prediction errors obtained for various sample sizes and species are presented in figures \ref{fig:Simulation_results:Uncorrelated_variables:full_covariance:prediction_error} and \ref{fig:Simulation_results:Uncorrelated_variables:diagonal_covariance:prediction_error}, on 100 replications. We observe that when the sample size is small, prediction errors for PLN and SICPLN  are lower compared to GLMNET.
However, as the sample size increases, the performance of SICPLN improves significantly compared to PLN and GLMNET in terms of prediction accuracy. Additionally, we observe that for a fixed sample size $n\;=\;1000$, variations of the number of columns in the counting matrix yield better prediction performances for SICPLN compared to PLN and GLMNET when the covariance structure is in the Full case. Nonetheless, when counts are independent
, SICPLN often shrinks to zero the coefficients associated with species that have multiple non-relevant variables. This leads to a slight increase in the estimation error of SICPLN coefficients compared to PLN, but the prediction errors between SICPLN and PLN remain similar.




\section{Real data analyses}
\label{sect-real_data}

We illustrate the proposed method through the analysis of two real abundance datasets. Our aim is to identify the relevant environmental variables that describe or control the abundances observed in the count matrix.

\subsection{Genus data}
\label{sect-real_data-genus}

{\bf Data description.} We first consider the genus data set from the \textsf{R} package \texttt{SCGLR}, 
recently reanalysed by \citet{doi:10.1080/10618600.2019.1598870}. 
The data set describes the abundance of 15 tree genera common to Congo Basin forests, measured on 1000 parcels. 
To explain these genera, 46 environmental variables are available. 
Among these, 21 variables describe the physical characteristics of the environment, 
such as \verb|altitude|, \verb|rainfall|, soil conditions, and so on. 
Vegetation properties are represented by 25 photosynthetic activity indices obtained by teledetection, 
including \verb|EVI| (Enhanced Vegetation Indexes), \verb|MIR| (Middle InfraRed) and \verb|NIR| (Near-InfraRed). 
Some categorical variables such as \verb|inventory|, \verb|mois_sec_50|, \verb|mois_sec_100|, \verb|geology| 
and spatial coordinates, longitude (\verb|center_x|) and latitude (\verb|center_y|), are also available.
A preliminary analysis examined the correlations between the variables. 
This was done by calculating the variance inflation factor (VIF) for each variable. 
We then grouped all vegetation indices into a new variable called \verb|vegetation_index| 
and aggregated all rainfall variables from \verb|pluvio_1| to \verb|pluvio_12| into a single variable called \verb|rainfall|. 
Based on the VIF results in Table \ref{tab:Real_data_analyses:Genus_data:Results:VIF:all}, we only consider 4 variables: 
\verb|vegetation_index|, \verb|wetness|, \verb|center_x|, \verb|center_y| and \verb|surface|. Following the approach of \citet{doi:10.1080/10618600.2019.1598870}, we treat the variable \verb|surface| as an offset variable and then apply PLN, SICPLN and GLMNET.
        \begin{table}[!h]
            \centering
            \resizebox{1\linewidth}{!}{
            \begin{tabular}{c|cccccccccc}
              Variables&vegetation index  &rainfall & altitude & awd & mcwd & wetness & CWD &center x& center y &pluvio an\\
              \hline
              VIF &1.92 &14588.77&7.27 &42.28 & 43.17& 2.02&29.40&10.89&21.77 &14690.40
            \end{tabular}
            }
            \caption{\textsc{VIF with all variables.}}
            \label{tab:Real_data_analyses:Genus_data:Results:VIF:all}
        \end{table}  
        \begin{table}[!h]
            \centering
            \begin{tabular}{c|cccc}
              Variables&vegetation index  & wetness &center x& center y \\
              \hline
              VIF & 1.36 &1.75&1.46 &1.51 
            \end{tabular}
            \caption{ \textsc{VIF with selected variables.}}
            \label{tab:Real_data_analyses:Genus_data:Results:VIF:select}
        \end{table}    

\begin{figure}
\centering
\begin{subfigure}{0.7\textwidth}
   \includegraphics[clip, trim=0mm 4cm 0mm 4cm, width=\textwidth]{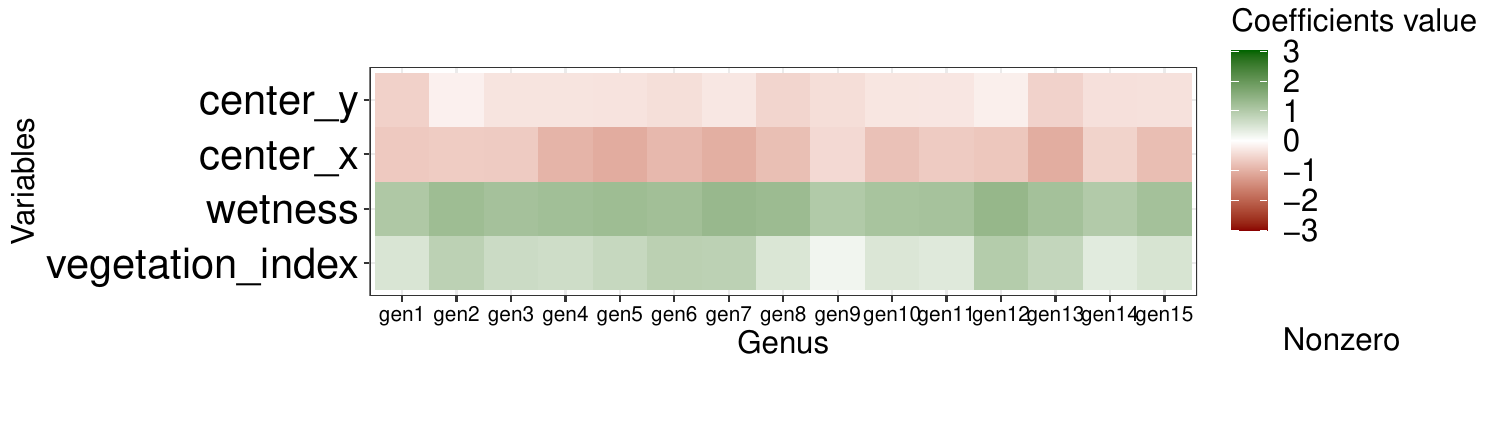}
 \caption{Coefficients estimation with PLN}
    \label{fig:Real_data_analyses:Genus_data:Results:Coefficients_estimation:PLN}
\end{subfigure}
\hfill
\begin{subfigure}{0.7\textwidth}
    \includegraphics[clip, trim=0mm 4cm 0mm 4cm, width=\textwidth]{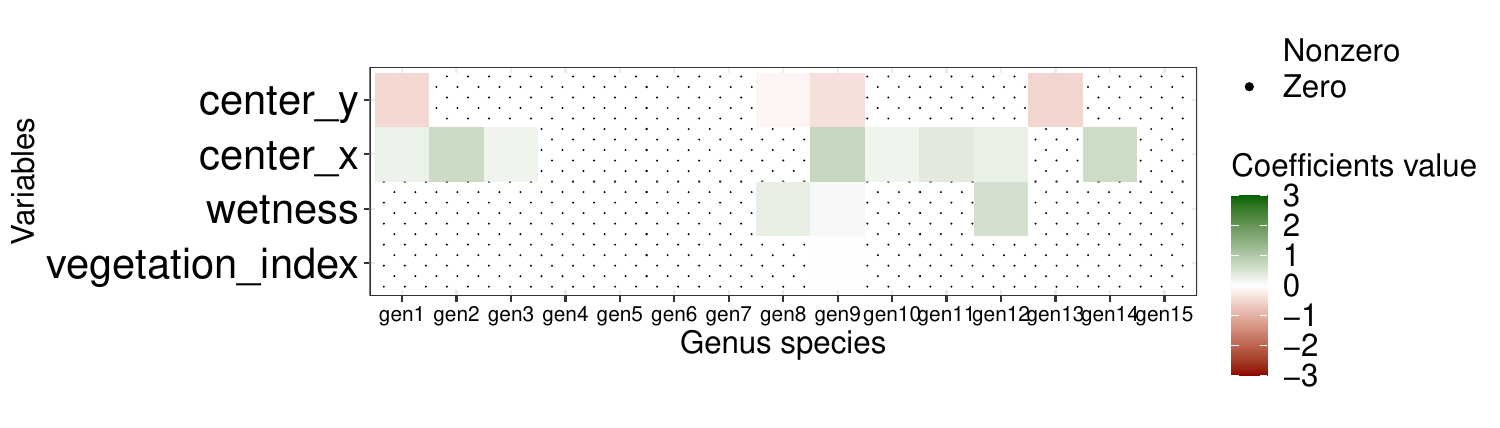}
    \caption{Coefficients estimation with GLMNET}
    \label{fig:Real_data_analyses:Genus_data:Results:Coefficients_estimation:GLMNET}
\end{subfigure}
\hfill
\begin{subfigure}{0.7\textwidth}
    \includegraphics[clip, trim=0mm 4cm 0mm 4cm, width=\textwidth]{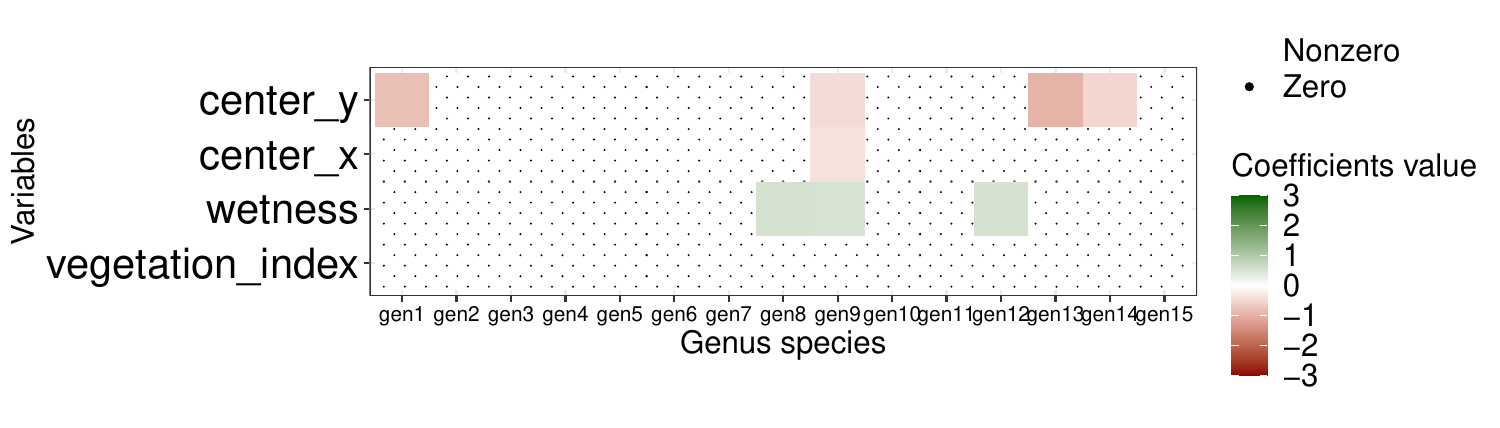}
    \caption{Coefficients estimation with SICPLN}
    \label{fig:Real_data_analyses:Genus_data:Results:Coefficients_estimation:SICPLN}
\end{subfigure}   
\caption{\textsc{Coefficients estimation with PLN, GLMNET and SICPLN.}}
\label{fig:Real_data_analyses:Genus_data:Results:Coefficients_estimation}
\end{figure}

\noindent {\bf Results.} 
The estimated coefficients for each species are shown in Figure
\ref{fig:Real_data_analyses:Genus_data:Results:Coefficients_estimation}.
This figure first includes the results obtained with the simple PLN model 
(Figure \ref{fig:Real_data_analyses:Genus_data:Results:Coefficients_estimation:PLN}). 
Next, Figure \ref{fig:Real_data_analyses:Genus_data:Results:Coefficients_estimation:GLMNET}
shows the coefficients obtained with GLMNET, where the abundance of each species is modeled in a univariate way 
(without taking into account the dependency structure) using a LASSO-based penalized Poisson regression.
Finally, Figure \ref{fig:Real_data_analyses:Genus_data:Results:Coefficients_estimation:SICPLN}
shows the results obtained with the SICPLN approach.

The regression coefficients estimated by the PLN model are non-zero, 
posing challenges in identifying relevant explanatory variables that significantly influence the abundance of species.
The estimates derived from GLMNET are close to those obtained with SICPLN in terms of sparsity. 
This closeness is certainly due to the lack of dependence between the genus species, as illustrated in 
Figure \ref{fig:Real_data_analyses:Results:precision_matrix:genus}. 
Specifically, 16 out of 60 coefficients were estimated as non-zero with GLMNET, compared to only 8 out of 60 for SICPLN. 
The abundance of most species does not seem to be related to environmental variables. 
Only the abundance of \verb|genus 8|, \verb|genus 9|, \verb|genus 12|, \verb|genus 13|, and \verb|genus 14| 
appears to be explained by one or more environmental variables.

\begin{figure}
\centering
\begin{subfigure}{0.47\textwidth}
   \includegraphics[width=\textwidth]{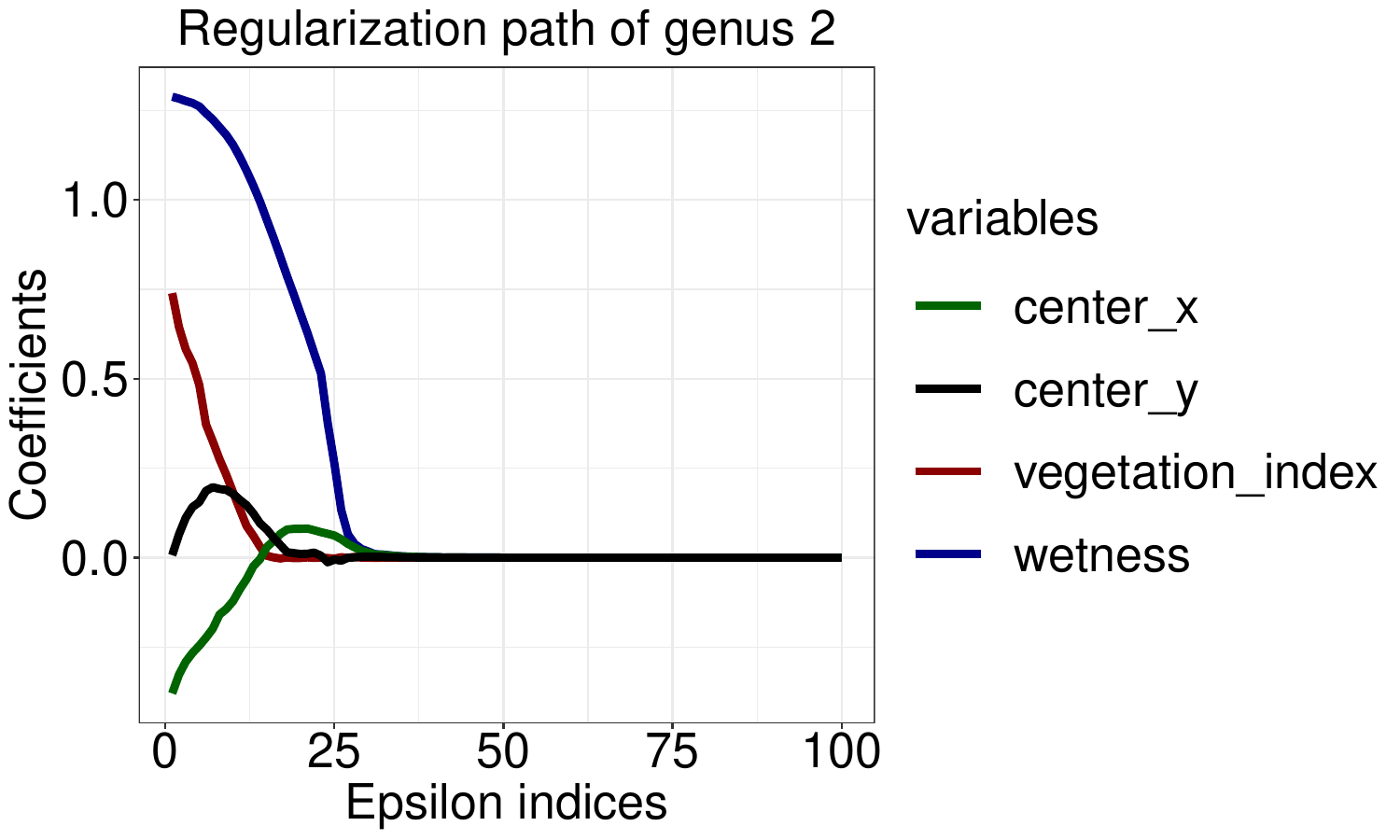}
 \caption{}
\label{fig:Real_data_nalyses:Genus_data:Results:regularisation_path:SICPLN:specie_2}
\end{subfigure}
\begin{subfigure}{0.47\textwidth}
   \includegraphics[width=\textwidth]{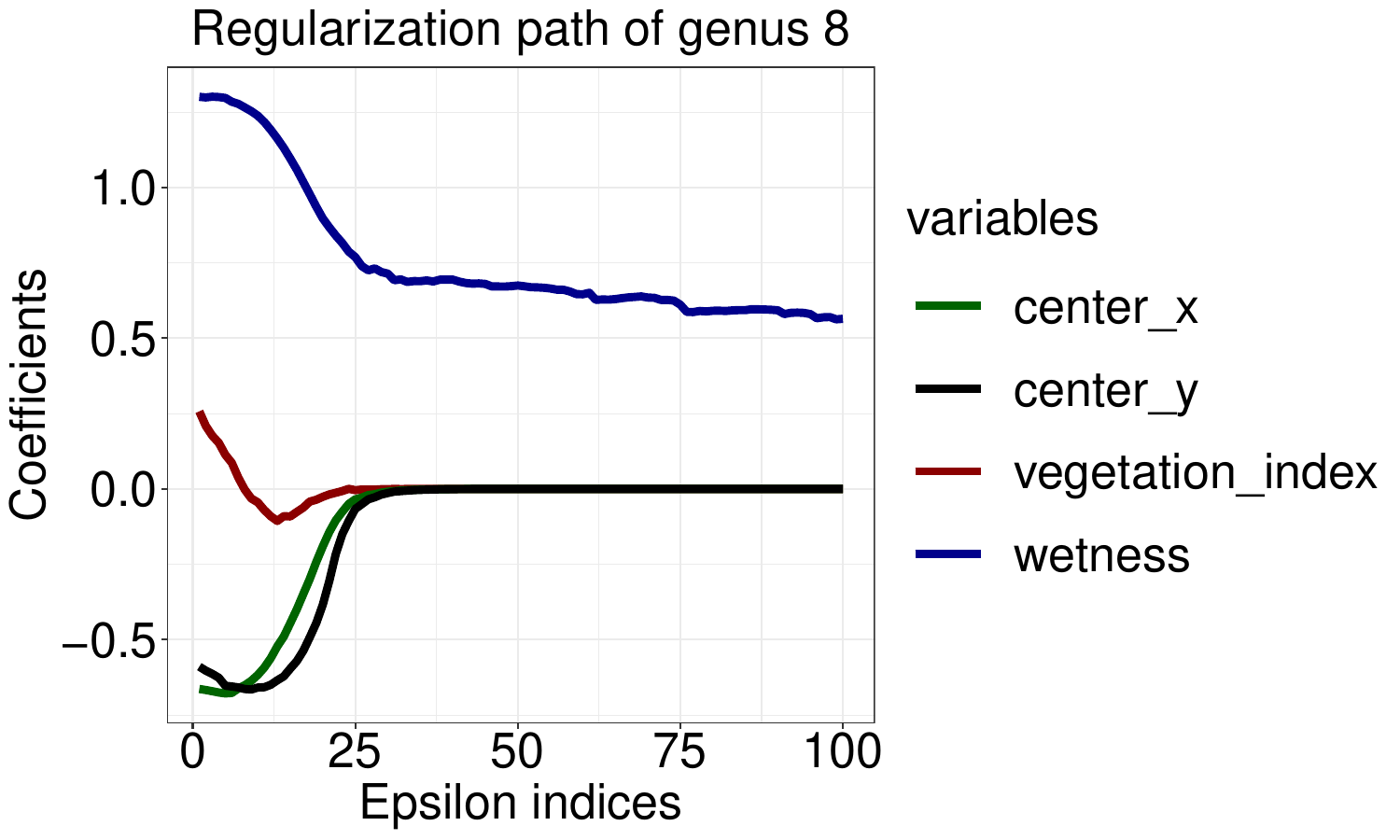}
 \caption{}
    \label{fig:Real_data_analyses:Genus_data:Results:regularisation_path:SICPLN:specie_8}
\end{subfigure}
\hfill
\begin{subfigure}{0.47\textwidth}
    \includegraphics[width=\textwidth]{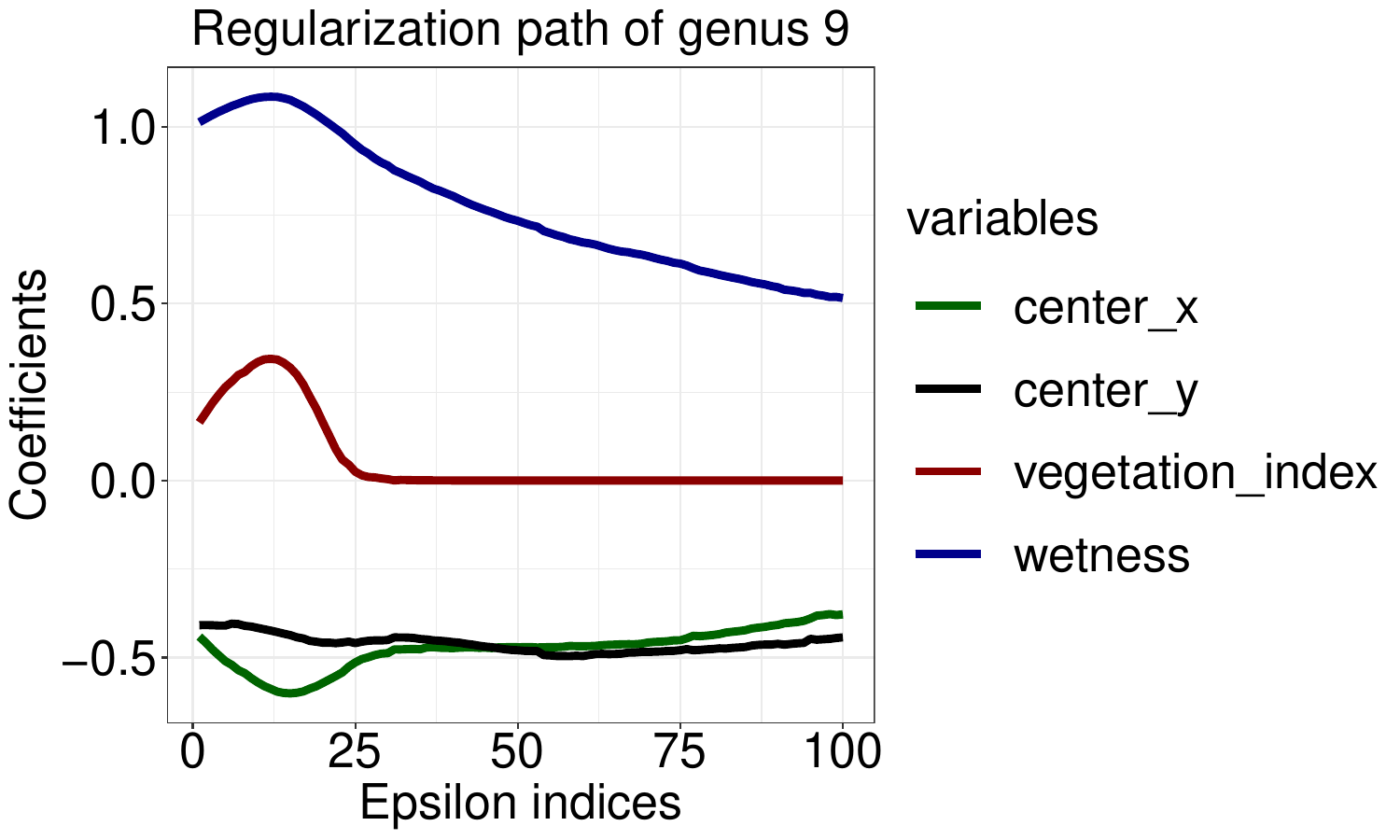}
    \caption{}
    \label{fig:Real_data_analyses:Genus_data:Results:regularisation_path:SICPLN:specie_9}
\end{subfigure}
\begin{subfigure}{0.47\textwidth}
    \includegraphics[width=\textwidth]{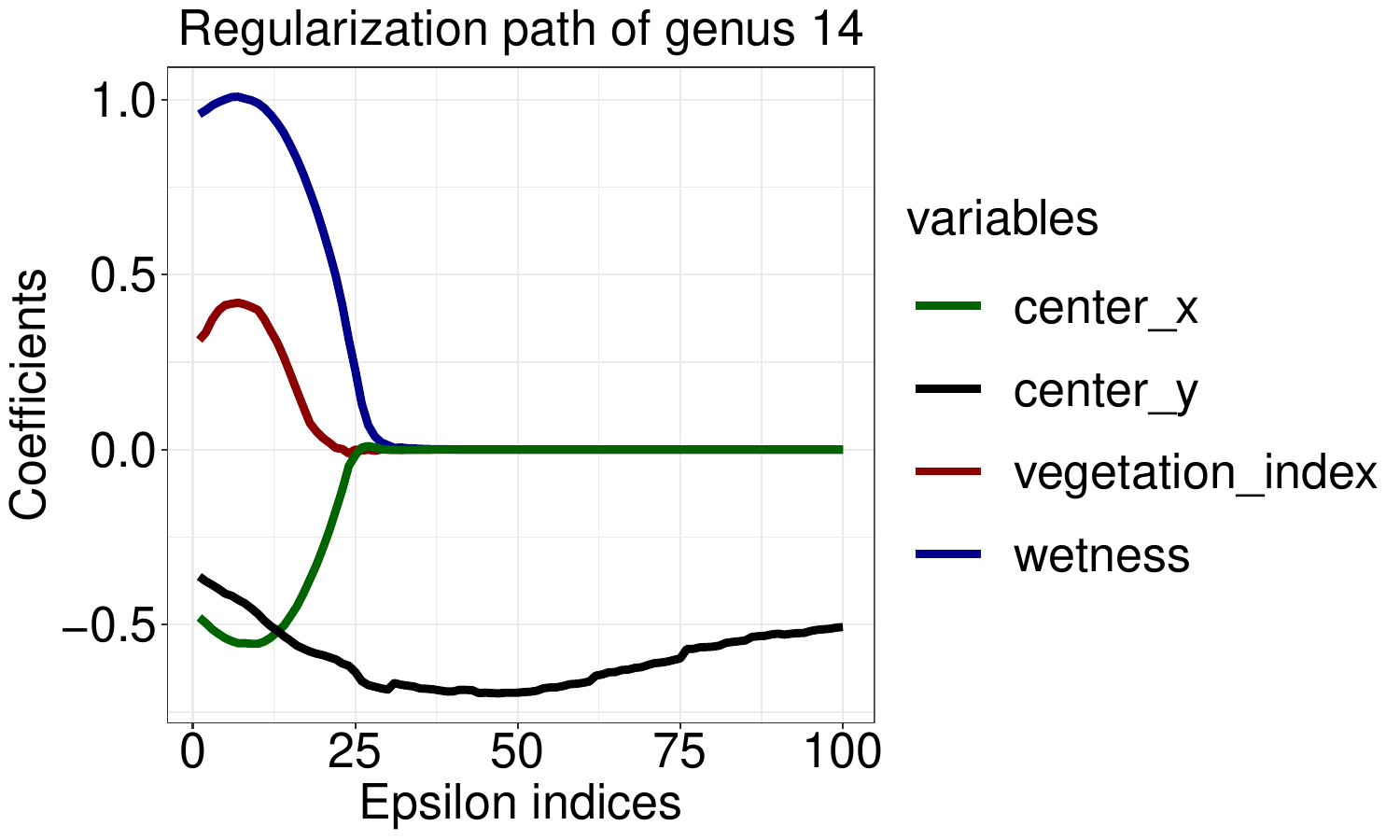}
    \caption{}
    \label{fig:Real_data_analyses:Genus_data:Results:regularisation_path:SICPLN:specie_14}
\end{subfigure}   
\caption{\textsc{Regularization of some species in genus data.}}
\label{fig:Real_data_analyses:Genus_data:Results:regularisation_path}
\end{figure}

Figure \ref{fig:Real_data_analyses:Genus_data:Results:regularisation_path} 
shows the regularization path using the SIC criterion. 
Unlike the LASSO \citep{tibshirani1996regression} or elastic net \citep{zou2005regularization} regularization path, 
it does not show the evolution of the coefficient values 
as a function of parameter $\lambda$ driving the intensity of regularization.
Instead, it shows the evolution of the estimated coefficients during the optimization step using the telescoping approach, 
which aims to move the problem closer to the $L_0$ penalty. 
This eventually sets the coefficient of the irrelevant variables to zero 
and stabilizes the coefficient of the relevant variables at a non-zero value.
It is worth noting that relevant variables may vary from one species to another. 
For instance, no variable is selected for \verb|genus 2| 
(Figure \ref{fig:Real_data_nalyses:Genus_data:Results:regularisation_path:SICPLN:specie_2}), 
only \verb|wetness| is selected for \verb|genus 8| 
(Figure \ref{fig:Real_data_analyses:Genus_data:Results:regularisation_path:SICPLN:specie_8}), 
and variables \verb|wetness|, \verb|center_x| and \verb|center_y| are selected for \verb|genus 9| 
(Figure \ref{fig:Real_data_analyses:Genus_data:Results:regularisation_path:SICPLN:specie_9}).

\subsection{The hunting spider dataset}

{\bf Data description.}
The hunting spider dataset used in this study, initially introduced by \citet{smeenk1974correlations}, 
comes from the \textsf{R} package \texttt{VGAM}. 
These data give the abundance of 12 spider species captured in traps across 28 sites. 
To characterize the observed abundances, six environmental variables are considered, namely 
\verb|WaterCon| (log percentage of soil dry mass), \verb|BareSand| (log percentage cover of bare sand), 
\verb|FallTwig| (log percentage cover of fallen leaves and twigs), \verb|CoveMoss| (log percentage cover of the moss layer),
\verb|CoveHerb| (log percentage cover of the herb layer), and 
\verb|ReflLux| (Reflection of the soil surface with cloudless sky). 

\begin{table}[!h]
            \centering
            \begin{tabular}{c|c|c|c|c|c|c}
              Variables&WaterCon  & BareSand &FallTwig& CoveMoss & CoveHerb& ReflLux\\
              \hline
              VIF & 4.34  &2.28&5.87&2.67&2.48&6.04 
            \end{tabular}
            \caption{\textsc{VIF for spiders environmental characteristics.}}
            \label{tab:Real_data_analyses:Spiders:Results:VIF}
        \end{table} 

\begin{figure}
\centering
\begin{subfigure}{0.70\textwidth}
    \includegraphics[clip, trim=0mm 3cm 0mm 3cm, width=\textwidth]{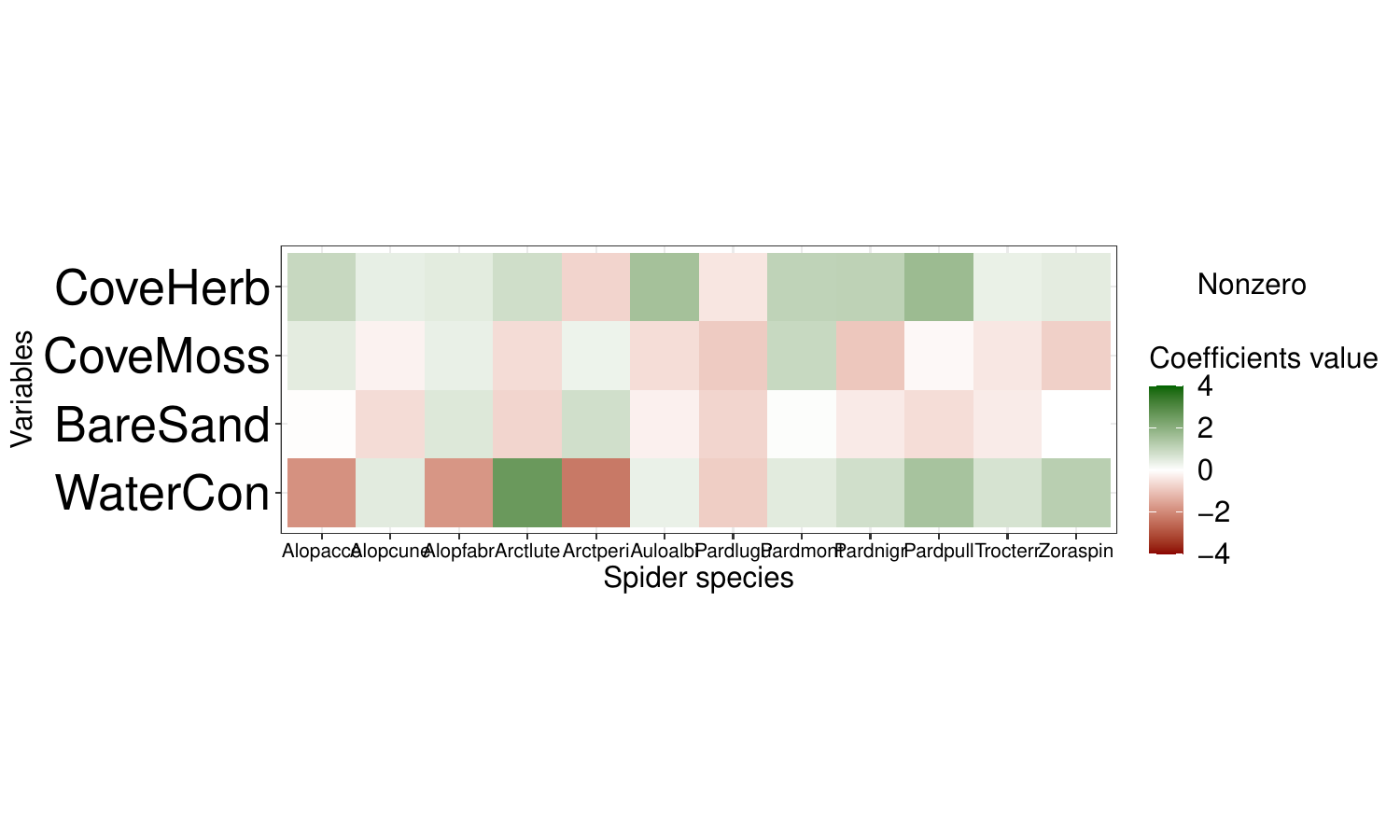}
 \caption{Coefficients estimation with PLN}
    \label{fig:Real_data_analyses:Spiders:Results:Coefficients_estimation:genus}
\end{subfigure}
\hfill
\begin{subfigure}{0.70\textwidth}
    \includegraphics[clip, trim=0mm 3cm 0mm 3cm, width=\textwidth]{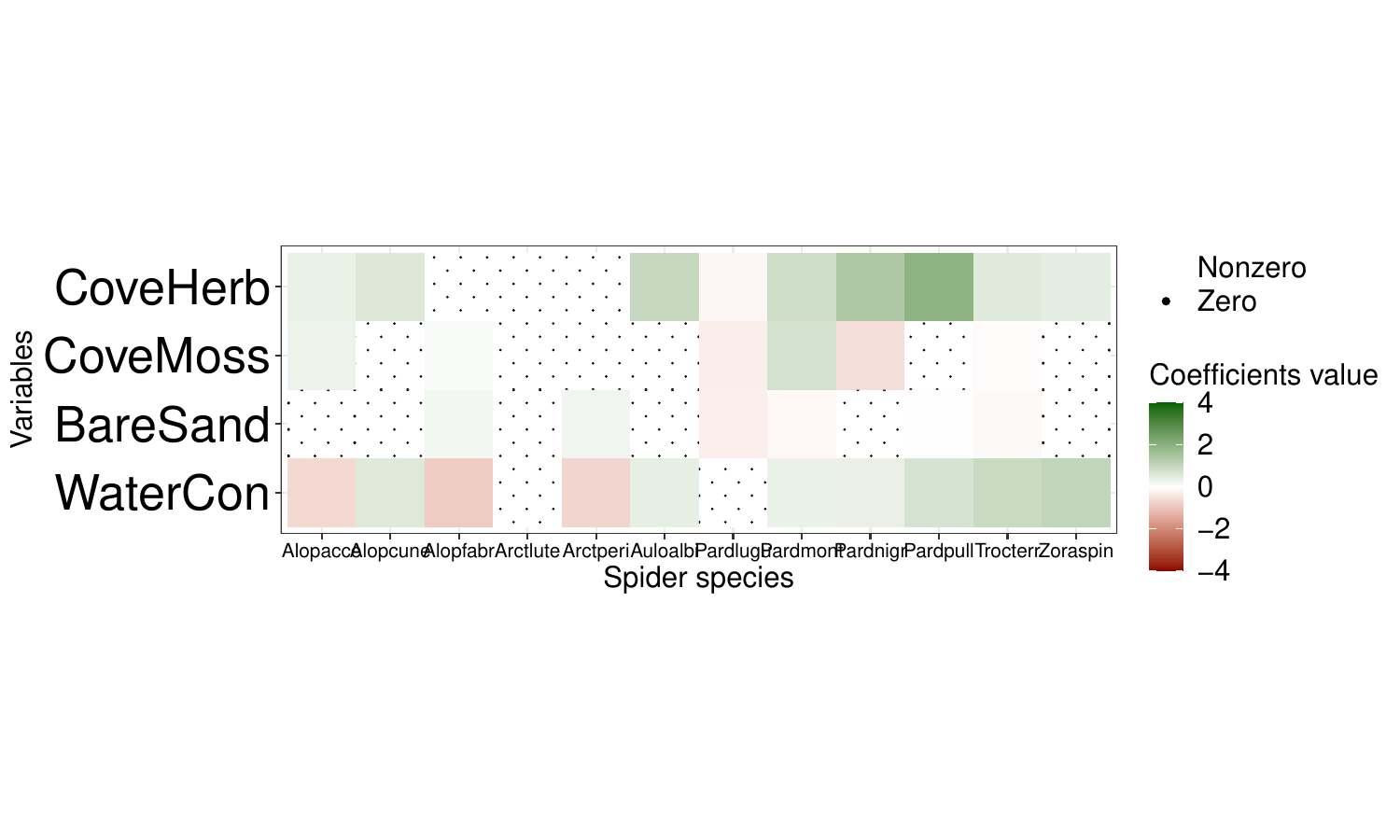}
    \caption{Coefficients estimation with GLMNET}
    \label{fig:Real_data_analyses:Spiders:Results:Coefficients_estimation:GLMNET}
\end{subfigure}
\begin{subfigure}{0.70\textwidth}
    \includegraphics[clip, trim=0mm 3cm 0mm 3cm, width=\textwidth]{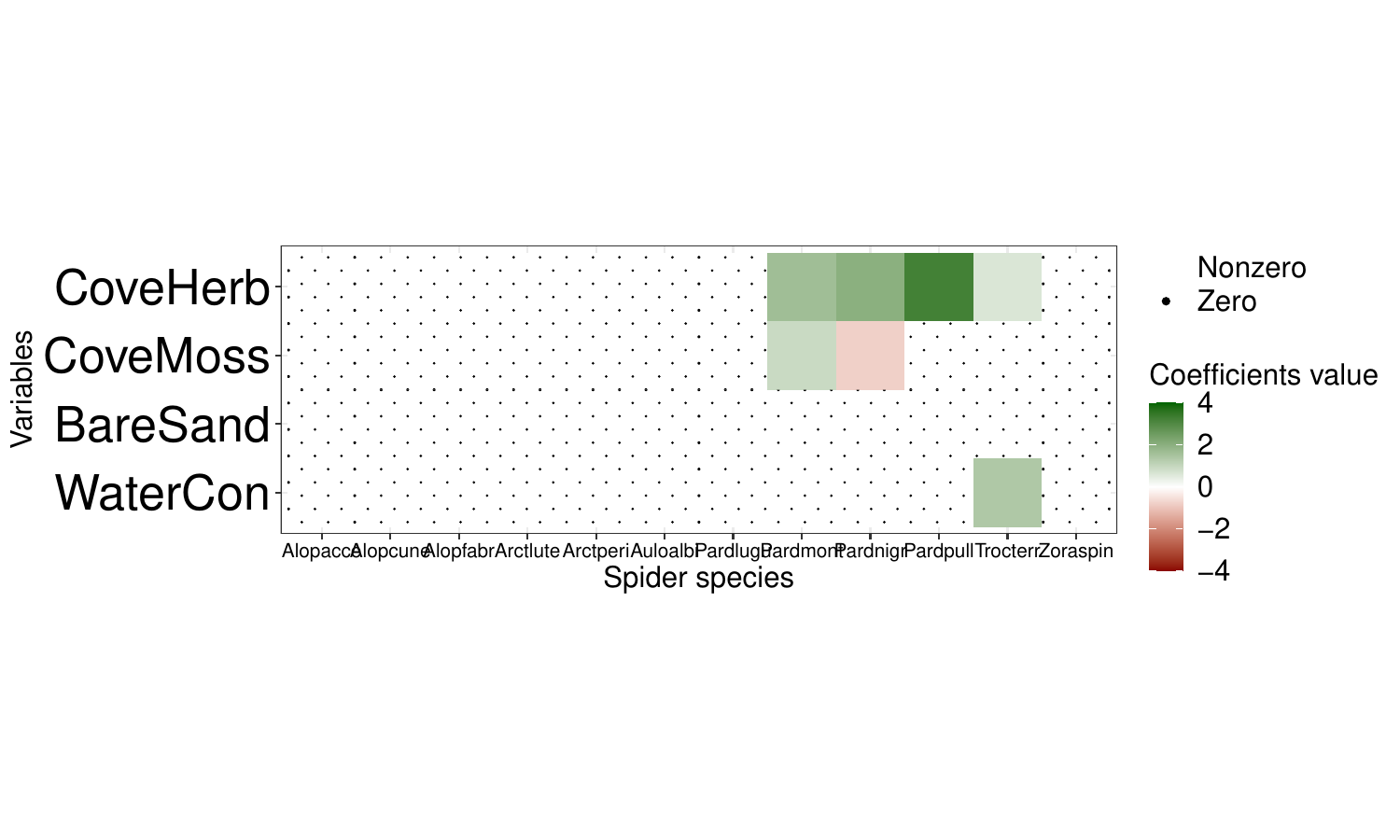}
    \caption{Coefficients estimation with SICPLN}
    \label{fig:Real_data_analyses:Spiders:Results:Coefficients_estimation:SICPLN}
\end{subfigure}   
\caption{\textsc{Coefficients estimation with PLN, GLMNET and SICPLN.}}
\label{fig:Real_data_analyses:Spiders:Results:Coefficients_estimation}
\end{figure}

\begin{figure}
\centering
\begin{subfigure}{0.47\textwidth}
   \includegraphics[width=\textwidth]{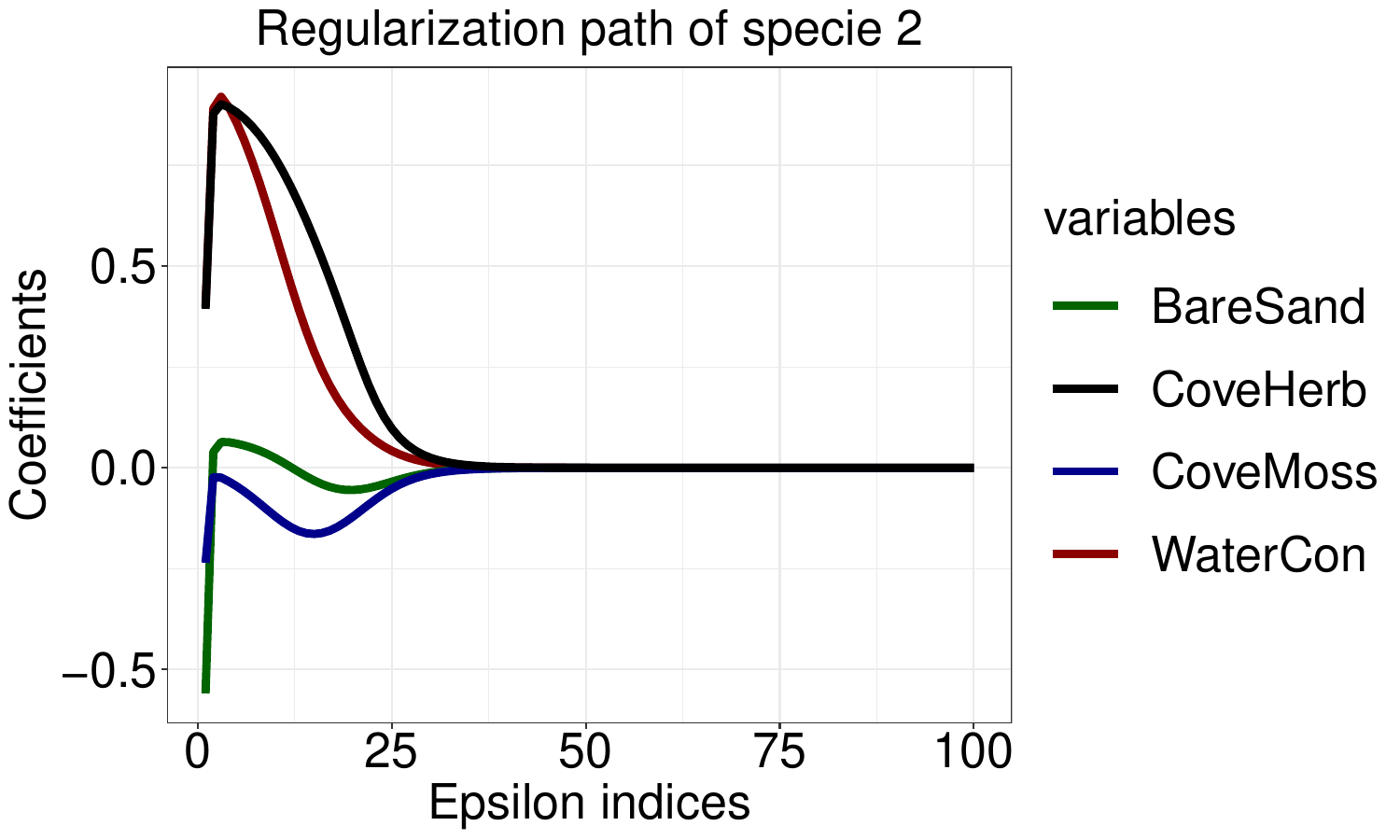}
 \caption{}
\label{fig:Real_data_nalyses:Spider:Results:regularisation_path:SICPLN:specie_2}
\end{subfigure}
\begin{subfigure}{0.47\textwidth}
   \includegraphics[width=\textwidth]{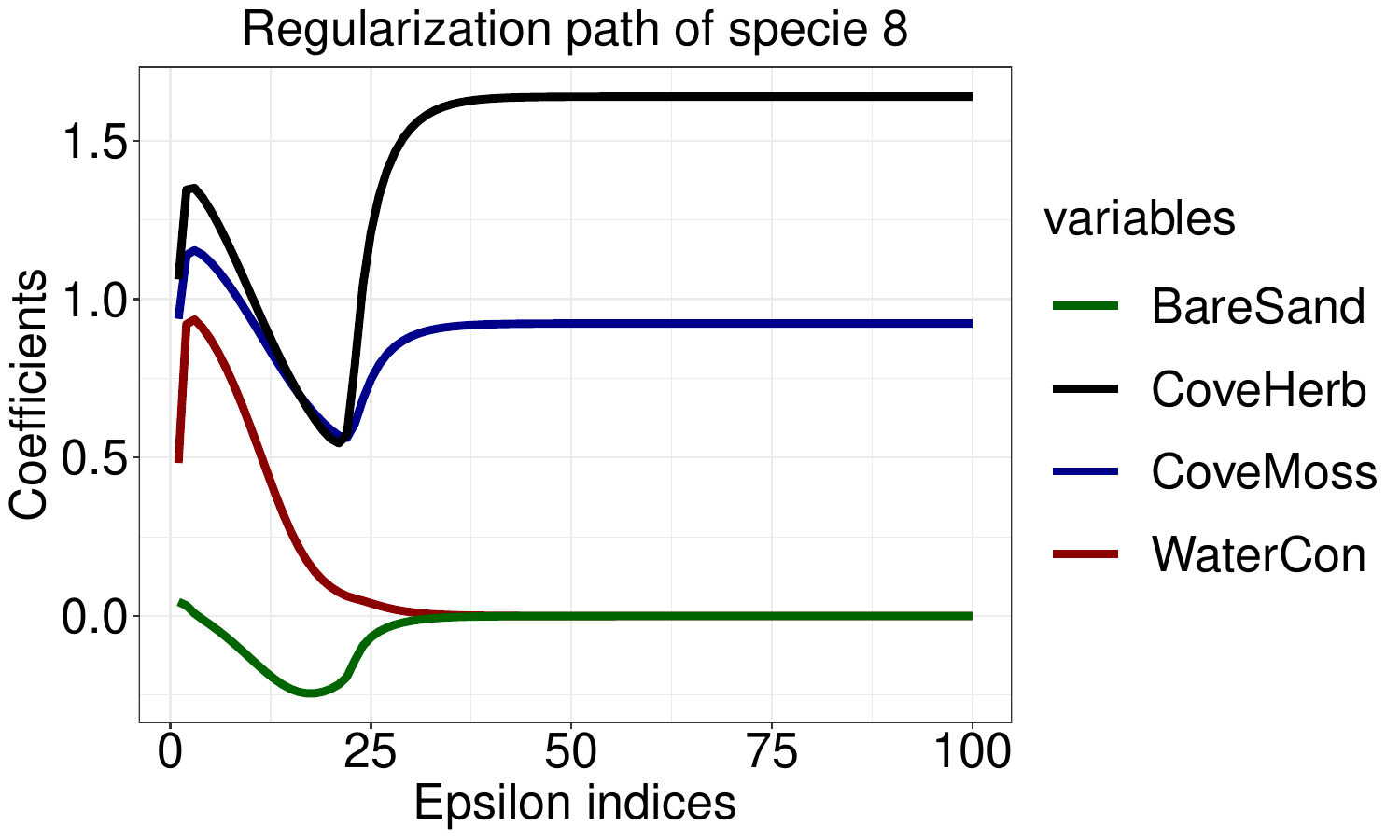}
 \caption{}
    \label{fig:Real_data_analyses:Spider:Results:regularisation_path:SICPLN:specie_8}
\end{subfigure}
\hfill
\begin{subfigure}{0.47\textwidth}
    \includegraphics[width=\textwidth]{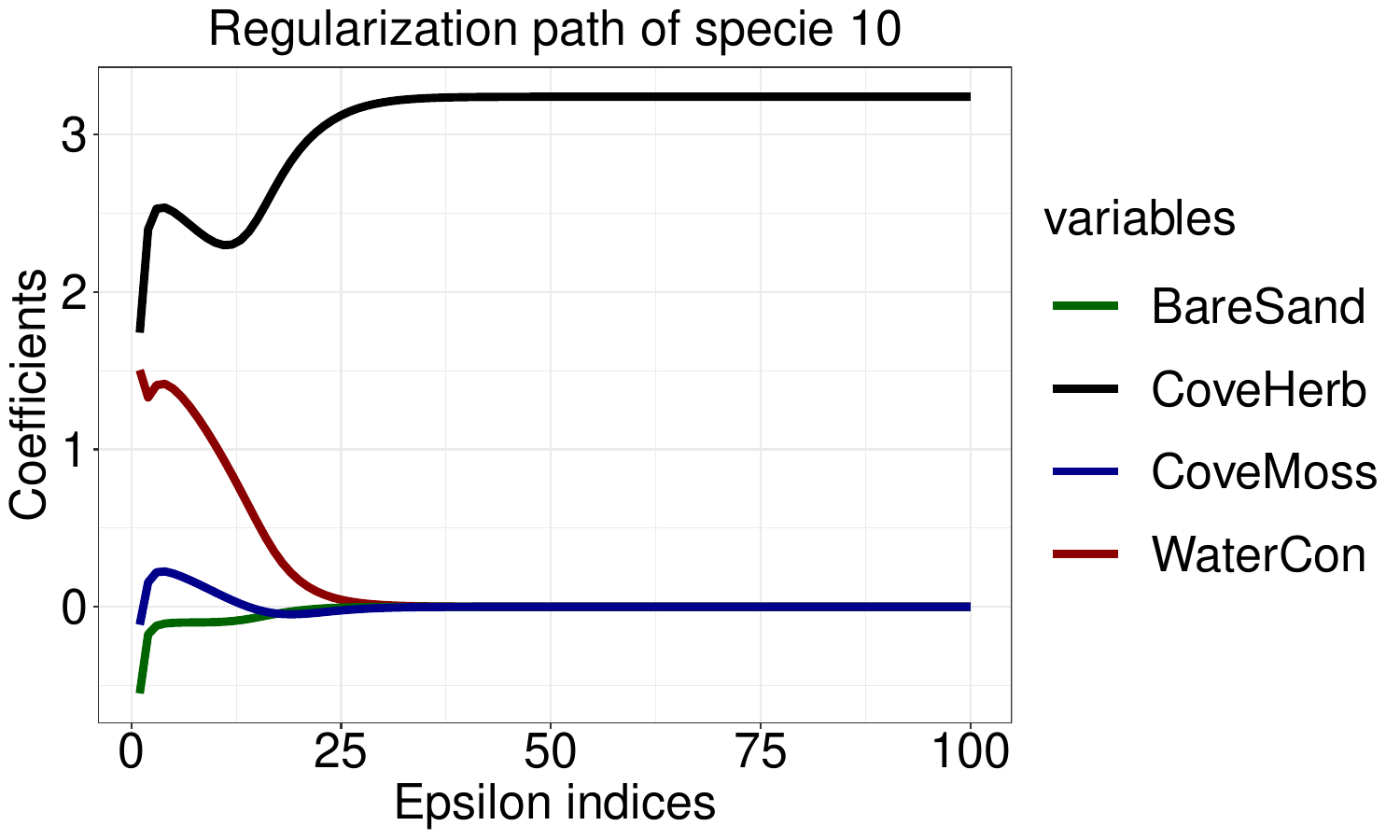}
    \caption{}
    \label{fig:Real_data_analyses:Spider:Results:regularisation_path:SICPLN:specie_10}
\end{subfigure}
\begin{subfigure}{0.47\textwidth}
    \includegraphics[width=\textwidth]{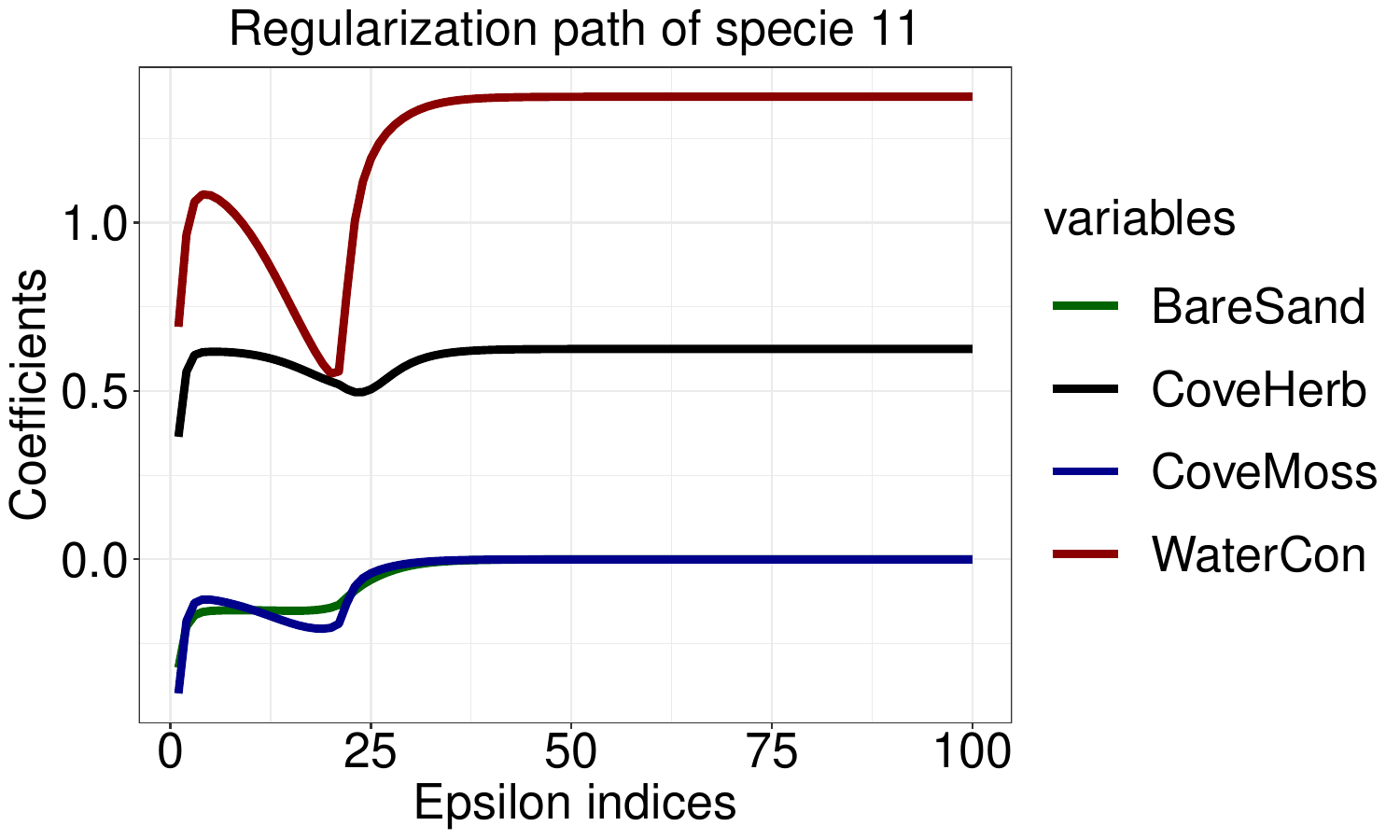}
    \caption{}
    \label{fig:Real_data_analyses:Spider:Results:regularisation_path:specie_11}
\end{subfigure}   
\caption{\textsc{Regularization of certain species in the hunting spider dataset.}}
\label{fig:Real_data_analyses:Spider:Results:regularisation_path}
\end{figure}

\begin{figure}
\centering
\begin{subfigure}{0.47\textwidth}
   \includegraphics[width=\textwidth]{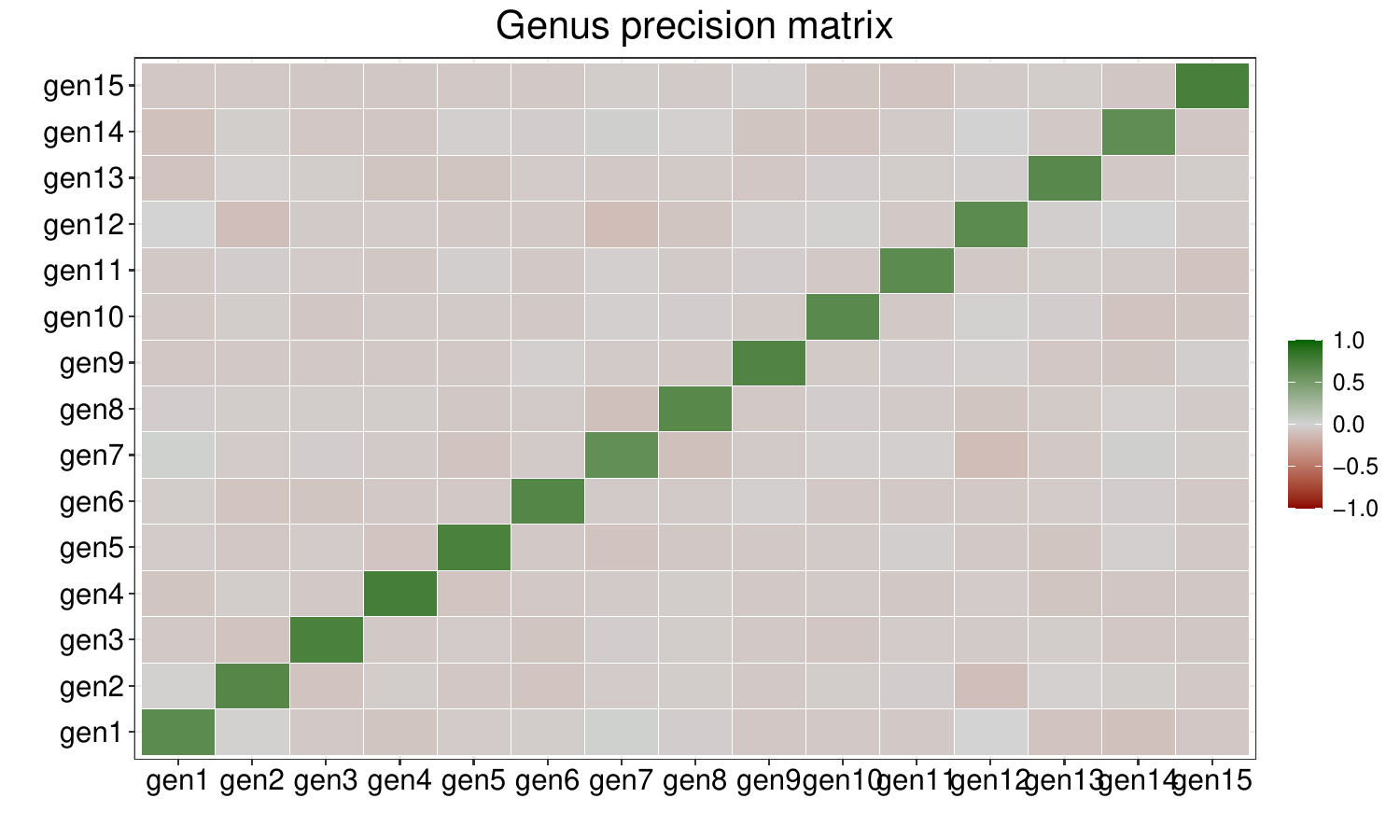}
 \caption{}
    \label{fig:Real_data_analyses:Results:precision_matrix:genus}
\end{subfigure}
\begin{subfigure}{0.47\textwidth}
   \includegraphics[width=\textwidth]{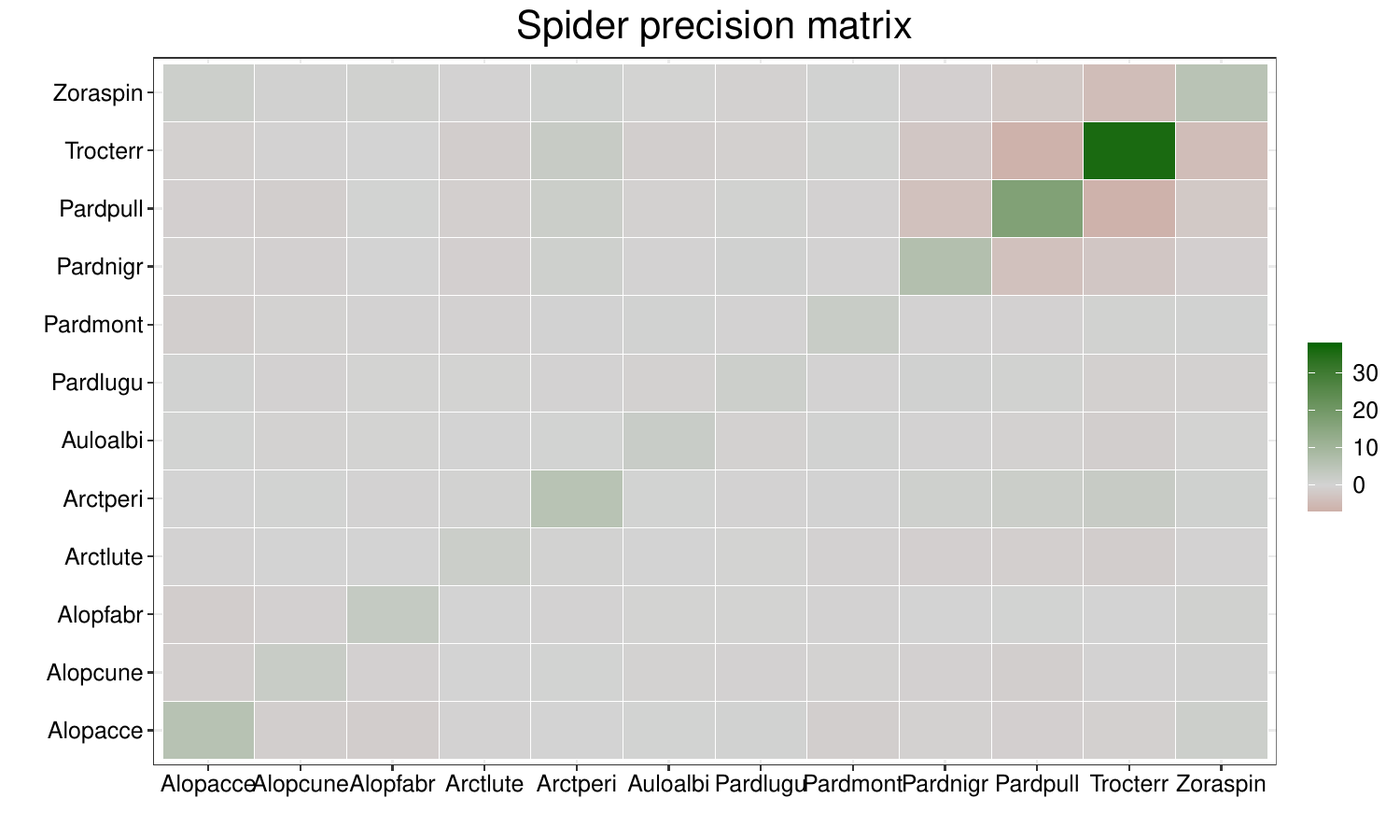}
    \caption{}
    \label{fig:Real_data_analyses:Results:precision_matrix:spider}
\end{subfigure} 
\caption{\textsc{Genus and spider species precision matrix}}
\label{fig:Real_data_analyses:Results:precision_matrix}
\end{figure}

\noindent {\bf Results.}   
After performing the VIF analysis (see Table \ref{tab:Real_data_analyses:Spiders:Results:VIF}), 
we selected the variables with a VIF coefficient lower than $5$. 
These variables include \verb|WaterCon|, \verb|BareSand|, \verb|CoveMoss|and \verb|CoveHerb|. 
Figure \Ref{fig:Real_data_analyses:Spiders:Results:Coefficients_estimation} shows that 
the estimated coefficients with PLN are all non-zero, while 
those obtained with GLMNET (univariate Poisson regression) contained $31$ non-zero coefficients out of $48$.
Conversely, our approach SICPLN provides the sparsest coefficients, with only $7$ non-zero coefficients out of $48$. 
The better performance of SICPLN compared to GLMNET in terms of sparsity 
is due to the existence of dependencies between species 
(see Figure \ref{fig:Real_data_analyses:Results:precision_matrix:spider} for the precision matrix), 
which are taken into account by SICPLN but not by GLMNET.
Figure \ref{fig:Real_data_analyses:Spider:Results:regularisation_path} 
shows the SICPLN regularization path used to select the most relevant variables 
to explain the abundance of four spider species. 
These results highlight that variables \verb|CoveHerb| and \verb|CoveMoss| is selected for species 8 (\verb|Pardmont|), 
variables \verb|CoveHerb| and \verb|WaterCon| is selected to explain the abundance of specie 11 (\verb|Trocterr|), 
while only \verb|CoveHerb| are selected for specie 10 (\verb|pardpull|). 
No variable is considered relevant to explain the abundance of specie 2 (\verb|Alopcune|).

%
\section{Discussion}
\label{sect-discussion}



In this article, we present a method for variable selection in a PLN model, providing both theoretical and practical assurances regarding its effective application.
In particular, a classical frequentist theoretical analysis is given  by presenting an oracle inequality, highlighting the asymptotic behavior as the sample size increases.
Concerning the proposed implementation, the estimation procedure involves an algorithm ($\varepsilon$-telescoping) that we combine with the PLN fitting procedure, and that progressively reduces $\varepsilon$ to 0, aiming to closely approximate the $L_0$-penalty and thereby achieve good variable selection performance.
To the best of our knowledge, it is not entirely clear theoretically what ensures variable selection as $\varepsilon$ tends to 0. One would expect to obtain a theoretical result indicating that variable selection performance increases as $\varepsilon$ approaches 0.
To begin addressing this issue, we present first results regarding this penalization approach, from a Bayesian analysis point of view. The results does not fully elucidate what happens as $\varepsilon$ tends to 0 since the prior distribution should be defined based on a fixed $\varepsilon$ in a Bayesian framework and cannot evolve, as the $\varepsilon$-telescoping algorithm does in the frequentist approach \cite{o2023variable}.
Nevertheless, in the case of a straightforward transposition of this penalization approach into a Bayesian model, Proposition~\ref{prop-SIC-formulations-bayes-epsilon} shows that if $\varepsilon$ is too close to 0, then the prior distribution reduces to a flat prior. However, when studying the prior distribution behavior with regards to the joint evolution of $\lambda$ and $\varepsilon$, with Proposition~\ref{prop-SIC-formulations-bayes-lambda_and_epsilon}, we observe a convergence rate at which the prior distribution converges to a mixture of a Dirac measure and a flat prior. Although this result corresponds to a very specific case, it demonstrates a first connection between the reduction of $\varepsilon$ towards 0 and a Bayesian mechanism of variable selection.
In practice, from Section~\ref{sect-simulation}, we observe that the proposed procedure yields results with comparable performance to PLN, and even outperforms it. Additionally, the resulting inference provides sparse estimates, which are highly beneficial in real-world contexts, as demonstrated in Section~\ref{sect-real_data}.

Following the work in this article, a potential future work may involve implementing this penalization method in the case of a zero-inflated PLN model, which can cover a broader range of database configurations requiring such modeling.
In the same vein, it could be possible to extend this penalization method to the precision matrix as well, with the hope of improving estimation performance.
Next, another research initiative involves developing an approach for a modification of cross-validation based on robust predictions of new data. From the results presented above, we find that prediction errors using formula "PLN prediction" in Section~\ref{sect-PLN-sparse}, are too high to ensure proper tuning parameter calibration via cross-validation. Additionally, using formula "Variational PLN prediction" in cross-validation would require obtaining variational parameters related to validation data. However, these variational parameters, for the $i$-th individual, depend, among other, on count data $Y_{1,1}, \dots, Y_{1,d}$. Therefore, it's not possible to obtain these variational parameters if the $i$-th data point is not in the training sample. Furthermore, based on our current investigations, inferring these variational parameters from the variational parameters estimated on the training sample is not feasible, as they exhibit significant variation depending on the count data $Y$.

\bibliography{SICPLN_arXiv}


\appendix

\section{Proofs and mathematical details}
\label{sect:appendix:proofs}

\subsection*{About the geometry of SIC}
\label{sect-appendix-geometry}
Consider here the case of a 2-dimensional balls related to the SIC function $\phi_\varepsilon$, so that results correspond to Figure~\ref{fig:point_of_view:geometry:balls}.
The contours of the balls associated with a value $k$ are defined by the following constraint:
$\phi_\varepsilon(y) = k - \phi_\varepsilon(x) $
and by using that the reciprocal expression of $\phi_\varepsilon$ given by: $x = \pm \varepsilon \sqrt{\frac{z}{1-z}}$, 
it follows the result that:
\begin{equation*}
    y = 
    \begin{cases}
        0 & \text{ if } x= \pm \varepsilon \sqrt{\frac{k}{1-k}}, \\
        \pm \varepsilon \sqrt{ \frac{k-\phi_\varepsilon(x)}{1-k+\phi_\varepsilon(x)} } & \text{ otherwise.}
    \end{cases}    
\end{equation*}

\subsection*{Proof of Proposition~\ref{prop-SIC-formulations-bayes-lambda}}
\label{sect-appendix-proposition-lambda}
 When $\lambda_n$ goes to $+\infty$, $c(\lambda_n, \varepsilon) $ converges to $0$ by applying the monotone convergence theorem. This means that the  weight of the flat part in  the prior $\pi_\text{SIC}^{(n)}$   goes to $0$. Let $f$ be a continuous real-valued function with compact support and $a_n=1$, we have from (\ref{eq:theoritical_result:bayes:mixture_decomposition})
 $$
  a_n\;\int f(\beta_j)\;\pi^{(n)}_\text{SIC}(\beta_j) \;d\beta_j
 =c(\lambda_n,\varepsilon) \;\int   f(\beta_j)  \;d\beta_j 
 +\int f(\beta_j)\; \tilde\pi^{(n)}(\beta_j) \;d\beta_j.
 $$
 Since $f$ is bounded with compact support, the first part converges to $0$.
 Since  $\tilde\pi^{(n)}$ converges narrowly  to $\delta_0$, the second part converges to $f(0)$.    Therefore,   $\pi_\text{SIC}(\beta_j)$ converges q-vaguely to a flat prior. Therefore,   $\pi_\text{SIC}^{(n)}(\beta_j)$ converges q-vaguely to a Dirac measure in $0$.  
 
\subsection*{Proof of Proposition~\ref{prop-SIC-formulations-bayes-epsilon}}

Put $a_n=1/c(\lambda, \varepsilon_n)$.  When $\varepsilon_n$ converge to $0$, $c(\lambda, \varepsilon_n) $ converges to  $+\infty$ and therefore $a_n$ converge to $0$. 
We have  
 
$$
 a_n\;\int f(\beta_j)\;   \pi^{(n)}_\text{SIC}(\beta_j) \;d\beta_j
=\int   f(\beta_j)\,  d\beta_j 
+a_n\;\int f(\beta_j)\; \tilde\pi^{(n)}(\beta_j) \;d\beta_j .
$$
 
Since $f(\beta)$ is bounded and $\tilde \pi $ is a probability distribution, the second part goes to $0$. The first part is the integral of $f$ w.r.t. the flat prior $\pi(\beta_j)\propto 1$. Therefore,   $\pi_\text{SIC}^{(n)}(\beta_j)$ converges q-vaguely to a flat prior.

\subsection*{Proof of Proposition~\ref{prop-SIC-formulations-bayes-lambda_and_epsilon}}
\label{sect-appendix-proposition-lambda_and_epsilon}

 Denote $l_n=\lambda_n/2\sigma^2$.
 
\noindent\textit{\underline{Part 1}} : we prove that 
$$
 c(l_n, \varepsilon_n) \xrightarrow[n \rightarrow +\infty]{}\frac{1}{\sqrt{\pi}K}.
$$
We have
\begin{align*}
 c(\lambda_n, \varepsilon_n)^{-1} =& \int_{-\infty}^{+\infty} \left(  \exp\left\{  l_n   \times\frac{\varepsilon_n^2}{ \beta^2 + \varepsilon_n^2} \right\}-1 \right)d\beta  
 =   \int_{-\infty}^{+\infty}\varepsilon_n\;\left(\exp\left\{l_n \frac{1}{1+y^2}\right\}-1\right) d y\quad (\textrm{with } y=\beta/\varepsilon)
 \\
 =&    {K}{ \sqrt{l_n}}\int_{-\infty}^{+\infty}\left(\exp\left\{-l_n \frac{y^2}{1+y^2}\right\}-\exp\{-l_n)\}\right) d y 
 \\
 =& 2   {K}{ \sqrt{l_n}}\int_{0}^{+\infty}\left(\exp\left\{-l_n \frac{y^2}{1+y^2}\right\}-\exp\{-l_n)\}\right) d y
 \\
 =&   {K}{ \sqrt{l_n}}\int_{0}^{l_n}l_n^{-1/2} (1-z/l_n)^{-3/2} z^{-\frac{1}{2}}
 l_n^{-1/2}\left(\exp\left\{-z\right\}-\exp\{-l_n\}\right) d z \quad (\textrm{with } z=\frac{l_n y^2}{1+y^2})\\
  =&    {K} \int_{0}^{l_n/2} (1-z/l_n)^{-3/2} z^{-\frac{1}{2}}
\left(\exp\left\{-z\right\}-\exp\{-l_n\}\right) d z \\
 &+  {K}  \int_ {l_n/2}^{l_n}(1-z/l_n)^{-3/2} z^{-\frac{1}{2}}
\left(\exp\left\{-z\right\}-\exp\{-l_n\}\right) d z 
  \\
   =&K (A+B)
   \end{align*}    
Now, we show that the second integral $B$ converge to $0$. Note that for  $l_n/2<z<l_n$, we have $\left|\frac{\exp\{-z\}-\exp\{-l_n\}}{z-l
_n}\right|\leq \exp\{-\lambda/2\}$ and then, with $w=z/l_n$, 

 \begin{align*}
B= \int_ {l_n/2}^{l_n}(1-z/l_n)^{-3/2} z^{-\frac{1}{2}}
\left(\exp\left\{-z\right\}-\exp\{-l_n\}\right) d z &\leq  l_n\exp\{-l_n/2\}\int_ {l_n/2}^{l_n} (1-z/l_n)^{-1/2} z^{-\frac{1}{2}}dz  \\
& =l_n^{\frac{3}{2}}\exp\{-\frac{l_n}{2}\} \int_{1/2}^1\frac{1}{(1-w)w} dw\\
&\xrightarrow[l_n\rightarrow +\infty]{} 0
 \end{align*} 
  Nxt, we show that the first integral $A$ converge to $\sqrt{\pi}$. For $0\leq z \leq l_n/2$ we have 
$\left|\frac{\exp\{-z\}-\exp\{-l_n\}}{\left(1-\frac{z}{l_n}\right)^{3/2}z^{1/2}} \right|\leq \frac{2^{3/2}\exp\{-z\}}{z^{1/2}}$. Therefore, we can apply the dominated convergence theorem: 

$$
A=\int_{0}^{l_n/2} (1-z/l_n)^{-3/2} z^{-\frac{1}{2}}
\left(\exp\left\{-z\right\}-\exp\{-l_n\}\right) d z \xrightarrow[l_n \rightarrow +\infty]{}\int_0^{+\infty} z^{-1/2} \exp\{-z\} dz =\sqrt{\pi} 
$$

\noindent\textit{\underline{Part 2}} : we prove that 
$\tilde \pi^{(n)}$ converge narrowly to the Dirac measure $\delta_0$. It is sufficient to prove that, for any $c>0$,  $\int_{|\beta| > c} \tilde\pi^{(n)} (\beta_j) d(\beta_j)\rightarrow 0$ when $n\rightarrow +\infty$.   Similarly to Part 1, after the same change of variables, we have :

\begin{align*}
 \int_{|\beta| > c} \tilde\pi^{(n)}(\beta_j) d(\beta_j)=&    c(\lambda_n,\varepsilon_n)    \int_{|\beta| >c} \left(  \exp\left\{  l_n   \times\frac{\varepsilon_n^2}{ \beta^2 + \varepsilon_n^2} \right\}-1 \right)d\beta \\
 =& K  c(\lambda_n,\varepsilon_n)  \int_{l_n\frac{c^2}{\varepsilon^2+c^2}}^{l_n}    (1-z/l_n)^{-3/2} z^{-\frac{1}{2}}
\left(\exp\left\{-z\right\}-\exp\{-l_n\}\right) d z 
\end{align*}
For $l_n\frac{c^2}{\varepsilon_n^2+c^2}<z<l_n$, we have $\left|\frac{\exp\{-z\}-\exp\{-l_n\}}{z-l
_n}\right|\leq \exp\{-l_n \frac{c^2}{c^2+\varepsilon_n^2}\}$.  Then
\begin{align*}
   \int_{l_n\frac{c^2}{\varepsilon_n^2+c^2}}^{l_n}  &   (1-z/l_n)^{-3/2} z^{-\frac{1}{2}}
\left(\exp\left\{-z\right\}-\exp\{-l_n\}\right) d z
\\
&\leq l_n\exp\{-l_n\frac{c^2}{c^2+\varepsilon_n^2} \} \int_{l_n\frac{c^2}{\varepsilon_n^2+c^2}}^{l_n}    (1-z/l_n)^{-1/2} z^{-\frac{1}{2}}dz
\\
&\leq l_n\exp\{-l_n\frac{c^2}{c^2+\varepsilon_n^2} \} \int_{0}^{l_n}    (1-z/l_n)^{-1/2} z^{-\frac{1}{2}}dz
\\
&= l_n^{3/2} \exp\{-l_n\frac{c^2}{c^2+\varepsilon_n^2} \} \int_0^1 (1-w)^{-1/2}w^{-1/2} dw\quad (\textrm{with } w=z/l_n) 
\\
& \xrightarrow[n\rightarrow + \infty]{} 0
\end{align*}
\noindent\textit{\underline{Part 3}:} 
let $f$ be a continuous real-valued function with compact support and $a_n=1$. From (\ref{eq:theoritical_result:bayes:mixture_decomposition}): 

\begin{align*}
    a_n\;\int f(\beta_j)\;\pi^{(n)}_\text{SIC}(\beta_j) \;d\beta_j
 &=c(\lambda_n,\varepsilon_n) \;\int   f(\beta_j)  \;d\beta_j 
 +\int f(\beta_j)\; \tilde\pi^{(n)}(\beta_j) \;d\beta_j \\
 &\xrightarrow[n\rightarrow + \infty]{} \frac{1}{K\sqrt \pi}\int   f(\beta_j)  \;d\beta_j  +  f(0)
\end{align*} 
 The result follows.

\subsection*{Proof of Lemma \ref{lemma:theoritical_result:equivalence_norms}}
\label{sect-appendix-lemma-norm}

   \noindent \textit{First inequality.} Since $\varepsilon^2 \leq x_j^2 + \varepsilon^2$ we have 
   \begin{equation*}
       \frac{1}{x_j^2+ \varepsilon^2} \leq \frac{1}{ \varepsilon^2} \quad \text{ and then} \quad \frac{x_j^2}{x_j^2+ \varepsilon^2}  \leq \frac{x_j^2}{ \varepsilon^2} 
   \end{equation*}
   which implies that $\|x \|_{0,\varepsilon} \leq \frac{1}{\varepsilon^2} \| x\|^2_2$.
   
   \noindent \textit{Second inequality.} Consider the target inequality: $\|x \|_{0,\varepsilon} \leq k \| x\|_1$ for a value $k$, which does not depend on $x$. Then, notice that this inequality holds if for each coordinate $x_j$, and if $x_j \neq 0$: 
   \begin{equation*}
       \frac{x_j^2}{x_j^2+\varepsilon^2} \leq k |x_j| 
       \quad \Leftrightarrow \quad 
       \frac{1}{|x_j| + \frac{\varepsilon^2}{|x_j|}} \leq k. 
   \end{equation*}
   Consider the function $f(y) = (y + \frac{\varepsilon^2}{y})^{-1}$, for $y> 0$, for which the first derivative is $f'(y) = (\varepsilon^2-y^2) (y^2+\varepsilon^2)^{-2}$. It is straightforward to see that for all $y>0$, $f(y) \leq f(\varepsilon) = \frac{1}{2\varepsilon}$, and then the result follows with $k = \frac{1}{2\varepsilon}$.
   
   \noindent \textit{Third inequality.} Note that we can write $\| x\|_{0, \varepsilon} = a^T b$, where $a$ and $b$ are $p$-dimensional vectors, those the $j^\text{th}$ elements are $a_j = |x_j|$ and $b_j = (|x_j| + \frac{\varepsilon^2}{|x_j|})^{-1}$, if $x_j \neq 0$. In the case of $x_j =0$, then $a_j = 0$ and $b_j =0$. First, consider the $p$-dimensional vector $c$ those the $j^\text{th}$ element is $c_j  = \frac{1}{2\varepsilon}$ and remark that $\| b \|_2 \leq \| c \|_2$, and that $\| c \|_2 = \frac{\sqrt{p}}{2\varepsilon}$. 
   Second, note that $\| a\|_2 = \|x \|_2$.
   Then, using the Cauchy-Schwarz inequality, we have that:
   \begin{equation*}
       \| x\|_{0, \varepsilon} \leq \|a \|_2 \|b \|_2 \leq \|x \|_2 \|c \|_2 \leq \frac{\sqrt{p}}{2\varepsilon} \|x \|_2
   \end{equation*}

\subsection*{Proof of Basic Inequality \ref{theorical:Basic_inequality}}
\label{sect-appendix-lemma-basic_inequality}
 By definition of $\widehat{\m{\beta}}$ from Equation~\eqref{SIC:lagrange}, we have the following inequality:
    \begin{align}
        \frac{\lVert \m{y}-\m{X}\widehat{\m{\beta}} \rVert _{2}^{2}}{n}+ \frac{\lambda}{n} \lVert\widehat{\m{\beta}} \rVert _{0,\varepsilon}  \le \frac{\lVert \m{y}-\m{X}{\m{\beta}^0} \rVert _{2}^{2}}{n}+ \frac{\lambda}{n} \lVert{\m{\beta^{0}}} \rVert _{0,\varepsilon} 
        \label{proof:basic_inequality}
    \end{align}
    Rearranging  and by substituting $ \m{y}=\m{X \beta}^{0}+ \m{\omega}$ gives 
    \begin{align*}
        \lVert \m{y}-\m{X}\widehat{\m{\beta}} \rVert _{2}^{2}-{\lVert \m{y}-\m{X}{\m{\beta}^0} \rVert _{2}^{2}}= \lVert \m{X}(\widehat{\m{\beta}}-\m{\beta}^0) \rVert _{2}^{2}-2 \m{\omega}^{\top}\m{X}(\widehat{\m{\beta}}-\m{\beta}^{0}),
    \end{align*}
and we finally have 
    
    \begin{equation*}
        \frac{\lVert \m{X}(\widehat{\m{\beta}}-\m{\beta}^0) \rVert _{2}^{2}}{n}+ \frac{\lambda}{n} \lVert\widehat{\m{\beta}} \rVert _{0,\varepsilon}  \le \frac{2 \m{\omega}^{\top}\m{X}(\widehat{\m{\beta}}-\m{\beta}^{0})}{n}+\frac{\lambda}{n} \lVert{\m{\beta^{0}}} \rVert _{0,\varepsilon}.
      \end{equation*}

\subsection*{Proof of Lemma \ref{lemme:theoritical_results:gram_condition}}
\label{sect-appendix-lemma-gram_condition}
Posing $\kappa_j= \frac{\left\lvert \m{\omega}^{\top}\m{X}^{j}\right\rvert }{\sqrt{n \sigma^2 }}$, $\kappa_j \sim \mathcal{N}(0,1)$. We know that $\m{\omega} \sim \mathcal{N}(0,\sigma^2 I )$, then  $\mathbb{E}(\kappa_j)$ and $\mathbb{V}(\kappa_j)$ are:
    \begin{equation*}
        \mathbb{E}(\kappa_j)= \frac{1}{\sqrt{n \sigma^2}}\mathbb{E}(\m{\omega}^\top)\m{X}^{j}=0 \;  \text{and} \; \mathbb{V}(\kappa_j)=  \frac{1}{n \sigma^2} (\m{X}^{(j)})^{\top}\mathbb{V}(\m{\omega^\top})\m{X}^{(j)}= \frac{1}{n \sigma^2}(\m{X}^{(j)})^{\top} \sigma^2 I_n\m{X}^{(j)} = \widehat{\sigma}^2=1.
    \end{equation*}
By Gaussian concentration inequality we know that if $\m{X} \sim \mathcal{N}(\mu,\tau^2 )$, then for all $x >0$,
\begin{equation*}
    \mathbb{P}\left(\lvert\m{X}-\mu\rvert>x \right) \le 2 \exp \left(-\frac{x^2}{2 \tau^2}\right),
\end{equation*} 
applying this to $\kappa_j \sim \mathcal{N}(0,1)$ yields to 
\begin{equation*}
    \mathbb{P}\left(\lvert \kappa_j \rvert > x \right)=\mathbb{P}\left(\lvert \kappa_j-\mathbb{E}(\kappa_j)\rvert>x \right) \le 2 \exp \left(-\frac{x^2}{2}\right), \; \text{for} \;j=1,\dots,d.
\end{equation*}
Then, we can derive the following inequality:
\begin{equation*}
    \mathbb{P}\left(\max_{j=1,\dots,d}\lvert \kappa_j \rvert > x \right)=\mathbb{P}\left(\bigcup \lbrace\lvert \kappa_j \rvert > x \rbrace \right)=\sum_{j=1}^{d}\mathbb{P}\left(\lvert \kappa_j \rvert>x \right) \le 2p\,  \exp \left(-\frac{x^2}{2}\right), \; \text{for} \;j=1,\dots,d.
\end{equation*}
For $x=\sqrt{t^2 + 2 \log (p)} >0$ we obtain for each $j=1,\dots,d$:
\begin{align*}
     &\mathbb{P}\left(\| \frac{ \m{\omega}^{\top}\m{X}^{j} }{\sqrt{n \sigma^2 }}\Vert_{\infty}  > \sqrt{t^2 + 2 \log (p)} \right)  \le 2p \exp \left(-\frac{{t^2 + 2 \log (p)} }{2}\right), \\
     \Leftrightarrow  \;&\mathbb{P}\left(2 \| \frac{ \m{\omega}^{\top}\m{X}^{j} }{{n}}\Vert_{\infty}  > 2 \sigma\sqrt{\frac{{t^2 + 2 \log (p)}}{n}} \right)  \le 2 \exp \left(-\frac{{t^2 } }{2}\right), \\
     \Leftrightarrow \; &\mathbb{P}\left(2 \| \frac{ \m{\omega}^{\top}\m{X}^{j} }{{n}}\Vert_{\infty}  \le \lambda_0\right)  \ge 1-2 \exp \left(-\frac{{t^2 } }{2}\right).
\end{align*}

\subsection*{Proof of Corollary \ref{cor:theoritical_results:consistency}}
\label{sect-appendix-corollary-consistency}
Using Basic Inequality \ref{theorical:Basic_inequality}, for $\lambda> 2\lambda_0$, with a probability of at least $1-2 \exp \left(-\frac{{t^2 } }{2}\right)$ we have:
    
    \begin{align*}
        \frac{\lVert \m{X}(\widehat{\m{\beta}}-\m{\beta}^0) \rVert _{2}^{2}}{n}+ \frac{\lambda}{n} \lVert\widehat{\m{\beta}} \rVert _{0,\varepsilon}  &\le \lVert\frac{2  \m{\omega}^{\top}\m{X}  }{n}\rVert_{\infty}\rVert \widehat{\m{\beta}}- \m{\beta}^{0} \rVert_{1}+\frac{\lambda}{n} \lVert{\m{\beta^{0}}} \rVert _{0,\varepsilon} 
        \le {\lambda_0 \rVert \widehat{\m{\beta}}- \m{\beta}^{0} \rVert_{1}}+\frac{\lambda}{n} \lVert{\m{\beta^{0}}} \rVert _{0,\varepsilon} \\
        & 
        \le \frac{\lambda}{2} \left(\lVert{\m{\widehat{\beta}}} \rVert _{1}+\lVert{\m{{\beta}^0}} \rVert _{1}\right)+\frac{\lambda}{n} \lVert{\m{\beta^{0}}} \rVert _{0,\varepsilon} \\
        \frac{2\lVert \m{X}(\widehat{\m{\beta}}-\m{\beta}^0) \rVert _{2}^{2}}{n} & \le {\lambda} \left(\lVert{\m{\widehat{\beta}}} \rVert _{1}+\lVert{\m{{\beta}^0}} \rVert _{1}\right)+ \frac{\lambda}{n \varepsilon}\left(\lVert{\m{\beta^{0}}} \rVert _{1}- \lVert\widehat{\m{\beta}} \rVert _{1}\right) \\
        & \le \lambda\left[\left(1-\frac{1}{n\varepsilon}\right)\lVert\widehat{\m{\beta}} \rVert _{1}+ \left(1+\frac{1}{n\varepsilon}\right)\lVert{\m{\beta}^0} \rVert _{1}\right] 
        \le \left(1+\frac{1}{n\varepsilon}\right)\lVert{\m{\beta}^0} \rVert _{1}
      \end{align*}
      
To prove the large probability at least $1-\alpha$ we use
    \begin{equation*}
        \m{Y}=\frac{2\lVert \m{X}(\widehat{\m{\beta}}-\m{\beta}^0) \rVert _{2}^{2}}{n \left(1+\frac{1}{n\varepsilon}\right)\lVert{\m{\beta}^0} \rVert _{1}}, 
    \end{equation*}
thus
    \begin{align*}
        \mathbb{P}(\m{Y} \ge \lambda)&=\mathbb{P}(\m{Y} \le \lambda | \lambda >2\lambda_0)\mathbb{P}( \lambda >2\lambda_0)+ \mathbb{P}(\m{Y} \le \lambda | \lambda \le 2\lambda_0)\mathbb{P}( \lambda \le 2\lambda_0)\\
        & \le \mathbb{P}(\m{Y} \le \lambda | \lambda >2\lambda_0)+\mathbb{P}( \lambda \le 2\lambda_0) \\
        &\le 2 \exp \left(-\frac{{t^2 } }{2}\right)+\mathbb{P}( \lambda \le 2\lambda_0) 
        = 2 \exp \left(-\frac{{t^2 } }{2}\right)+\mathbb{P}( \widehat{\sigma} \le \sigma) 
    \end{align*}
Moreover, by choosing $\alpha = 2 \exp \left(-\frac{{t^2 } }{2}\right)+\mathbb{P}( \widehat{\sigma} \le \sigma)$, we have 
    \begin{align*}
        \mathbb{P}(\m{Y} \le \lambda) =1-\mathbb{P}(\m{Y} \ge \lambda)=1-2 \exp \left(-\frac{{t^2 } }{2}\right)-\mathbb{P}( \widehat{\sigma} \le \sigma) =1- \alpha.
    \end{align*}

\subsection*{Proof of Lemma \ref{lemma:theoritical_results:sparsity}}
\label{sect-appendix-corollary-sparsity}
Assume that $ \zeta = \left\{\lambda_0 \ge \left\lVert 2\m{\omega}^{\top}\m{X}\right\rVert_{\infty} \right\} $ and $\lambda \ge \lambda_0$ By application of Basic Inequality \ref{theorical:Basic_inequality} we can write,
    \begin{align}
        \label{rewrite_basic_inq}
        \frac{2\lVert \m{X}(\widehat{\m{\beta}}-\m{\beta}^0) \rVert _{2}^{2}}{n}+ \frac{\lambda}{n \varepsilon}\lVert \widehat{\m{\beta}}\rVert_1 \le \lambda\lVert{\m{\widehat{\beta}}-\m{{\beta}}^0} \rVert _{1}+\frac{\lambda}{n\varepsilon}\lVert {\m{\beta}^0}\rVert_1.
      \end{align}
      By following a similar procedure to \citet{buhlmann2011statistics}, triangle inequality enables us to write:
      \begin{align*}
        \lVert\m{\widehat{\beta}}\rVert_{1} = \lVert\m{\widehat{\beta}}_{\mathcal{S}_0}\rVert_{1}+\lvert\m{\widehat{\beta}}_{\mathcal{S}_0^c}\rVert_{1} \ge \lVert\m{\beta}^0_{\mathcal{S}_0}\rVert_{1}- \lVert \m{\widehat{\beta}}_{\mathcal{S}_0}-\m{\beta}_{\mathcal{S}_0}^0\rVert_{1} +\lVert\widehat{\m{\beta}}_{\mathcal{S}_0^c}\rVert_{1},\; 
      \end{align*}
      \begin{equation*}
        \label{beta_sparse}
        \lVert{\m{\widehat{\beta}}-\m{{\beta}}^0} \rVert _{1}=\lVert\widehat{\m{\beta}}_{\mathcal{S}_0}-\m{\beta}^0_{\mathcal{S}_0}+ \rVert _{1}+\lVert\widehat{\m{\beta}}_{\mathcal{S}_0^c}\rVert_{1}
      \end{equation*}
Then \eqref{rewrite_basic_inq} becomes: 
\begin{align*}
    \frac{2\lVert \m{X}(\widehat{\m{\beta}}-\m{\beta}^0) \rVert _{2}^{2}}{n}+\frac{ \lambda}{n\varepsilon}\left[\lVert\m{\beta}^0_{\mathcal{S}_0}\rVert_{1}- \lVert \m{\widehat{\beta}}_{\mathcal{S}_0}-\m{\beta}_{\mathcal{S}_0}^0\rVert_{1} +\lVert\widehat{\m{\beta}}_{\mathcal{S}_0^c}\rVert_{1}\right] & \le \lambda\left[\lVert\widehat{\m{\beta}}_{\mathcal{S}_0}-\m{\beta}^0_{\mathcal{S}_0}+ \rVert _{1}+\lVert\widehat{\m{\beta}}_{\mathcal{S}_0^c}\rVert_{1}\right] +\frac{\lambda}{n\varepsilon}\lVert {\m{\beta}^0}_{\mathcal{S}_0}\rVert_1 \\
     \frac{2\lVert \m{X}(\widehat{\m{\beta}}-\m{\beta}^0) \rVert _{2}^{2}}{n} +\lambda \left(\frac{1}{n\varepsilon}-1\right)\lVert\widehat{\m{\beta}}_{\mathcal{S}_0^c}\rVert_{1} & \le \lambda\left(\frac{1}{n\varepsilon}+1\right) \lVert \m{\widehat{\beta}}_{\mathcal{S}_0}-\m{\beta}_{\mathcal{S}_0}^0\rVert_{1}
\end{align*}
and the result follows thanks to Cauchy-Schwarz inequality.

\subsection*{Proof of Theorem \ref{theorem:theoritical_results:sparsity}}
\label{sect-appendix-thm-sparsity}
From Basic Inequality \ref{theorical:Basic_inequality} and Lemma \ref{lemma:theoritical_results:sparsity}, we have:
    \begin{align*}
    \frac{2\lVert \m{X}(\widehat{\m{\beta}}-\m{\beta}^0) \rVert _{2}^{2}}{n} +\lambda \left(\frac{1}{n\varepsilon}-1\right)\lVert\widehat{\m{\beta}}_{\mathcal{S}_0^c}\rVert_{1} \le \lambda\left(\frac{1}{n\varepsilon}+1\right) \lVert \m{\widehat{\beta}}_{\mathcal{S}_0}-\m{\beta}_{\mathcal{S}_0}^0\rVert_{1} 
    \end{align*}
    this is equal to:
    \begin{align*}
        \frac{2\lVert \m{X}(\widehat{\m{\beta}}-\m{\beta}^0) \rVert _{2}^{2}}{n} + \lambda \left(\frac{1}{n\varepsilon}-1\right)\lVert \m{\widehat{\beta}}-\m{\beta}^0\rVert_{1} \le
        & \lambda \left(\frac{1}{n\varepsilon}-1\right)\lVert \m{\widehat{\beta}}-\m{\beta}^0\rVert_{1}+ \lambda\left(\frac{1}{n\varepsilon}+1\right) \lVert \m{\widehat{\beta}}_{\mathcal{S}_0}-\m{\beta}_{\mathcal{S}_0}^0\rVert_{1} \\
        & -\lambda \left(\frac{1}{n\varepsilon}-1\right)\lVert\widehat{\m{\beta}}_{\mathcal{S}_0^c}\rVert_{1}
    \end{align*}
     By replacing $\lVert{\m{\widehat{\beta}}-\m{{\beta}}^0} \rVert _{1}=\lVert\widehat{\m{\beta}}_{\mathcal{S}_0}-\m{\beta}^0_{\mathcal{S}_0}+ \rVert _{1}+\lVert\widehat{\m{\beta}}_{\mathcal{S}_0^c}\rVert_{1}$ and using the compatibility condition in the right term we have
    \begin{align*}
        \frac{2\lVert \m{X}(\widehat{\m{\beta}}-\m{\beta}^0) \rVert _{2}^{2}}{n} + \lambda \left(\frac{1}{n\varepsilon}-1\right)\lVert \m{\widehat{\beta}}-\m{\beta}^0\rVert_{1} 
        &\le \lambda \left(\frac{1}{n\varepsilon}-1\right)\lVert \m{\widehat{\beta}}_{\mathcal{S}_0}-\m{\beta}_{\mathcal{S}_0}^0\rVert_{1}+ \lambda\left(\frac{1}{n\varepsilon}+1\right) \lVert \m{\widehat{\beta}}_{\mathcal{S}_0}-\m{\beta}_{\mathcal{S}_0}^0\rVert_{1} \\
       & \le \frac{2\lambda}{n\varepsilon}\lVert \m{\widehat{\beta}}_{\mathcal{S}_0}-\m{\beta}_{\mathcal{S}_0}^0\rVert_{1} 
       \le  \frac{2\lambda}{n\varepsilon}\frac{\lVert \m{X}(\m{\widehat{\beta}}_{\mathcal{S}_0}-\m{\beta}_{\mathcal{S}_0}^0) \rVert _{2} \sqrt{s_0}}{\sqrt{n}\eta}
    \end{align*}
By using $2uv \le u^2+2v^2$ and by posing, 
 $u=\frac{\lVert \m{X}(\m{\widehat{\beta}}_{\mathcal{S}_0}-\m{\beta}_{\mathcal{S}_0}^0) \rVert _{2}}{\sqrt{n}}$ 
    and $v=\frac{\lambda \sqrt{s_0}}{n\varepsilon \eta}$, we have 
    \begin{equation*}
        \frac{\lVert \m{X}(\widehat{\m{\beta}}-\m{\beta}^0) \rVert _{2}^{2}}{n} + \lambda \left(\frac{1}{n\varepsilon}-1\right)\lVert \m{\widehat{\beta}}-\m{\beta}^0\rVert_{1} 
        \le
        \frac{2\lambda^2{s_0}}{(n\varepsilon \eta)^2}.
    \end{equation*}

\subsection*{Proof of Oracle Inequality \ref{cor:theortical_result:oracle}}
\label{sect-appendix-oracle_inequality}
From
    \begin{align*}
        & \frac{\lVert \m{X}(\widehat{\m{\beta}}-\m{\beta}^0) \rVert _{2}^{2}}{n} + \lambda \left(\frac{1}{n\varepsilon}-1\right)\lVert \m{\widehat{\beta}}-\m{\beta}^0\rVert_{1} 
        \le
        \frac{2\lambda^2{s_0}}{(n\varepsilon \eta)^2}, 
    \end{align*}
we can write
    \begin{align*}
        & \frac{\lVert \m{X}(\widehat{\m{\beta}}-\m{\beta}^0) \rVert _{2}^{2}}{n} 
        \le
        \frac{2\lambda^2{s_0}}{(n\varepsilon \eta)^2}
    \end{align*}
    for $\lambda=4\widehat{\sigma}\sqrt{\frac{t^2+2 \log (p)}{n}}$, and $t=\sqrt{2 \log(p)}$
    \begin{align*} 
        \frac{\lVert \m{X}(\widehat{\m{\beta}}-\m{\beta}^0) \rVert _{2}^{2}}{n} &
        \le
        \frac{1}{\varepsilon^2}\frac{2\lambda^2{s_0}}{ (n\eta)^2} 
        \le
        \frac{1}{\varepsilon^2}\frac{128 \sigma^2 \log(p){s_0}}{ (n\eta)^2}
    \end{align*}
By posing $const=\frac{128}{\eta^2} \in \mathbb{R}$ we find
    \begin{equation*}
        \frac{\lVert \m{X}(\widehat{\m{\beta}}-\m{\beta}^0) \rVert _{2}^{2}}{n} 
        \le
        const. \frac{ \sigma^2 \log(p){s_0}}{ n^3 \varepsilon^2}
    \end{equation*}

\end{document}